\documentclass[showpacs,amssymb,preprint,preprintnumbers,nofootinbib,superscriptaddress]{revtex4-2}
\usepackage{amsmath}
\usepackage{amssymb}
\usepackage{graphicx}
\graphicspath{{Figures/}}
\usepackage{latexsym}
\usepackage{amsfonts}
\usepackage{url}
\usepackage{hyperref}
\usepackage{bm}
\usepackage{textcomp}
\usepackage{color}
\usepackage{bbm}
\usepackage{slashed}
\usepackage{caption}
\usepackage{epstopdf}
\usepackage{subcaption}
\usepackage{placeins}
\usepackage{xcolor}
\usepackage{cancel} 
\usepackage[normalem]{ulem}
\usepackage{tcolorbox}
\usepackage{soul}

\begin{document}
\title{Black Hole Thermodynamics via Tsallis Statistical Mechanics}

\author{Phuwadon Chunaksorn \footnote{Email: maxwelltle@gmail.com}}
\affiliation{The Institute for Fundamental Study, Naresuan University, Phitsanulok, 65000, Thailand}

\author{Ratchaphat Nakarachinda \footnote{Email: tahpahctar\_net@hotmail.com}}
\affiliation{Department of Mathematics and Computer Science, Faculty of Science, Chulalongkorn University, Bangkok 10330, Thailand}

\author{Pitayuth Wongjun \footnote{Email: pitbaa@gmail.com}}
\affiliation{The Institute for Fundamental Study, Naresuan University, Phitsanulok, 65000, Thailand}

	
\begin{abstract}
An investigation of black hole thermodynamics based on Tsallis statistical mechanics is explored through the study of the thermodynamics of a gas system located near the horizon of a black hole. In spite of the difficulty in exploring black hole thermodynamics through statistical mechanics, the entropy of the nearby gas system is found to be proportional to the black hole's horizon area using Gibbs-Boltzmann statistical mechanics. This allows us to study black hole thermodynamics by using statistical mechanics through the thermodynamic behaviors of the gas system. Since the entropy of the black hole is proportional to the horizon area, it is more suitable to use non-extensive statistical mechanics instead of the usual Gibbs-Boltzmann ones. In this work, the black hole entropy is derived based on Tsallis statistical mechanics, one of well-known non-extensive statistical mechanics. It is found that the black hole entropy gets a modification due to non-extensivity. By using such an entropy, the black hole can be stabilized due to the non-extensivity, and the bound on the non-extensive parameter is also determined.
\end{abstract}
\maketitle{}
\newpage


\section{Introduction}
\par
One of the important properties of black holes is the laws of black hole mechanics \cite{Bardeen:1973gs}, such as Hawking's area theorem, etc. Based on these laws, their formulae are similar to those of thermodynamic laws and then they inspire physicists to study black holes in terms of thermodynamics. Bekenstein had considered a black hole as just a thermodynamic system and found that the black hole entropy must be proportional to the black hole's surface area \cite{Bekenstein:1973ur,Bekenstein:1974ax}. By considering the quantum effect, a black hole can behave like a black body emitting radiation with a certain temperature, $T_{H}$, known as Hawking temperature \cite{Hawking:1975vcx}. Based on the first law of black hole mechanics, the corresponding entropy is also obtained as $S_{BH}$, called the Bekenstein-Hawking entropy. As a result, it is possible to interpret the black hole as a thermodynamic system carrying certain entropy and temperature.  The study of thermodynamic properties of black holes has been intensively investigated in order to explore characteristic phenomena and the quantum nature of spacetime.
\par
One of the important issues in the study of black hole thermodynamics is the physical interpretation of the black hole entropy or what the black hole entropy actually measures. By considering the same manner of the interpretation of matter entropy which is quantified by using statistical mechanics, it might be worthwhile to interpret the black hole entropy from a description of statistical mechanics. However, one of the obstructions to this manner is the lack of knowledge of the gravitational entropy \cite{Wallace:2009dc,Padmanabhan:2010xe,Gregoris:2021fiw}. Moreover, for the thermodynamic system in curved spacetime, it is unclear how the usual thermodynamic entropy of matter precisely relates to the gravitational entropy. Note that for a specific example in string theory such that a collection of branes turns into an extremal black hole, the Bekenstein-Hawking entropy can be derived by counting the degeneracy of BPS soliton bound states \cite{Strominger:1996sh}. Remarkably, the entropy of a system of gas near the black hole's horizon can be calculated by using statistical mechanics and found to be proportional to the horizon area \cite{Kolekar:2010py,Mirza:2011dd,Bhattacharya:2017bpl,Li:2021zuw,Sourtzinou:2022xhd}. In this work, we investigate the possible way to interpret such matter entropy as the black hole entropy \cite{Pretorius:1997wr,Oppenheim:2001az}.   
\par
The black hole entropy is proportional to its area; $S_{BH} \propto A_{h}$ rather than its volume; $S_{BH} \propto V$. Accordingly, the black hole entropy is non-extensive and non-additive. The effect of non-extensivity and non-additivity on entropy can be characterized by the emergence of long-range interaction.
Since Gibbs-Boltzmann (GB) statistical mechanics does not rely on long-range interaction, the black hole entropy should be evaluated by using other statistical mechanics associated with long-range interaction. 
\par
One of the non-extensive entropies is Tsallis entropy \cite{Tsallis:1987eu}, compatible with thermodynamic systems including gravitational interaction. Tsallis entropy obeys the pseudo-additive composition rule; $S_{q (1+2)} = S_{q (1)} + S_{q (2)} + \big[(1-q)/k_{B}\big] S_{q (1)} S_{q (2)}$ where $q>0$ is a non-extensive parameter. With the pseudo-additivity, the empirical temperature obtained from Tsallis entropy is not compatible with the zeroth law of thermodynamics. This implies that a system described by Tsallis statistical mechanics must belong to one with hydrostatic equilibrium instead of thermal equilibrium. Consequently, Tsallis statistical mechanics has been widely applied to investigate physical systems with long-range interactions, including self-gravitating systems \cite{Louis-Martinez:2010jvm,jiulin2004nonextensive,silva2005relativity,jiulin2007nonextensivity,liu2011nonextensivity,zheng2017limit}, and cosmology \cite{Sheykhi:2018dpn,Saridakis:2018unr,Lymperis:2018iuz,Jizba:2023fkp,Sheykhi_2020,Dehpour:2023dfo}. Moreover, the applications of Tsallis entropy in black hole physics have been intensively investigated. By treating Bekenstein-Hawking entropy as Tsallis entropy, the black holes' thermodynamic properties have been explored \cite{Biro:2013cra,Czinner:2015eyk,Czinner:2017tjq,Tannukij:2020njz,Promsiri:2020jga,Promsiri:2021hhv,Nakarachinda:2021jxd,Hirunsirisawat:2022fsb,Chunaksorn:2022whl}.
Black hole thermodynamics with Tsallis entropy including quantum effect have been investigated via the generalized uncertainty principle \cite{Alonso-Serrano:2020hpb,Cimidiker:2023kle}.  
Using Tsallis entropy (instead of Bekenstein-Hawking entropy) to characterize the black hole's entropy may lead to a thermodynamic inconsistency. It is based on the argument that a black hole emits Hawking radiation while maintaining a Hawking temperature \cite{Nojiri:2021czz}.
However, in the framework of generalized statistical mechanics, the temperature does not necessarily correspond to the Hawking temperature \cite{_imdiker_2023}.
The black hole's temperature associated with the generalized entropy should be defined in such a way that the algebraic relation among geometric quantities (Smarr formula) and the first law of black hole thermodynamics can be satisfied
\cite{Nakarachinda:2022gsb}.
Besides the Tsallis entropy, other generalized entropies have been extensively investigated \cite{Nojiri:2022aof,Nojiri:2022dkr,Nojiri:2022sfd,Nojiri:2022ljp,Odintsov:2022qnn,DAgostino:2024sgm,volovik2024tsalliscirtoentropyblackhole,manoharan2025reconcilingfractionalentropyblack,Lu:2024ppa}.
\par
As suggested in Refs.~\cite{Pretorius:1997wr,Oppenheim:2001az}, the entropy of the black hole may be investigated by using matter entropy near the black hole's horizon. This allows us to explore the black hole entropy via statistical mechanics. In fact, the matter entropy near the horizon has been investigated by using GB statistical mechanics \cite{Kolekar:2010py,Mirza:2011dd,Bhattacharya:2017bpl,Li:2021zuw,Sourtzinou:2022xhd}, while there has not been any investigation of non-extensive statistical mechanics in such a context. 
Among the various types of generalized entropies, e.g.,  R\'enyi \cite{Rnyi1959OnTD}, Barrow \cite{Barrow:2020tzx}, and Sharma-Mittal entropies \cite{sharma1975new}, Tsallis entropy is a suitable candidate for non-extensive entropy since it is directly compatible with non-extensive statistical mechanics.
Note that the Barrow entropy is relevant to quantum correction, rather than derived from a statistically mechanical point of view.
For R\'enyi entropy, it is a formal logarithm map of Tsallis one \cite{Biro:2011ncf}, and then they are in fact based on the same statistical mechanics.
The Sharma-Mittal entropy is more complicated than Tsallis and R\'enyi ones since it is the two-parameter extension of the Gibbs-Boltzmann entropy.

Therefore, in this work, we restrict our consideration to the one-parameter extension of Gibbs-Boltzmann statistical mechanics and aim to explore the black hole entropy through the matter entropy near the black hole's horizon by using Tsallis statistical mechanics. Note that, by considering a system of the interacting bosonic gas near the horizon via Tsallis statistical mechanics, the black hole entropy still be in the form of the Bekenstein-Hawking entropy \cite{doi:10.1142/S0217751X24500192}. This suggests that for quantum gas, the non-extensivity does not alter the form of black hole entropy. In our case, we found that the black hole entropy will get the correction due to the non-extensivity. 
\par
For GB statistical mechanics, the entropy of the gas system is claimed to be the Bekenstein-Hawking entropy of the black hole by requiring that the number of particles corresponds to the area at the black hole's horizon in the unit of the Planck area and the gas should also be sufficiently closed to the horizon. We discuss this issue in  Sec.~\ref{GB section}. In Sec.~\ref{Tsallis section}, we investigate the thermodynamic properties of classical gas by using Tsallis statistical mechanics. 
To achieve this, we have to propose a suitable relation between the internal energies for in the GB and Tsallis gas.
The black hole's $q$-entropy can be obtained by the similar manner as the GB case.
In Sec.~\ref{thermodynamics of bh}, 
the black hole thermodynamic phase space is constructed based on Tsallis statistical mechanics.
The thermodynamic stability is also investigated for various processes. 
We find that 
the black hole can be stable in a certain situation.
It may imply that there exists some classical correlation of the gas near the horizon in order to maintain the black hole to be thermodynamically stable. In Sec.~\ref{conclusion}, we summarize important results and discuss the effect of the non-extensivity on black hole thermodynamics. The units used in this work are set as the Boltzmann constant, the universal constant of gravitation, the reduced Planck constant, and the speed of light being unity, $k_{B} = G = \hbar = c = 1$, only in Sec.~\ref{thermodynamics of bh}.

\section{Black hole entropy based on Gibbs-Boltzmann statistical mechanics}
\label{GB section}
\par
In this section, we briefly review and devote to summarize the important results of the thermodynamic properties of the gas system described by usual GB statistical mechanics in curved spacetime \cite{Kolekar:2010py}.
Let us begin by assuming that there is a fancy isolated system with two subsystems, namely, a gas system and a black hole, undergoing the quasi-static process. 
In addition, they are in thermal contact with each other so there is heat transfer between them. This implies that the considered isolated system belongs to the canonical ensemble, in which all possible states of the system are in thermodynamic equilibrium. To demonstrate the aforementioned system, we impose that $N$-indistinguishable classical gas particles with a temperature $T=1/(k_B\beta)$ form as a thin spherical shell with a thickness of $H$. It is located in a static and spherically symmetric spacetime characterized by the interval:
\begin{equation}
\displaystyle
ds^{2} = -g(r) c^{2} dt^{2} + g^{-1}(r) dr^{2} + r^{2} (d\theta^{2} + \sin^{2}\theta d\phi^{2}),
\label{ds}
\end{equation}
where $g(r)$ is a horizon function. 
Note that the interval can be taken in the form of Eq.~(\ref{ds}) (i.e., the (00) and (11) components of metric tensor are dependent) when the energy-momentum tensor $T^\mu_{\,\,\,\,\nu}$ satisfies a condition: $T^{0}_{\,\,\,\,0} = T^{1}_{\,\,\,\,1}$. 
\par
A phase-space volume element for a single particle of the gas with mass $m$ and energy $E$ in the curved spacetime can be expressed as \cite{Padmanabhan:1989qn}
\begin{equation}
\displaystyle
d\mathcal{V}(E)
= \frac{(4\pi)^2}{3h^{3}}\left[\frac{r^2}{\sqrt{g(r)}}\left(\frac{E^{2}}{c^{2}g(r)}-m^{2} c^{2}\right)^{3/2}\right]dr,
\end{equation}
where $h$ is the Planck constant.
It is noted that in the above expression, the volume elements for both 3-position and 3-momentum spaces obey the spherical symmetry so that the angular sectors can be integrated out.
The leftover part of the position space is the radial one described by the coordinate $r$. 
For the (leftover) radial sector of the momentum space, the volume element is equivalent to that of the spherical hypersurface with a certain radius associated with the 4-momentum constraint: $p^2=E^{2}/(c^{2}g(r))-m^2c^2$ where $p^2$ is the square norm of the spatial components of the 4-momentum.
In addition, this volume element is invariant under the general coordinate transformation.
\par
We are interested in the scenario that the shell of the gas with a thickness of $H$ is located close to the the black hole's horizon.
The distance between the black hole's horizon and the inner radius of the shell is $L$ which is assumed to be very small compared to the thickness of the shell $H$ and horizon radius $r_{bh}$. Very close to the horizon; $r\to r_{bh}$, the horizon function $g(r)$ can be approximated:
\begin{equation}
\displaystyle
g(r\to r_{bh})\approx\frac{dg(r)}{dr}\bigg|_{r_{bh}}(r-r_{bh})=\frac{2\kappa}{c^2}(r-r_{bh}),\label{g approx}
\end{equation}
where $\kappa$ is the surface gravity of the static and spherically symmetric black hole \cite{Poisson:2009pwt}. As a result, the phase-space volume of the gas particle with the energy $E$ near the black hole's horizon can be approximately computed as
\begin{equation}
\displaystyle
\mathcal{V}(E)
\approx\frac{4\pi^2c}{3h^{3}\kappa^2}\int_{r_{bh}+L}^{r_{bh}+L+H}dr\frac{r^2}{\sqrt{r-r_{bh}}}\left(\frac{E^{2}}{r-r_{bh}}-2m^{2} c^{2}\kappa\right)^{3/2}
\approx\frac{\pi c E^3 A_h}{3h^{3}\kappa^2 L},\label{vol in L}
\end{equation}
where $A_{h} = 4 \pi r_{bh}^2$ is the black hole's surface area at its horizon. The last approximation is obtained from the series expansion of the result with the small distance $L$.
\par
Based on GB statistical mechanics, the partition function of the $N$-indistinguishable particles associated with the total energy of the system, denoted by $E_{tot}$ can be written as
\begin{eqnarray}
\displaystyle
Z = \frac{1}{N!} \int d\mathcal{V} \exp (-\beta E_{tot})=\frac{1}{N!} \prod_{i = 1}^{N} \left[\int d\mathcal{V} _{i} \exp (-\beta E_{i})\right]=\frac{1}{N!} (Z^{(1)})^{N},\label{z}
\end{eqnarray}
where the quantities with a subscript $i$ denote those belonging to the $i$-th particle and $Z^{(1)}$ is the one-particle partition function defined by
\begin{equation}
\displaystyle
Z^{(1)} = \int d\mathcal{V} \exp (-\beta E).
\end{equation}
It is important to emphasize that the partition function and the one-particle  partition function are also invariant under general coordinate transformation since the quantities $\beta E$ and $\beta E_{tot}$ are invariant.
According to the expression of the phase-space volume in Eq.~\eqref{vol in L}, the integration over phase-space can be done for all possible states of energy.
Thus, the one-particle partition function is further expressed as
\begin{eqnarray}
\displaystyle
Z^{(1)} 
=\frac{\pi c A_h}{h^3\kappa^2 L}\int^{\infty}_{0} dE E^{2} \exp(-\beta E) 
=\frac{\pi c}{h^3} \frac{A_h}{\beta^3\kappa^2 L}\Gamma(3)\label{1-par z}
\end{eqnarray}
where $\Gamma(z)$ is the gamma function.
It is also noticed that we keep the above expression in terms of the gamma function for the sake of considering generalized statistical mechanics as will be discussed in the next section.

In the rest frame of the gas particle, the distance between the inner shell and the horizon $L$ can be measured as the particle's proper length:
\begin{eqnarray}
\displaystyle
l_{loc}=\int^{r_{bh}+L}_{r_{bh}}\frac{dr}{\sqrt{g(r)}}
\approx\sqrt{\frac{2c^2L}{\kappa}}.\label{l loc}
\end{eqnarray}
The approximation for the horizon function~\eqref{g approx} is employed in the above result.
Another quantity that depends on the frame of reference is the gas temperature.
The relation between the temperature measured by a local observer in the gravitational field $T_{loc}$ and the temperature measured by an observer located far from the source of the gravitational field $T$ is described by Tolman's law \cite{PhysRev.36.1791}: 
$T_{loc}=T_{loc}(r)=T/\sqrt{-g_{00}(r)}$.
In other words, at the inner shell of the gas system, the position-independent (local) inverse temperature of static fluid $\beta_{loc}$ can be expressed in terms of the position-independent one $\beta$ as
\begin{equation}
\displaystyle
\beta_{loc}(r_{bh}+L)=\beta\sqrt{g(r_{bh}+L)}\approx\beta\sqrt{\frac{2\kappa L}{c^{2}}}.\label{beta loc}
\end{equation}
Again, Eq.~\eqref{g approx} is used in the last approximation. Using all of the approximations in Eqs.~\eqref{1-par z}--\eqref{beta loc}, the partition function in Eq.~\eqref{z} is then written in terms of the local quantities as
\begin{equation}
\displaystyle
Z = \frac{1}{N!}\left(\frac{2\pi}{h^3 c^3} \frac{A_hl_{loc}}{\beta_{loc}^3}\,\Gamma(3)\right)^N.
\end{equation} 
According to the fact that the partition function is the same for all observers, we can choose to consider it as the quantity measured by the local observer.
In the thermodynamic limit, i.e., the number of the gas particles $N$ is extremely large, the logarithmic function of $Z$ can be obtained by using Stirling's approximation as
\begin{equation}
\displaystyle
\ln Z \approx N\ln\left(\frac{2\pi}{h^3 c^3} \frac{A_hl_{loc}}{\beta_{loc}^3N}\,\Gamma(3)\right)+N.\label{z GB}
\end{equation}
Following the framework for the canonical ensemble, $\ln Z$ is thought of as a function of the inverse temperature $\beta_{loc}$ and a volume $V_{loc}$; $\ln Z=\ln Z(\beta_{loc}, V_{loc})$. To obtain the standard result for an ultra-relativistic gas \cite{pathria2021}, the volume $V_{loc}$ must be taken in the form of $V_{loc}=A_hl_{loc}/2$. One can notice that the volume $V_{loc}$ is a very thin shell with a thickness of $l_{loc}/2$, and the inner and outer radii are approximately the black hole's horizon radius. It can be interpreted that when the gas particles closely approach the black hole's horizon, their volume becomes very small as $V_{loc}$ instead of the shell one $4\pi\big[(r_{bh}+L+H)^3-(r_{bh}+L)^3\big]/3$.
This feature arises due to the gravitational field near the horizon, affecting how to count the number of microstates. 
As a result, the partition in Eq.~\eqref{z GB} can be rewritten as 
\begin{equation}
\displaystyle
\ln Z(\beta_{loc}, V_{loc})
= N\ln\left(\frac{4\pi}{h^3 c^3} \frac{V_{loc}}{\beta_{loc}^3N}\,\Gamma(3)\right)+N.\label{z GB in V}
\end{equation}
The thermodynamic mean energy, mean pressure, and entropy are, respectively, given by
\begin{eqnarray}
\displaystyle
U_{loc}
&=&-\frac{\partial\ln Z}{\partial\beta_{loc}}
=\frac{3N}{\beta_{loc}}, \label{UGB}\\
P_{loc}
&=&\frac{1}{\beta_{loc}}\frac{\partial\ln Z}{\partial V_{loc}}
=\frac{N}{V_{loc} \beta_{loc}},\\
S_{loc}
&=&k_B\left(1-\beta_{loc}\frac{\partial}{\partial\beta_{loc}}\right)\ln Z 
= k_BN\left[\ln\left(\frac{4\pi}{h^3 c^3} \frac{V_{loc}}{\beta_{loc}^3N}\,\Gamma(3)\right)+4\right].
\end{eqnarray}
It is very important to note that the above $U_{loc}$ and $P_{loc}$ depend on the frame of references while the entropy $S_{loc}$ does not.
This is reasonable because the entropy in GB statistical mechanics is proportional to the number of microstates which should be the same for all observers.
These quantities obey the first law of thermodynamics:
\begin{equation}
\displaystyle
dU_{loc}=\frac{1}{k_B\beta_{loc}}dS_{loc}-P_{loc}dV_{loc}.
\end{equation}
In the frame of the observer at the asymptotically far region, the gas system also satisfies the first law of the quantities measured by that observer.
\par
From the fact that the asymptotically far observer cannot know whether the gas crosses the black hole's horizon.
Consequently, due to the very small (proper) distance between the inner radius of the gas' shell and the event horizon (supposed to be in the order of Planck length), it is not possible to distinguish them \cite{Kolekar:2010py,Li:2021zuw}. This may imply that the gas particles of the gas system can be thought of as sitting on the black hole's surface area. However, the number of the gas particles is constrained by the argument that each particle can occupy the smallest area of the Planck area $l^{2}_{P}$.
The maximum number of gas particles is then counted as the surface area of the black hole in the unit of the Planck area:
\begin{equation}
\displaystyle
N=\frac{A_{h}}{l_{P}^{2}}.
\end{equation}
Using this argument, we can establish a link between the black hole and its surroundings (i.e., the gas shell).
Via the thermal contact, these two systems can be in thermal equilibrium so that the temperature of the gas equals the Hawking temperature $T_H=h\kappa/(4\pi^2ck_B)$ of the black hole \cite{Hawking:1975vcx}, or equivalently,
\begin{equation}
\displaystyle
\beta=\frac{4\pi^2c}{h\kappa}.
\end{equation}
Note that the Hawking temperature is measured by the observer at an asymptotically far region.
As a result, the entropy of the gas in equilibrium $S^{(0)}_{loc}$ is expressed as
\begin{equation}
\displaystyle
S^{(0)}_{loc}=\frac{k_{B}A_{h}}{l^{2}_{P}}\left[\ln\left(\frac{l_P^2\Gamma(3)}{32\pi^5l_{loc}^2}\right)+4\right].\label{S equi}
\end{equation}
As the entropy of the gas shell is identical to that of the system located at the horizon, it is then equal to the black hole entropy $S^{(0)}_{bh}$ by setting the proper length being extremely small as
\begin{equation}
\displaystyle
l_{loc}
=\frac{e^{15/8}\sqrt{\Gamma(3)}}{4\sqrt{2}\pi^{5/2}}l_{P}
=\frac{e^{15/8}}{4\pi^{5/2}}l_{P}
\sim10^{-1}l_{P}.
\end{equation}
Note that this length scale can be promoted as a cut-off scale, at which the quantum effect is significant so that quantum description of gravitation has to be required \cite{Calmet:2004mp}. In other words, the entropy in Eq.~\eqref{S equi} becomes the Bekenstein-Hawking entropy $S_{BH}$ \cite{Bekenstein:1973ur,Hawking:1975vcx};
\begin{equation}
\displaystyle
S_{BH} \equiv S^{(0)}_{bh}=k_{B} \frac{A_{h}}{4 l^{2}_{P}}.
\label{BH-entropy from GB}
\end{equation}
\par
It has been seen that the Bekenstein-Hawking entropy can be derived as the statistical entropy of the black hole based on GB statistical mechanics. In a standard manner,  the GB entropy at equilibrium is written as $S^{(0)}_{bh}=k_{B}\ln\Omega$, where $\Omega$ is the number of microstates. The number of microstates is straightforwardly obtained as
\begin{equation}
\displaystyle
\Omega_{bh} = \exp \left(\frac{A_{h}}{4 l_{P}^{2}}\right).
\end{equation}
Remarkably, one can only derive the entropy via the thermal equilibrium condition.
Based on this point of view, the other thermodynamic quantities for black holes cannot be defined as those belonging to the gas system.
This may imply that to obtain the complete set of thermodynamic variables, the geometric relation of the black hole needs to be realized.
For example, it is not possible to define the thermodynamic volume and pressure for the Schwarzschild black hole with GB statistical mechanics \cite{Bardeen:1973gs}.
However, the two aforementioned quantities can be constructed for the black holes with the cosmological constant \cite{Kubiznak:2016qmn}.
Consequently, after defining the entropy of black holes, the thermodynamic phase-space can be constructed by adopting the relations among the black holes' geometric quantities (it will be discussed in Sec.~\ref{thermodynamics of bh}).
\par
From Eq.~\eqref{BH-entropy from GB}, it is worthwhile to point out that the significant feature of the black hole entropy is scaled by its area instead of its volume. 
This property, however, contradicts the extensive (and also volume-scaling) entropy in conventional GB statistical mechanics. 
Therefore, the black hole entropy is not suitable to be described by the conventional GB entropy.
A framework of statistical mechanics based on generalized entropy may be more appropriate to study the thermodynamic properties of black holes instead of using GB statistical mechanics.
Since the derivation of the black hole entropy is general and can be applied to other types of statistical mechanics, the black hole entropy will be derived by following the procedure discussed in this section.
The purpose of the next section is to achieve this objective.

\section{Generalized entropy of black holes based on Tsallis statistical mechanics}\label{Tsallis section}
\par
In this section, by using a similar manner to the previous section, we will determine the black hole entropy in the aspect of statistical mechanics associated with one of the generalized entropies namely, Tsallis entropy ($q$-entropy). The $q$-entropy with non-extensive parameter, $q > 0$, can be written in terms of one-particle distribution function $f$ as \cite{Tsallis:1987eu}
\begin{equation}
\displaystyle
S_{q} = - k_{B} \int d \mathcal{V} f^{q} \ln_{q} f,
\label{Tsallis entropy}
\end{equation}
where $\ln_{q} f$ is the $q$-logarithmic function of $f$. In fact, the $q$-entropy is generalized from GB entropy by replacing normal logarithm with $q$-logarithm and promoting $f$ to be $f^q$. The $q$-logarithm and its inverse function (i.e., the $q$-exponential) are, respectively, expressed as \cite{2003cond.mat..9093T,gell2004nonextensive}
\begin{eqnarray}
\displaystyle  \ln_{q} Y = \frac{1}{1-q} (Y^{1-q} - 1),\qquad
\exp_{q} Y = \big[1 + (1-q) Y\big]^{\frac{1}{1-q}}.\label{Eq-ln-q-exp-q}
\end{eqnarray}
The Tsallis entropy obeys the non-additive composition rule, for example,  $S_{q (1+2)} = S_{q (1)} + S_{q (2)} + \big[(1-q)/k_{B}\big] S_{q (1)} S_{q (2)}$ where the subscript $``(1+2)"$ refers to the combined system of two subsystems, while subscript $``(1)"$ and $``(2)"$ refer to each subsystem. Moreover, at the limit $q \rightarrow 1$, the Tsallis entropy is reduced to the GB entropy, where the additive composition rule: $S_{(1+2)} = S_{(1)} + S_{(2)}$ is restored.
\par
For GB statistical mechanics, the distribution function can be obtained by maximizing the entropy, subjecting to the normalization condition: $\int d \mathcal{V} f = 1$, and then thermodynamic quantities, for example, the internal energy can be obtained by using the expectation value. For Tsallis statistical mechanics, the distribution function can be calculated by using the same manner as performed in the GB case. However, it is not unique to evaluate the expectation value in Tsallis statistical mechanics. There are various choices for the expectation value of internal energy \cite{Tsallis:1987eu,Curado:1991jc,Tsallis:1998ws,Taruya:2001mv,Taruya:2002bh,Taruya:2002jz,Sakagami:2003qs,Zavala:2006de,Coraddu:1998yb,Chakrabarti_2010,Chandrashekar_2011,Tsallis:2002tp,Wilk:2002uf,Wilk:2014zka,Jiulin:2004fi,Jiulin:2004bg,Mart_nez_2000,Lenzi:2000gj,Ferri:2005yf,Abe_2001}. However, there exists a transformation to link between such the various definitions \cite{Ferri_2005}. In this work, we choose the choice, for convenience, by following \cite{Curado:1991jc} as
\begin{equation}
\displaystyle
\begin{array}{lcr}
\displaystyle \int d \mathcal{V} f = 1, \qquad
\displaystyle \int d \mathcal{V} f^{q} E = U_{q}. \label{normal}
\end{array}
\end{equation}
\par
To obtain the $q$-distribution function, one can extremize the Tsallis entropy, $\delta S_{q} = 0$, subjecting to the above constraints by using the Lagrange multiplier, $a_{1}$ and $a_{2}$ as follows
\begin{eqnarray}
\displaystyle
\delta \left[\frac{k_{B}}{1-q} \int d \mathcal{V} (f^{q} - f) - a_{1} \left(\int d \mathcal{V} f - 1\right) - a_{2} \left(\int d \mathcal{V} f^{q} E - U_{q}\right) \right] &=& 0,
\label{cons 1}\\
\int d \mathcal{V} \left[\frac{k_{B}}{1-q} (q f^{q-1} - 1) - a_{1} - a_{2} q f^{q-1} E\right] \delta f &=& 0.
\label{cons 2} 
\end{eqnarray}
Since Eq. (\ref{cons 2}) does not depend on the choice of $\delta f$, one obtains
\begin{equation}
\displaystyle
\frac{k_{B}}{1-q} (q f^{q-1} - 1) - a_{1} - a_{2} q f^{q-1} E = 0,
\end{equation}
and then the $q$-distribution function can be expressed as
\begin{equation}
\displaystyle
f = B_{q} \left(1 - \frac{(1-q) a_{2}}{k_{B}} E\right)^{\frac{1}{1-q}}, \label{fBq}
\end{equation}
where $B_{q}$ is a constant determined from the normalization condition found in Eq. \eqref{normal}. As a result, the constant $B_{q}$ can be obtained in terms of partition function, $Z_q$, as 
\begin{eqnarray}
\displaystyle
B_{q} = \frac{1}{\displaystyle \int d \mathcal{V} \left(1 - \frac{(1-q) a_{2}}{k_{B}}E\right)^{\frac{1}{1-q}}} = \frac{1}{Z_q}, \label{Bq}
\end{eqnarray}
where the $q$-partition function plays the role of a constant for tuning the probability as in the appropriate range $[0, 1]$ defined explicitly by 
\begin{equation}
\displaystyle
Z_{q} \equiv  \int d \mathcal{V} \left(1 - \frac{(1-q) a_{2}}{k_{B}}E\right)^{\frac{1}{1-q}} = \int d \mathcal{V} \exp_{q} (-\beta_q E).
\label{q-partition}
\end{equation}
Note that we have set $a_{2} = 1/T_{q}$ in order to recover the GB partition function at limit $q\rightarrow 1$.
\par
Using Eq. (\ref{Bq}), Eq. (\ref{fBq}), the $q$-distribution function can be rewritten as 
\begin{equation}
\displaystyle
f = \frac{1}{Z_{q}} (1 - (1-q) \beta_{q} E)^{\frac{1}{1-q}} = \frac{1}{Z_{q}} \exp_{q} (-\beta_q E),
\label{q-distribution 1}
\end{equation}
where $\beta_{q} = 1/(k_{B} T_{q})$ and $T_{q}$   is $q$-temperature.
For the $N$-indistinguishable particles with the total energy $E_{tot}$, the $q$-partition function has to be scaled by $N!$:
\begin{equation}
\displaystyle
Z_{q} = \frac{1}{N!} \int d \mathcal{V} \exp_{q} (-\beta_q E_{tot}) = \frac{1}{N!} \prod_{i=1}^{N} \left[\int d \mathcal{V}_{i} \exp_{q}(\beta_{q} E_{i})\right] = \frac{1}{N!} (Z^{(1)}_{q})^{N},
\label{q-partition 1}
\end{equation}
where $Z_{q}^{(1)}$ is the one-particle partition function defined by
\begin{equation}
\displaystyle
Z_{q}^{(1)} = \int d \mathcal{V} (1 - (1-q) \beta_{q} E)^{\frac{1}{1-q}}.
\label{zq 3}
\end{equation}
Note that at the limit $q \rightarrow 1$, the $q$-partition function is reduced to the conventional GB partition function. It is important to emphasize that in order to write the partition function to be factorized form shown in Eq. (\ref{q-partition 1}), the total energy must be non-additive expressed as \cite{WANG2002131}
\begin{equation}
E_{tot} = \sum_{i=1}^{N} E_{i} + \sum_{j=2}^{N} [(q-1) \beta_{q}]^{j-1} \sum^{N}_{n_{1} < n_{2} <...< n_{j}} \prod^{j}_{l = 1} E_{l}.
\label{q-total energy}
\end{equation}
Note that the additive form of the total energy can be obtained by using the formal logarithm map given by \cite{Biro:2011ncf,Wang_2002}
\begin{equation}
\displaystyle
E_{add} = -\frac{1}{(1-q) \beta_{q}} \ln |1 - (1-q) \beta_{q} E_{tot}| = -\frac{1}{(1-q) \beta_{q}} \sum_{i=1}^{N} \ln |1 - (1-q) \beta_{q} E_{i}|.
\end{equation}
\par
From the phase-space volume in Eq. (\ref{1-par z}), one can see that the contribution from the non-extensive effect does not alter the phase-space volume. In fact, the phase-space volume depends only on spacetime geometry, i.e., $g(r)$. In addition, the partition function is invariant under the general coordinate transformation, let us choose to consider it in the local frame as done for the GB case in the previous section. Consequently, using Eqs. (\ref{vol in L}) and (\ref{l loc}), Eq. (\ref{zq 3}) can be rewritten as
\begin{equation}
\displaystyle
Z^{(1)}_{q} = \frac{2 \pi c^{3} A_{h}}{h^{3} \kappa^{3} l_{loc}^{2}} \int^{\infty}_{0}dE E^{2} (1-(1-q) \beta_{q} E)^{\frac{1}{1-q}}.
\label{Zq 1}
\end{equation}
Actually, the integration on the right-hand side of Eq. (\ref{Zq 1}) is the $q$-Laplace transform \cite{axioms5030024}:
\begin{equation}
\displaystyle
\mathcal{L}_{q} (f(E)) = \int^{\infty}_{0} dE f(E) (1-(1-q) \beta_{q} E)^{\frac{1}{1-q}},
\end{equation}
where $E > 0$ and $\beta_{q} \in \mathcal{C}$ with $\text{Re} (\beta_{q}) > 0$. For the algebraic function $f(E) = E^{\alpha - 1}$, the $q$-Laplace transform is given by $\mathcal{L}(E^{\alpha-1}) = \Gamma_{q}(\alpha)/\beta_{q}^{\alpha}$ where $\alpha, \beta_{q} \in \mathcal{C}$, $\text{Re} (\beta_{q}) > 0$ and $q \neq 1$, and $\Gamma_{q} (\alpha)$ is the $q$-gamma function given by \cite{axioms5030024}
\begin{equation}
\displaystyle
\Gamma_{q}(\alpha) = \left\{
\begin{array}{lcr}
\displaystyle
\frac{1}{(1-q)^{\alpha}} \frac{\Gamma(\alpha) \Gamma \left(\displaystyle \frac{1}{1-q} + 1\right)}{\Gamma\left(\displaystyle \frac{1}{1-q} + \alpha + 1\right)}, & \text{for} & q < 1 \\
\displaystyle
\Gamma (\alpha), & \text{for} & q = 1 \\
\displaystyle
\frac{1}{(q-1)^{\alpha}} \frac{\Gamma(\alpha) \Gamma \left(\displaystyle \frac{1}{q-1} - \alpha\right)}{\Gamma \left(\displaystyle \frac{1}{q-1}\right)}, & \text{for} & q > 1. \\
\end{array} 
\right.
\end{equation}
As a result, Eq. (\ref{Zq 1}) can be written as
\begin{equation}
\displaystyle
Z_{q}^{(1)} = \frac{4 \pi c^{3}}{h^{3}} \frac{V_{loc}}{(\beta_{q} \kappa l_{loc})^{3}} \Gamma_{q} (3),
\label{Zq-gamma}
\end{equation}
where an explicit form of $\Gamma_{q} (3)$  in terms of the non-extensive parameter is given by
\begin{eqnarray}
\displaystyle
\Gamma_{q} (3)
=\frac{2}{(2-q)(3-2q)(4-3q)}.\label{Gamma3}
\end{eqnarray}
Note that the above expression is valid for both $q<1$ and $q>1$.
Generally, we have $\beta_{q} \neq \beta$ for $q \neq 1$, hence it is worthwhile to introduce a mathematical relation as
\begin{equation}
\displaystyle
\beta_{q} = \mathcal{J}(q) \beta,
\label{fq1}
\end{equation}
where $\mathcal{J}(q)$ is a positive-valued function satisfied $\lim_{q \rightarrow 1} \mathcal{J}(q)=1$. 
Using this relation together with Tolman's law for $q$-thermodynamics $\beta_{q, loc} = \mathcal{J}(q) \beta_{loc}$ as performed in the GB case, the $q$-partition function for $N$  particles can be expressed as
\begin{equation}
\displaystyle
Z_{q} \approx \frac{1}{N!} \left[\frac{4 \pi}{h^{3} c^{3}} \frac{V_{loc}}{\beta^{3}_{q, loc}} \Gamma_{q}(3)\right]^{N}. \label{zq 10}
\end{equation}
At the thermodynamic limit: $N \gg 1$, using Stirling's approximation: $N! \approx (N/e)^{N}$, the $q$-logarithmic function of $Z_{q}$ can be obtained as
\begin{equation}
\displaystyle
\ln_{q} Z_{q} = \frac{1}{1-q} \left(Z_{q}^{1-q} - 1\right) \approx \frac{1}{1-q} \left[\left(\frac{4 \pi e}{h^{3} c^{3}} \frac{V_{loc}}{\beta^{3}_{q, loc} N} \Gamma_{q}(3)\right)^{N (1-q)} - 1\right].
\label{lnZq}
\end{equation}
Again, one can check that the above results reduce to that of the GB case in the limit $q\rightarrow 1$. To obtain the thermodynamic quantities of the gas system, by using Eq. (\ref{lnZq}), the mean energy and the mean pressure of the gas system can be computed as follows (see \ref{q-quantity}, for a detailed derivation)
\begin{equation}
\displaystyle
U_{q, loc} = -\frac{\partial}{\partial \beta_{q,loc}} \ln_{q} Z_{q} = \frac{3N}{\beta_{q,loc}} \left(\frac{4 \pi e}{h^{3} c^{3}} \frac{V_{loc}}{\beta^{3}_{q, loc} N} \Gamma_{q}(3)\right)^{N(1-q)},
\label{UTsallis}
\end{equation}
and
\begin{equation}
\displaystyle
P_{q, loc} = \frac{1}{\beta_{q,loc}} \frac{\partial}{\partial V_{loc}} \ln_{q} Z_{q} = \frac{N}{V_{loc} \beta_{q,loc}} \left(\frac{4 \pi e}{h^{3} c^{3}} \frac{V_{loc}}{\beta^{3}_{q, loc} N} \Gamma_{q}(3)\right)^{N(1-q)}.
\label{PTsallis}
\end{equation}
Furthermore, the gas $q$-entropy can be obtained as follows
\begin{eqnarray}
\displaystyle
S_{q, loc} &=& k_{B} \left(1 - \beta_{q, loc} \frac{\partial}{\partial \beta_{q, loc}}\right) \ln_{q} Z, \nonumber\\
\displaystyle
&=& \frac{k_{B}}{1-q} \left[\Big\{1 + 3 N (1-q)\Big\} \left(\frac{4 \pi e}{h^{3} c^{3}} \frac{V_{loc}}{\beta^{3}_{q, loc} N} \Gamma_{q}(3)\right)^{N (1-q)} - 1\right].
\label{q-entropy 1}
\end{eqnarray}
The $q$-thermodynamic quantities, namely, $P_{q, loc}$, $U_{q, loc}$ and $S_{q, loc}$, satisfy the first law of $q$-thermodynamics:
\begin{equation}
\displaystyle
dU_{q, loc} = \frac{1}{k_{B} \beta_{q, loc}} dS_{q, loc} - P_{q, loc} dV_{loc}.\label{1st q gas}
\end{equation}
\par
As seen in Eq.~(\ref{fq1}), we have introduced the auxiliary function $\mathcal{J}(q)$ characterizing how $\beta_q$ differs from $\beta$. 
It is not unique to choose this function. 
For example, one can choose the function $\mathcal{J}(q)$ satisfying a condition of identical internal energy $U_{q,loc} = U_{loc}$ and one can also find that the function $\mathcal{J}$ can be solved exactly.
Even though this choice is simple, it provides reasonable results since $\mathcal{J}(q)\rightarrow1$ at the limit $q\rightarrow1$ and $U_{q,loc}$ taken as a linear function of $U_{loc}$ is compatible with the description of Tsallis statistics.  
It should be noted that the function $\mathcal{J}(q)$ will, despite initially appearing to be arbitrary, actually reflect the thermodynamic properties of the black hole. 
For example, for the simple choice, $U_{q,loc} = U_{loc}$, the black hole entropy cannot be written in terms of a homogeneous function, and then the Smarr formula or the first law of black hole thermodynamics cannot be properly defined via a scaling law.
Therefore, the form of the function $\mathcal{J}$ is also constrained by the condition that the consequent black hole's entropy must be a homogeneous function.
In order to obtain a proper form of $\mathcal{J}(q)$ and the consistent thermodynamic properties of the black hole, let us consider a more general form as
\begin{eqnarray}
\displaystyle
U_{q, loc}=u_1\Gamma_q(3)^{u_2}U_{loc},
\label{Uq=U}
\end{eqnarray}
where $u_1$ and $u_2$ are arbitrary constants. As will be discussed in Sec.~\ref{thermodynamics of bh}, by choosing $u_2=-1/3$, $\Gamma_q(3)$ is eliminated in the $q$-entropy and then the $q$-entropy can be treated as a homogeneous function. As a result, black hole thermodynamics can be properly investigated.
Note also that the $q$-internal energy is additive.
As a result, using Eqs.~(\ref{UGB})~and~(\ref{UTsallis}), the function $\mathcal{J}(q)$ can be obtained as follows:
\begin{eqnarray}
\displaystyle
\mathcal{J}(q) = \left[\frac{\Gamma_{q}(3)^{1/3}}{u_1}
\left(\frac{4 \pi e}{h^{3} c^{3}} \frac{V_{loc}}{\beta^{3}_{loc} N} \Gamma_{q}(3)\right)^{N(1-q)}\right]^{\frac{1}{1+3N(1-q)}}.
\label{solfq1}
\end{eqnarray}
To satisfy the condition: $\displaystyle\lim_{q \rightarrow 1} \mathcal{J}(q) = 1$, the constant is set as $u_1=2^{1/3}$.   
Due to the fact that the factor $u_1\Gamma_q(3)^{u_2}$ in Eq.~\eqref{Uq=U} is always positive.
It leads to the constraint on the non-extensive parameter as follows:
\begin{eqnarray}
\displaystyle
0<q<\frac{4}{3},\qquad
\frac{3}{2}<q<2.
\label{bound J}
\end{eqnarray}
\par
Moreover, since the mean pressure in Eq.~(\ref{PTsallis}) can be written in terms of the $q$-internal energy in Eq.~(\ref{UTsallis}), the condition~\eqref{Uq=U} also leads to a relation between the mean pressures for both GB and Tsallis statistical mechanics as
\begin{eqnarray}
\displaystyle
P_{q, loc} 
=\frac{U_{q,loc}}{3V_{loc}}
=\frac{\left(\displaystyle \frac{2}{\Gamma_q(3)}\right)^{1/3}U_{loc}}{3V_{loc}} 
=\left(\frac{2}{\Gamma_q(3)}\right)^{1/3}P_{loc}.
\label{Pq}
\end{eqnarray}
According to Eq.~\eqref{Gamma3}, one can check that for $q>1$ (in fact, the range is $1 < q < 4/3$, and $3/2 < q < 2$, but let us keep this regime as $q>1$ for simplicity), the pressure of the non-extensive gas system is less than the pressure of the ideal gas one: $P_{q, loc} < P_{loc}$. 
We may argue that a gas molecule attracts other molecules due to its non-extensivity. 
This causes the gas molecules to effectively exhibit a lower force hitting the boundary of the system (corresponding to a lower pressure) compared to the GB case. 
In contrast, for $q < 1$, the pressure of the non-extensive gas system is less than that of the GB case, i.e., $P_{q, loc} > P_{loc}$ due to the gas molecules repelling each other in this regime of non-extensivity.
As a consequence of the first law~\eqref{1st q gas}, the heat term obeys
\begin{eqnarray}
\displaystyle
\delta Q_{q, loc} = T_{q, loc} dS_{q, loc} = \left(\frac{2}{\Gamma_q(3)}\right)^{1/3} T_{loc} dS_{loc} =  \left(\frac{2}{\Gamma_q(3)}\right)^{1/3} \delta Q_{loc}.
\end{eqnarray}
Again, the heat transfer also gets a modification due to the non-extensivity. 
The efficiency in transferring the heat by the molecules of the non-extensive gas system is less (more) than that of the traditional extensive gas system caused by the attractive (repulsive) behavior of the non-extensive gas molecules with $q>1$ ($q<1$). 
\par
\begin{figure}[ht]\centering
\includegraphics[width=7cm]{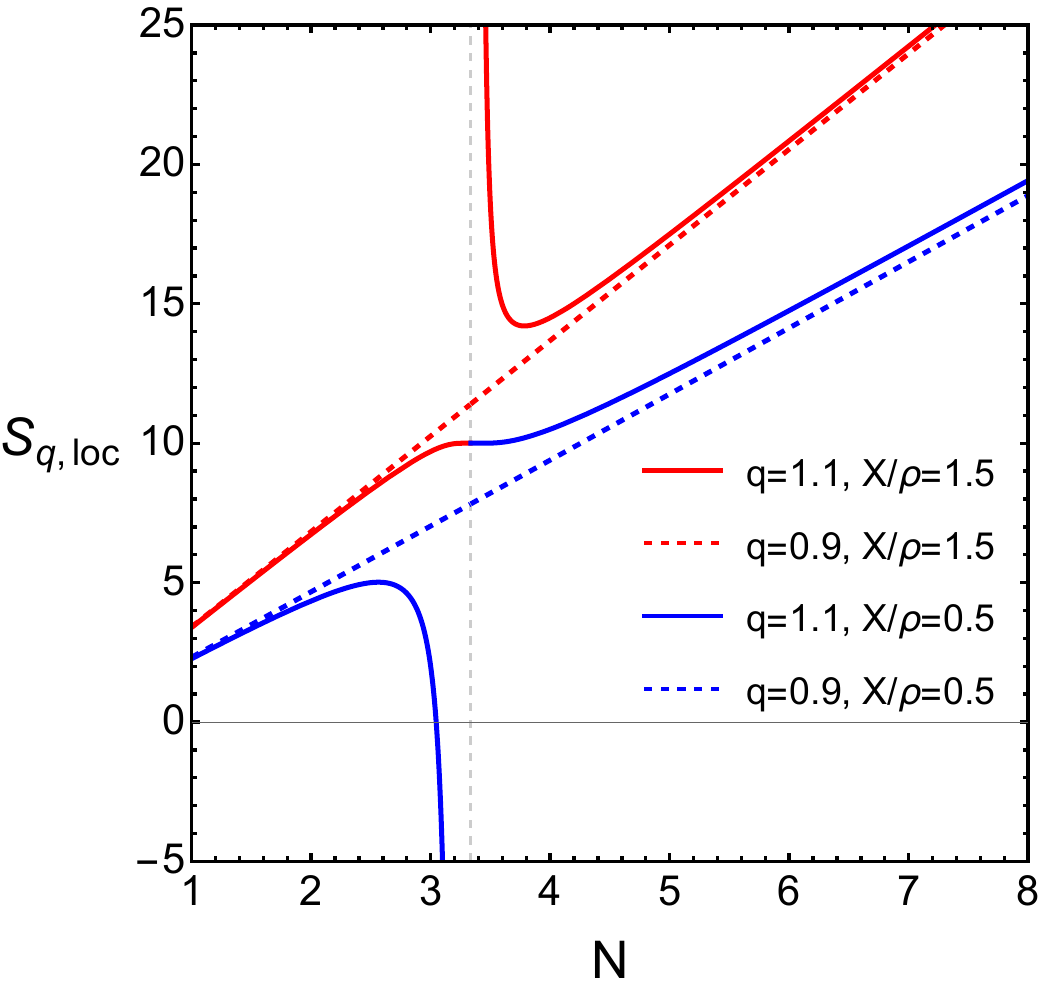}
\caption{The profiles of the gas entropy versus $N$ for various values of the non-extensive parameter and the base $X/\rho$.}
\label{gas q-entropy}
\end{figure}

Let us turn our attention back to the $q$-entropy.
This entropy can be expressed in terms of $\beta_{loc}$ by substituting Eq.~(\ref{solfq1}) into Eq.~(\ref{q-entropy 1}) as
\begin{equation}
\displaystyle
S_{q, loc} = \frac{k_{B}}{1-q}\left[\Big\{1 +3N(1-q)\Big\}\left(\frac{8\pi e}{h^{3} c^{3}}\frac{V_{loc}}{\beta^{3}_{loc}N}\right)^{\frac{N(1-q)}{1+ 3N(1-q)}}-1\right].
\label{gas q-entropy 1}
\end{equation}
As expected, this resulting entropy is independent of $\Gamma_q(3)$.
The behavior of the gas entropy can be illustrated in Fig.~\ref{gas q-entropy}. 
It can be seen that the behavior of the entropy depends on the parameter $q$ and the base of the exponential function $\frac{8\pi e}{h^{3} c^{3}}\frac{V_{loc}}{\beta^{3}_{loc}N}$. 
For $q<1$, the entropy grows smoothly as the number of particles increases (see the dashed lines in Fig.~\ref{gas q-entropy}).
However, divergent and local extremum points can emerge when $q>1$.
In the regime $q>1$, the key feature of the entropy is affected by the value of the base $X/\rho$ where $X\equiv8\pi e/(h^{3} c^{3}\beta^{3}_{loc})$ and $\rho\equiv N/V_{loc}$ describe how hot and dense the gas is, respectively.
The density $\rho$ of the gas is assumed to be a constant in this consideration (this assumption is applicable to the $q$-entropy of a black hole as will be discussed soon). 
It can be split into three regimes:
\begin{itemize}
\item[$i$)] Sufficiently cold or dense gas ($X/\rho<1$): as increasing $N$, $S_{q, loc}$ first reaches its local maximum and then drops to minus infinity (see the solid blue line in Fig.~\ref{gas q-entropy}).
\item[$ii$)] Sufficiently hot or loose gas ($X/\rho>1$): as increasing $N$, $S_{q, loc}$ diverges first to infinity, and then drops to its local minimum (see the solid red line in Fig.~\ref{gas q-entropy}).
\item[$iii$)] Critical point ($X/\rho=1$): the entropy does not depend on $q$, but linearly depends on $N$ as $S_{q, loc}=3k_BN$.
\end{itemize}
\par
The existence of the extremum and divergence of the entropy could be argued as a result of taking the self-gravitating effect (long-range interaction) into account, and this effect becomes significantly stronger. In fact, these results are in agreement with the ones of applying the Tsallis entropy in astrophysics (see, e.g., \cite{Taruya:2002bh,Lynden-Bell:1968awc}). With GB statistical mechanics, the thermodynamic description of the self-gravitating system leads to a conclusion that the system is isothermal with negative heat capacity, referring to thermodynamic instability. Therefore, GB statistical mechanics may not be properly described in such a system. One needs statistical mechanics compatible with a non-isothermal system in hydrostatic equilibrium instead of thermal equilibrium, for example, Tsallis statistical mechanics \cite{chunaksorn2024qequilibriumgasspacetimemultihorizon}. We will see in the next section that a thermal system associated with a black hole is not well-defined when the number $N$ is too large. This is because the system's temperature is diverging and becomes negative. Therefore, it is worthwhile to consider the system with a sufficiently small number of particles in order to restrict our consideration to the thermal system characterized by well-behaved temperatures.
\par
Following the argument mentioned in the previous section, the number of particles, the inverse temperature and volume of the gas shell are set as $N = A_{h}/l^{2}_{P}$, $\beta_{loc} = 4 \pi^{2} l_{loc}/hc$, $V_h=Al_{loc}/2$, respectively.
In addition, the proper distance between the inner shell of the gas and the black hole's horizon is also fixed as $l_{loc} = l_{P} \exp (15/8)/4 \pi^{5/2}$.
The generalized Bekenstein-Hawking entropy corresponding to the black hole entropy in $q$-thermodynamics can be obtained as
\begin{equation}
\displaystyle
S_{q, BH} \equiv S_{q, loc(bh)}^{(0)} = \frac{k_{B}}{1-q} \left[\left\{1 + 3(1-q) \frac{A_{h}}{l_{P}^{2}}\right\}\exp\left\{{-\frac{11(1-q)\frac{A_{h}}{l_{P}^{2}}}{4\big(1+3(1-q)\big)\frac{A_{h}}{l_{P}^{2}}}}\right\} - 1\right].
\label{BH q-entropy 1}
\end{equation}
One can straightforwardly check that the base $X/\rho$ is less than unity so that the behavior of the black hole entropy is similar to the entropy represented as the solid and dashed blue lines Fig.~\ref{gas q-entropy}. 
The form of the black hole entropy expressed in Eq.~(\ref{BH q-entropy 1}) is complicated, and not easy to capture its physical meaning. By keeping $q$ closed to $1$, the generalized Bekenstein-Hawking entropy can be approximated as the power series around $q=1$:
\begin{equation}
\displaystyle
S_{q, BH} \approx k_{B} \frac{A_{h}}{4 l_{P}^{2}} + k_{B} \frac{121 A_{h}^2}{32 l_{P}^{4}} (1-q) + \mathcal{O} [(1-q)^{2}].
\label{q-entropy use}
\end{equation}
From Eq. (\ref{q-entropy use}), it is seen that the black hole entropy is written as the usual Bekenstein-Hawking entropy with the correction term due to the non-extensivity. Furthermore, the Tsallis entropy at $q$-thermodynamic equilibrium is expressed as $S_{q}^{(0)} = k_{B} \ln_{q} \Omega$. Using Eq. (\ref{BH q-entropy 1}), we can calculate the number of microstates of the black hole as
\begin{equation}
\displaystyle
\Omega_{bh} = \left[\left\{1 + 3(1-q) \frac{A_{h}}{l_{P}^{2}}\right\}\exp\left\{{-\frac{11(1-q)\frac{A_{h}}{l_{P}^{2}}}{4\big(1+3(1-q)\big)\frac{A_{h}}{l_{P}^{2}}}}\right\}\right]^{\frac{1}{1-q}}.
\label{q-microstates}
\end{equation}
Expanding Eq. (\ref{q-microstates}) around $q=1$, one obtains
\begin{equation}
\displaystyle
\Omega_{bh} \approx \exp \left(\frac{A_{h}}{4l_{P}^{2}}\right) + \frac{15A_{h}^2}{4 l^{4}_{P}} \exp \left(\frac{A_{h}}{4l_{P}^{2}}\right)(1-q) + \mathcal{O} [(1-q)^{2}].
\label{no of micro q}
\end{equation}
From this expression, one can see that for $q >1$, the number of microstates is decreased due to the non-extensive nature of the system. Therefore, if black hole thermodynamics is supposed to rely on the non-extensive system, available states in which particles can occupy are less than one for the usual case. In this case, we can imagine that the Planck area units are partially overlapped and then there are over-counting states. As a result, the actual number of microstates is less than one counted from GB statistical mechanics. Oppositely, for $q <1$, the number of microstates will be increased due to the Planck area units moving away from each other.
\par
It is important to note that there are various applications of Tsallis entropy which take the form $S_\delta = \alpha_\delta A_h^\delta$ \cite{Tsallis_2013} where $A_h$ denotes the horizon area and, $\alpha_\delta$ and $\delta$ are constants. However, it is not perfectly clear what the form of the Tsallis entropy for the black hole is. This argument is still debatable \cite{article,Pessoa:2020cih,Tsallis:2021mvq,Pessoa:2022fwo}. Our result suggests that the form of Tsallis entropy does not satisfy the power-law form as argued in the literature. This may provide a hint to naturally include the non-extensive effect to study black hole thermodynamics as well as in a cosmological context.

\section{Thermodynamic properties of black hole}\label{thermodynamics of bh}
\par
In this section, we try to construct the thermodynamic quantities of the black hole by replacing the Bekenstein-Hawking entropy with the generalized Bekenstein-Hawking entropy, as well as investigate the thermodynamic properties of the black hole, e.g., the stability. In this section, we will work in the unit of $k_{B} = G = \hbar = c = 1$ (the Planck length $l_P$ is also unity in this unit) for convenience. The first law of thermodynamics can be obtained by treating the black hole mass as a homogeneous function. In addition, by employing Euler's theorem, one hence obtains the algebraic relation known as Smarr formula \cite{PhysRevLett.30.71}. Let us introduce the aforementioned procedure by starting with the definition of the homogeneous function. An arbitrary function $\Tilde{f}(x_{i})$, where $x_{i} \in R$, is said to be a homogeneous function of degree $n \in R$, if and only if it satisfies that $\Tilde{f}(I x_{i}) = I^{n} \Tilde{f}(x_{i})$, with $I \in R$. According to Euler's theorem for the homogeneous function, the function $\Tilde{f}(x_{i})$ satisfies the following relation:
\begin{equation}
\displaystyle
n \Tilde{f}(x_{i}) = x_{j} \frac{\partial \Tilde{f}(x_{i})}{\partial x_{j}}.
\end{equation}
\par
For the simple case in consideration, we choose the static and spherically symmetric black hole as the Schwarzschild (Sch) black hole. The interval is taken in the form found in Eq. (\ref{ds}) with the horizon function given by
\begin{equation}
\displaystyle 
g(r) = 1 - \frac{2M}{r},
\end{equation}
where $M$ is the black hole mass. The black hole mass can be obtained by solving $g(r_{bh}) = 0$ as
\begin{equation}
\displaystyle
M = \frac{r_{bh}}{2}.
\label{Sch mass}
\end{equation}
Furthermore, the Bekenstein-Hawking entropy can be written as
\begin{equation}
\displaystyle
S_{BH\text{(Sch)}} = \frac{A_{h}}{4} = \pi r^{2}_{bh}.
\label{Sch GB entropy}
\end{equation}
By substituting Eq. (\ref{Sch GB entropy}) into Eq. (\ref{Sch mass}), the black hole mass can be rewritten as 
\begin{equation}
\displaystyle
M = \frac{1}{2} \sqrt{\frac{S_{BH\text{(Sch)}}}{\pi}}.
\end{equation}
The Smarr formula for the black hole mass written in the form of a homogeneous function of degree $1/2$, i.e., $M(I S_{BH\text{(Sch)}}) = I^{1/2} M(S_{BH\text{(Sch)}})$, can be obtained by employing the Euler's theorem as
\begin{equation}
\displaystyle
M = 2 S_{BH\text{(Sch)}} T_{H\text{(Sch)}},
\label{Sch GB Smarr}
\end{equation}
where the conjugate quantity of $S_{BH\text{(Sch)}}$ is given by
\begin{equation}
\displaystyle
T_{H\text{(Sch)}} = \left(\frac{\partial M}{\partial S_{BH\text{(Sch)}}}\right) = \frac{1}{4 \pi r_{bh}}.
\label{Sch Smarr GB}
\end{equation}
In fact, the above equation is the Hawking temperature. Furthermore, the first law of thermodynamics can be written as
\begin{equation}
\displaystyle
dM = T_{H\text{(Sch)}} dS_{BH\text{(Sch)}}. \label{Sch GB 1law}
\end{equation}
\par
Here, we have employed Euler's theorem to the black hole mass regarded as a homogeneous function of degree $1/2$. By identifying the suitable expressions of thermodynamic quantities of the black hole, we obtain the thermodynamic relations that are equivalent to those obtained from the geometric approach. Moreover, the mentioned relations are obtained by considering the black hole entropy as the Bekenstein-Hawking entropy. In other words, the thermodynamic relations are rooted in GB statistical mechanics. However, as we have known, GB statistical mechanics is not suitable to describe the thermodynamic properties of the black hole due to the non-extensivity. In this work, in order to solve the aforementioned issue, the generalized Bekenstein-Hawking entropy is chosen to construct the thermodynamic relations instead of the usual one.
\par
The Smarr formula with the generalized Bekenstein-Hawking entropy can be constructed by following the same manner done in the GB framework. Unfortunately, the expression of $r_{bh}$ in terms of the generalized Bekenstein-Hawking entropy is not easy to display. We may write the generalized Bekenstein-Hawking entropy in terms of the black hole mass instead. As mentioned, by substituting Eq.~(\ref{Sch mass}) into Eq.~(\ref{BH q-entropy 1}), we can treat the generalized Bekenstein-Hawking entropy, with a re-defined non-extensive parameter: $\eta = (1-q)$:
\begin{eqnarray}
\displaystyle
S_{\eta, BH\text{(Sch)}} 
&=& \frac{1}{\eta} \left[(1+3\eta A_{h})\exp\left\{-\frac{11\eta A_{h}}{4(1+3\eta A_{h})}\right\}- 1\right], \nonumber\\
\displaystyle
&=& \frac{1}{\eta} \left[(1+48\pi M^2\eta)\exp\left\{-\frac{44\pi M^2\eta}{1+48\pi M^2\eta}\right\}- 1\right],
\label{Sbh 1}
\end{eqnarray}
as a homogeneous function of degree $1$, i.e., $S_{\eta, BH\text{(Sch)}} (I M^{2}, I \eta^{-1}) = I S_{\eta, BH\text{(Sch)}} (M^{2}, \eta^{-1})$. 
As mentioned in the previous section, this is the reason for choosing the entropy to be independent of $\Gamma_q(3)$.
If the term $\Gamma_q(3)$ appears in the black hole entropy, it is not possible to treat such an entropy as the homogeneous function.
Furthermore, by using Euler's theorem, the generalized Bekenstein-Hawking entropy can be expressed as
\begin{eqnarray}
\displaystyle
S_{\eta, BH\text{(Sch)}} = \frac{1}{2} M \left(\frac{\partial S_{\eta, BH\text{(Sch)}}}{\partial M}\right)_{\eta} - \eta \left(\frac{\partial S_{\eta, BH\text{(Sch)}}}{\partial \eta}\right)_{M}.
\end{eqnarray}
The generalized Smarr formula can be obtained by rearranging the above equation as
\begin{equation}
\displaystyle
M = 2 S_{\eta, BH\text{(Sch)}} \left(\frac{\partial M}{\partial S_{\eta, BH\text{(Sch)}}}\right)_{\eta} + 2\eta \left(\frac{\partial S_{\eta, BH\text{(Sch)}}}{\partial \eta}\right)_{M} \left(\frac{\partial M}{\partial S_{\eta, BH\text{(Sch)}}}\right)_{\eta}. \label{Sch modified Smarr 1}
\end{equation}
The last term on the right-hand side of Eq. (\ref{Sch modified Smarr 1}) can be rewritten by using the cyclical relation, and hence the generalized Smarr formula reads
\begin{equation}
\displaystyle
M = 2 S_{\eta, BH\text{(Sch)}} T_{\eta, H\text{(Sch)}} - 2 \eta \Phi_{\eta\text{(Sch)}},
\label{Sch modified Smarr 2}
\end{equation}
where the conjugate thermodynamic quantities are properly defined by
\begin{equation}
\displaystyle
\begin{array}{lcr}
\displaystyle
T_{\eta, H\text{(Sch)}} = \left(\frac{\partial M}{\partial S_{\eta, BH\text{(Sch)}}}\right)_{\eta},  \qquad \displaystyle \Phi_{\eta\text{(Sch)}} = \left(\frac{\partial M}{\partial \eta}\right)_{S_{\eta, BH\text{(Sch)}}}.
\end{array}
\label{Sch temp 1}
\end{equation}
Note that the black hole mass in Eq. (\ref{Sch modified Smarr 2}) is the same as the one in Eq. (\ref{Sch GB Smarr}). Furthermore, the first law of thermodynamics can be written as follows
\begin{eqnarray}
dM&=& \left(\frac{\partial M}{\partial S_{\eta, BH\text{(Sch)}}}\right)_{\eta} dS_{\eta, BH\text{(Sch)}} + \left(\frac{\partial M}{\partial \eta}\right)_{S_{\eta, BH\text{(Sch)}}} d\eta, \nonumber\\
&=& T_{\eta, H\text{(Sch)}} dS_{\eta, BH\text{(Sch)}} + \Phi_{\eta\text{(Sch)}} d\eta. \label{Sch Smarr 1 law}
\end{eqnarray}
Note that at the limit $\eta \rightarrow 0$, the generalized Smarr formula (\ref{Sch modified Smarr 2}) and the first law of thermodynamics (\ref{Sch Smarr 1 law}) can be, respectively, reduced to Eqs. (\ref{Sch GB Smarr}) and (\ref{Sch GB 1law}) as expected.
It is also important to note that the generalized Smarr formula found in Eq. (\ref{Sch modified Smarr 2}) is derived from the fact that black hole entropy is thought of as the homogeneous function instead of considering the black hole mass. However, it is still written in the appropriate form and reduced to one in the GB case at limit $\eta \rightarrow 0$. Furthermore, the thermodynamic relations obtained from the same procedure with the R\'{e}nyi entropy for the various kinds of black holes have been shown in Ref. \cite{Nakarachinda:2022gsb}.
\par
\begin{figure}[ht]\centering
\includegraphics[width=7cm]{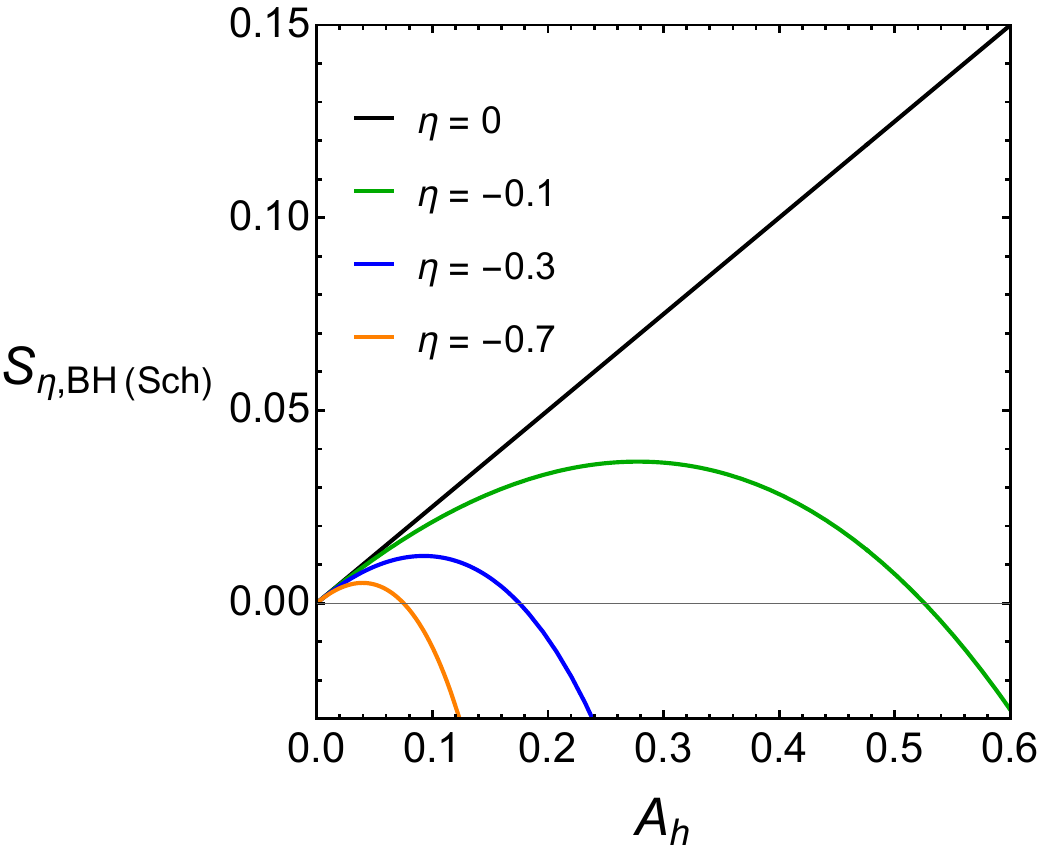}
\qquad
\includegraphics[width=7cm]{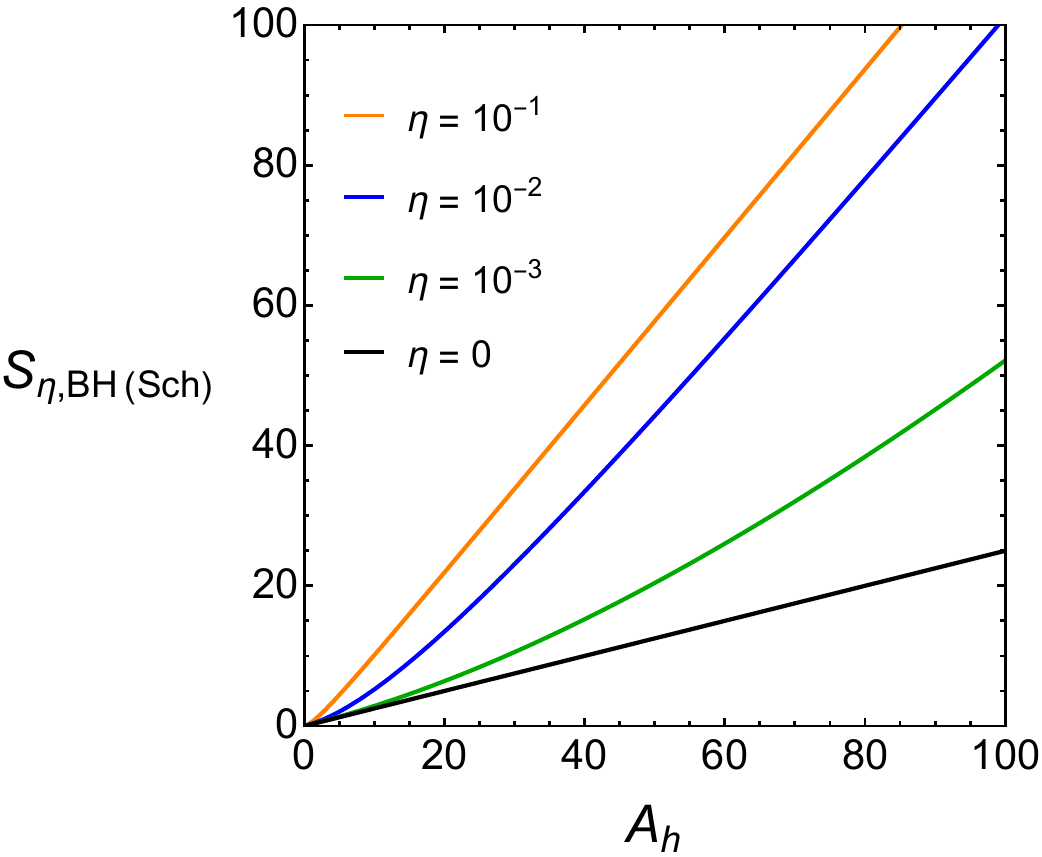}
\caption{The black hole entropy versus $A_h$ with negative $\eta$ (left) and positive $\eta$ (right).}
\label{Smax}
\end{figure}

In this part of our work, let us first consider the Sch black hole undergoing a process of fixing $\eta$. The first law of thermodynamics is, therefore, reduced to 
\begin{eqnarray}
dM = T_{\eta, H\text{(Sch)}} dS_{\eta, BH\text{(Sch)}}.
\end{eqnarray}
The behaviors of the black hole entropy in Eq.~\eqref{Sbh 1} for various values of $\eta$ can be illustrated in Fig.~\ref{Smax}.
It is very important to point out that the entropy with negative $\eta$ has an extremum point, while there is no such a point for one with $\eta \geq 0$.
In addition, the entropy with negative $\eta$ is unfortunately not well-behaved for the whole range of horizon radius, e.g. it becomes at a certain horizon radius as seen in the left panel of Fig.~\ref{Smax}.
It is also noticed that the greater the negative value of the non-extensive parameter, the smaller the area at which the entropy drops to zero.
On the other hand, the entropy with positive $\eta$ is a monotonically increasing function of the horizon area.
The contribution due to the non-extensivity makes the $q$-entropy grow faster than the GB one.
\par
For the temperature of the black hole, it can be straightforwardly computed by using the first relation in Eq.~(\ref{Sch temp 1}) as 
\begin{equation}
\displaystyle
T_{\eta, H(\text{Sch})} = \frac{(1+3\eta A_{h}) \exp\left[\displaystyle{\frac{11\eta A_{h}}{4(1+3\eta A_{h})}}\right]}{2 \sqrt{\pi A_{h}} (1+36\eta A_{h})}.
\label{TA 1}
\end{equation}
Again, it is easy to check that at limit $\eta \rightarrow 0$, Eq.~(\ref{TA 1}) can be reduced to the Hawking temperature:
$\displaystyle
\lim_{\eta \rightarrow 0} T_{\eta, H(\text{Sch})} = 1/(2\sqrt{\pi A_h})$.
According to the definition in the first relation of Eq.~\eqref{Sch temp 1}, the temperature with negative $\eta$ diverges when the entropy reaches its local maximum.
It implies that the black hole's temperature in this regime of $\eta$ cannot be defined for a sufficiently large size.
This issue is relevant to the argument mentioned in the previous section that for large enough $N$, the temperature associated with Tsallis entropy is not easily defined for the self-gravitating system in general since the system is no longer in thermodynamic equilibrium. 
In order to restrict our attention to the thermodynamic equilibrium system, we will set the upper bound limit of $N$ at the local maximum of the entropy.
Such an upper bound of $N$ corresponds to the largest surface area of the mentioned maximum point denoted as $A_\text{max}$, in which the temperature in the range $0<A_h<A_\text{max}$ is well-behaved.
\par
\begin{figure}[ht]\centering
\includegraphics[width=6.7cm]{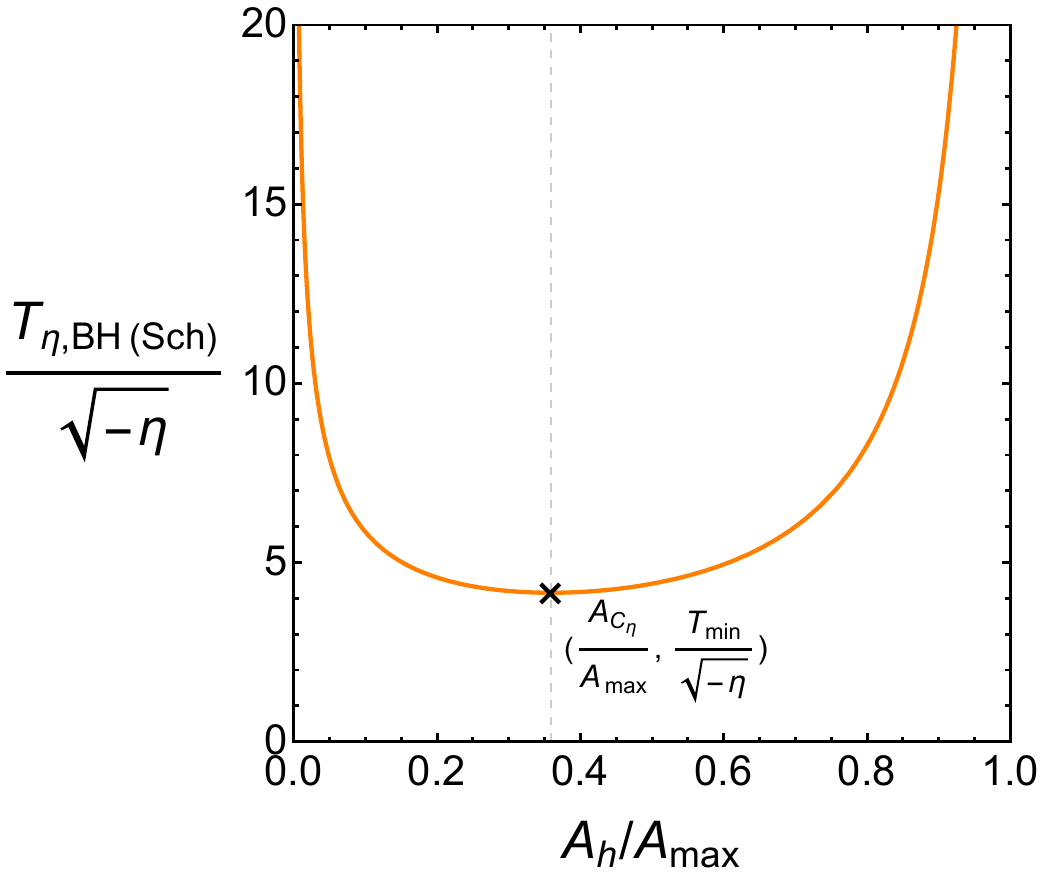}
\qquad
\includegraphics[width=7cm]{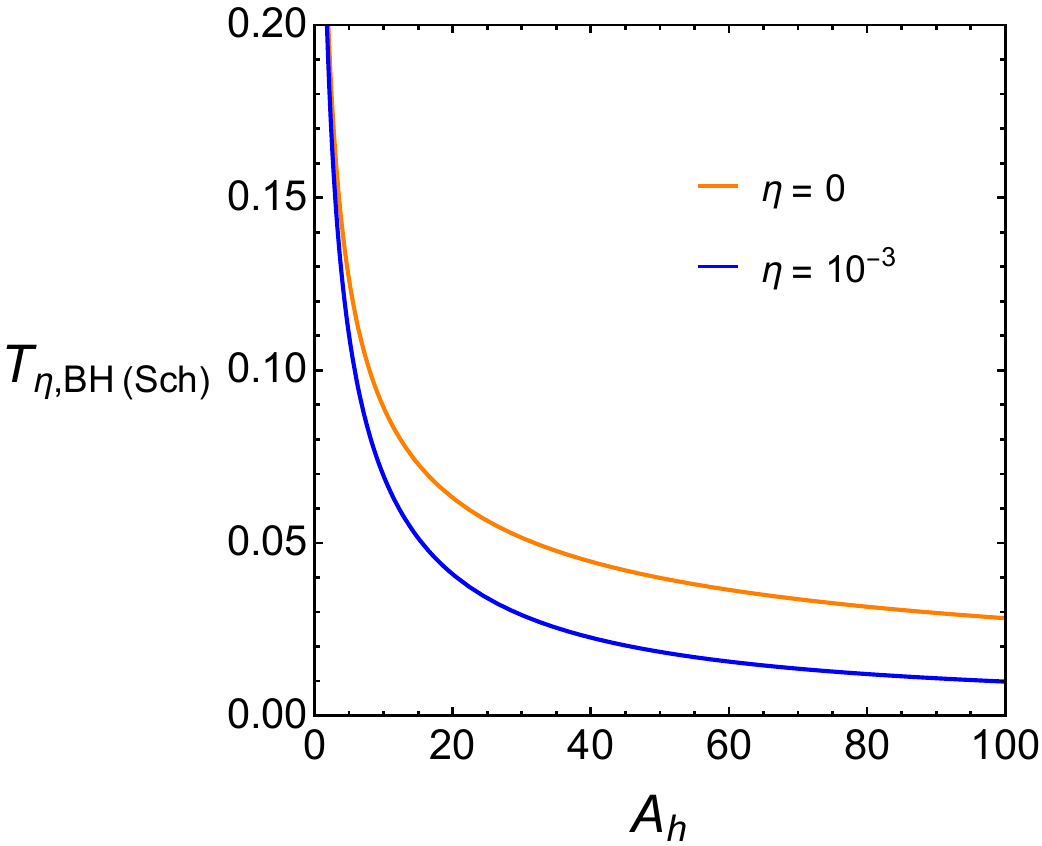}
\caption{Profiles of the temperature of the black hole with negative $\eta$ (left) and positive $\eta$ (right).}
\label{T}
\end{figure}

The local maximum of the entropy in Eq.~\eqref{Sbh 1} is simply obtained from the condition $dS_{\eta,BH\text{(Sch)}}/dA_h|_{A_\text{max}}=0$. 
As a result, the area at the local maximum is obtained as
\begin{eqnarray}
A_\text{max}=-\frac{1}{36\eta}.
\end{eqnarray}
The profiles of the temperature are illustrated in Fig.~\ref{T}.
According to the first relation in Eq.~\eqref{Sch temp 1} together with the fact that the derivative of the black hole's mass with respect to $A_h$ is always positive, the trend of the temperature depends only on the change in entropy.
For the asymptotic values of the temperature, one can also check that $dM/dA_h\propto1/\sqrt{A_h}$.
Therefore, the temperature approaches infinity as $A_h\to0$.
For negative $\eta$, the temperature drops to its minimum at a certain area denoted as $A_{C_\eta}$, then grows to infinity again due to the maximum of the entropy.
The area $A_{C_\eta}$ is indeed the point at which the heat capacity diverges as will be discussed soon.
By using the minimum condition: $dT_{\eta,BH\text{(Sch)}}/dA_h=0$, one can solve for an explicit form of $A_{C_\eta}$ as follows:
\begin{eqnarray}
\displaystyle
A_{C_\eta}&=&\frac{1}{108\eta}\left[11^{2/3} \left(3 \sqrt{7869}+236\right)^{1/3}-55 \left(\frac{11}{3 \sqrt{7869}+236}\right)^{1/3}-25\right], \nonumber\\
&\approx&0.359A_\text{max}.
\label{AC eta}
\end{eqnarray}
It is interestingly found that this behavior of the temperature with negative $\eta$ does not depend on the magnitude of $\eta$.
Therefore, as seen in the left panel of Fig.~\ref{T}, the temperature for negative $\eta$ profiles can be plotted without identifying the value of $\eta$.
For the positive $\eta$ regime, the temperature monotonically decreases from infinity to zero as the horizon area increases (see the right panel of Fig.~\ref{T}) straightforwardly caused by the behavior of the entropy.
In addition, since the area $A_h$ is indeed in the unit of the Planck area, the non-zero area should be at least one (the black hole's horizon area is equal to the Planck area).
It yields a bound to the magnitude of the negative $\eta$.
As a result, the constraint~\eqref{bound J} can be further restricted as
\begin{eqnarray}
\displaystyle
-\frac{1}{36}<\eta<1.
\end{eqnarray}
\par
\begin{figure}[ht]\centering
\includegraphics[width=5.5cm]{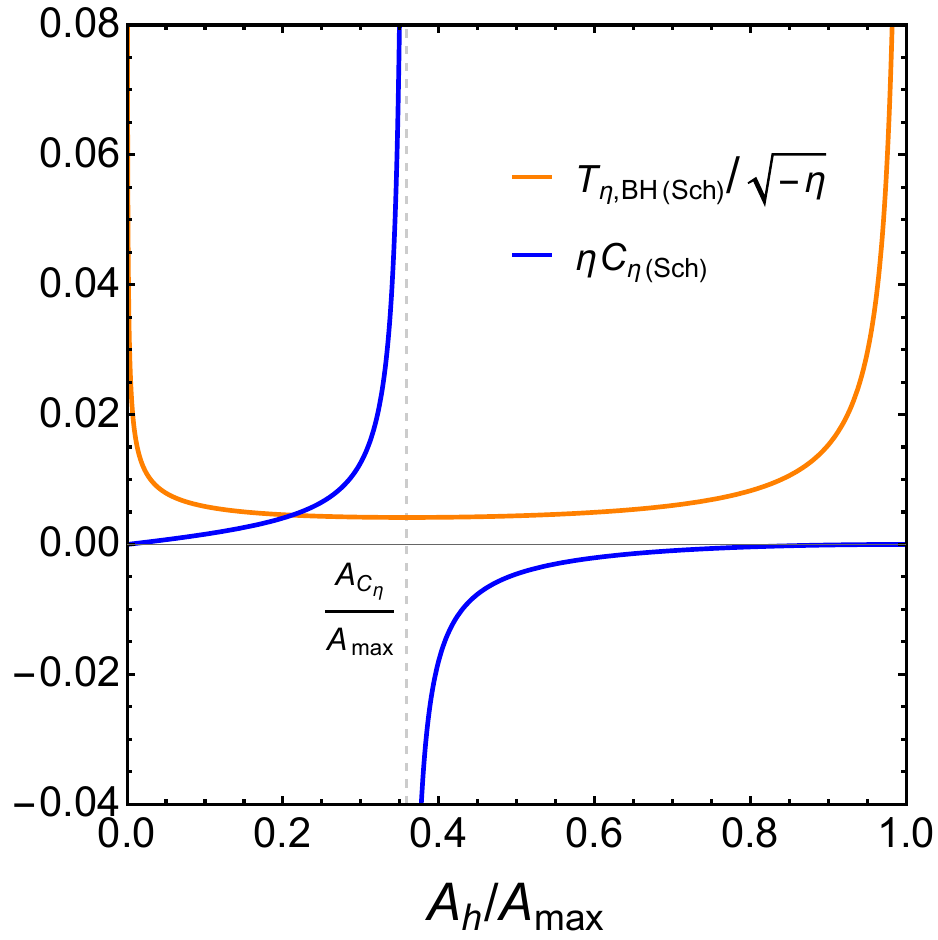}
\qquad
\includegraphics[width=6.5cm]{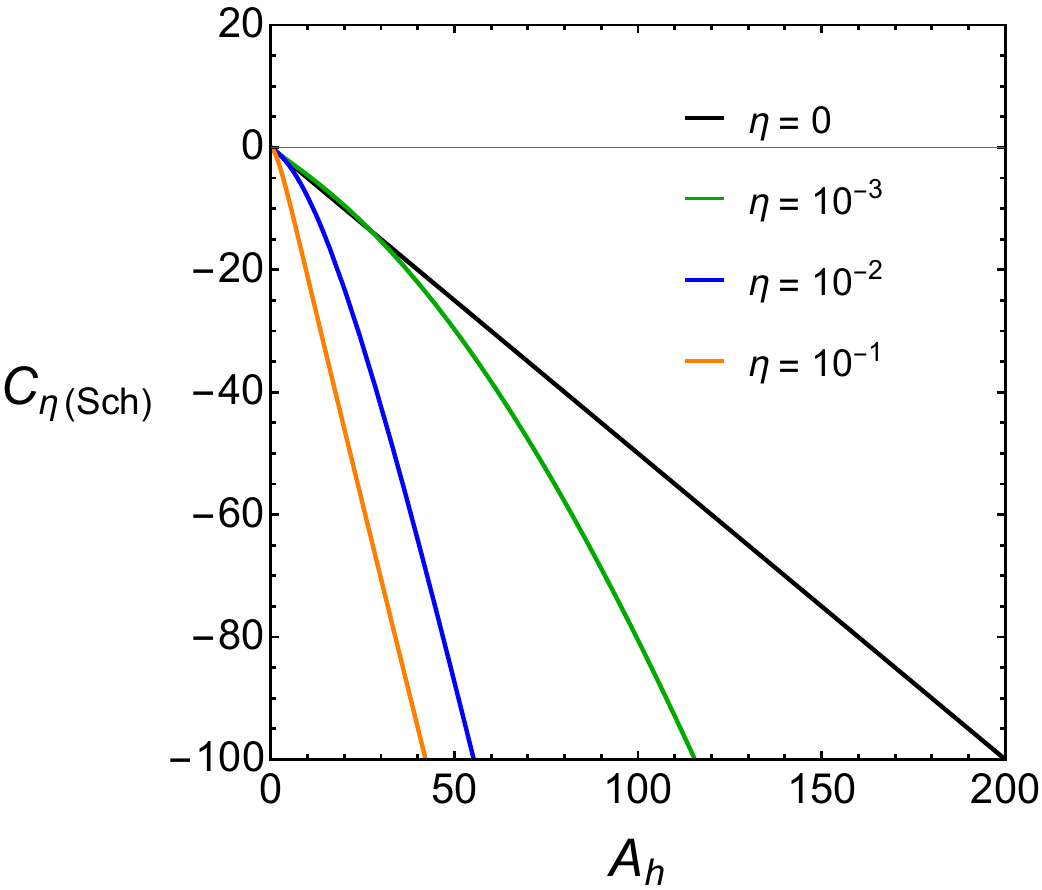}
\caption{Profiles of heat capacity for negative $\eta$ (left) and positive $\eta$ (right).}
\label{T C}
\end{figure}

Next, let us investigate the thermodynamic stability of the Sch black hole based on the Tsallis statistical mechanics.
A thermal system is said to be locally stable if its heat capacity is positive. 
In addition, the phrase ``local" in this context means that the black hole can exist by itself. 
When the black hole is locally unstable (characterized by negative heat capacity), it will evaporate due to thermal radiation and eventually disappear. For simplicity, we consider the process with fixing $\eta$, the heat capacity of the black hole undergoing this process is defined by
\begin{eqnarray}
\displaystyle
C_{\eta(\text{Sch})} &=& \left(\frac{\partial M}{\partial T_{\eta, H(\text{Sch})}}\right)_{\eta}
=-\frac{A_{h}\exp\left[\displaystyle{\frac{11\eta A_{h}}{4(1+3\eta A_{h})}}\right](1+3\eta A_{h}) (1 + 36 \eta A_{h})^{2}}{2+\eta A_{h} \Big[205 + 18 \eta A_{h} (25+36 \eta A_{h})\Big]}.
\label{C SdS rb}
\end{eqnarray}
The behavior of the heat capacity is illustrated in Fig.~\ref{T C}. It is found that the heat capacity in the GB limit is always negative, $\displaystyle{\lim_{\eta \rightarrow 0} C_{\eta(\text{Sch})} = -A_{h}/2}$, implying that the Sch black hole is locally unstable based on GB statistical mechanics.
Due to the fact that the change under the process of fixing $\eta$ is that with respect to the change of $A_h$.
As previously mentioned, the change of the black hole mass is always positive so that the sign of the heat capacity is identified by the sign of the slope of the temperature $dT_{\eta,H(\text{Sch})}/dA_h$ (see the first equality in Eq.~\eqref{C SdS rb}).
According to the analysis of the behavior of the temperature, the black hole with positive $\eta$ is always locally unstable.
On the other hand, the black hole with negative $\eta$ can be both locally unstable and stable for sufficiently small size and large size, respectively.
The transition between these two phases is at $A_{C_\eta}$.
One can conclude that the non-extensive parameter is constrained by the local stability condition as
\begin{eqnarray}
\displaystyle
-\frac{1}{36}<\eta<0.\label{bound eta C}
\end{eqnarray}
The black hole with the above negative $\eta$ is locally stable in the region: $A_{C_\eta}<A_h<A_\text{max}$ (large-sized black hole phase).
\par
\begin{figure}[ht]\centering
\includegraphics[width=6cm]{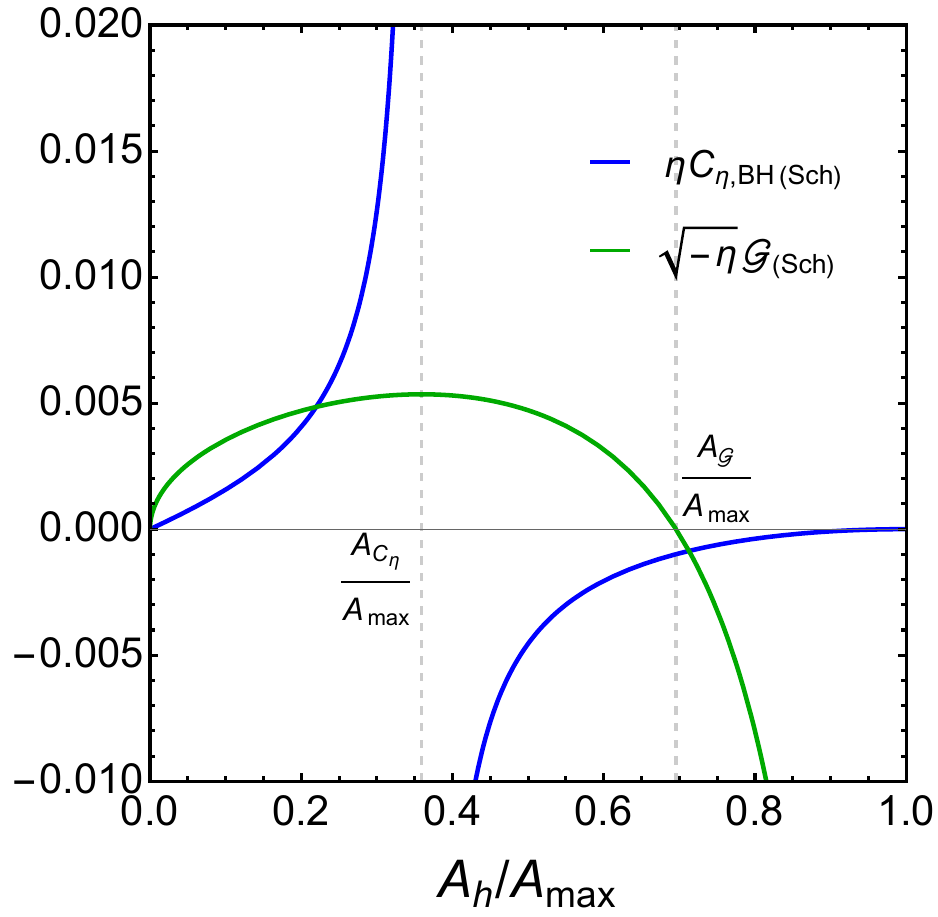}
\qquad
\includegraphics[width=7.3cm]{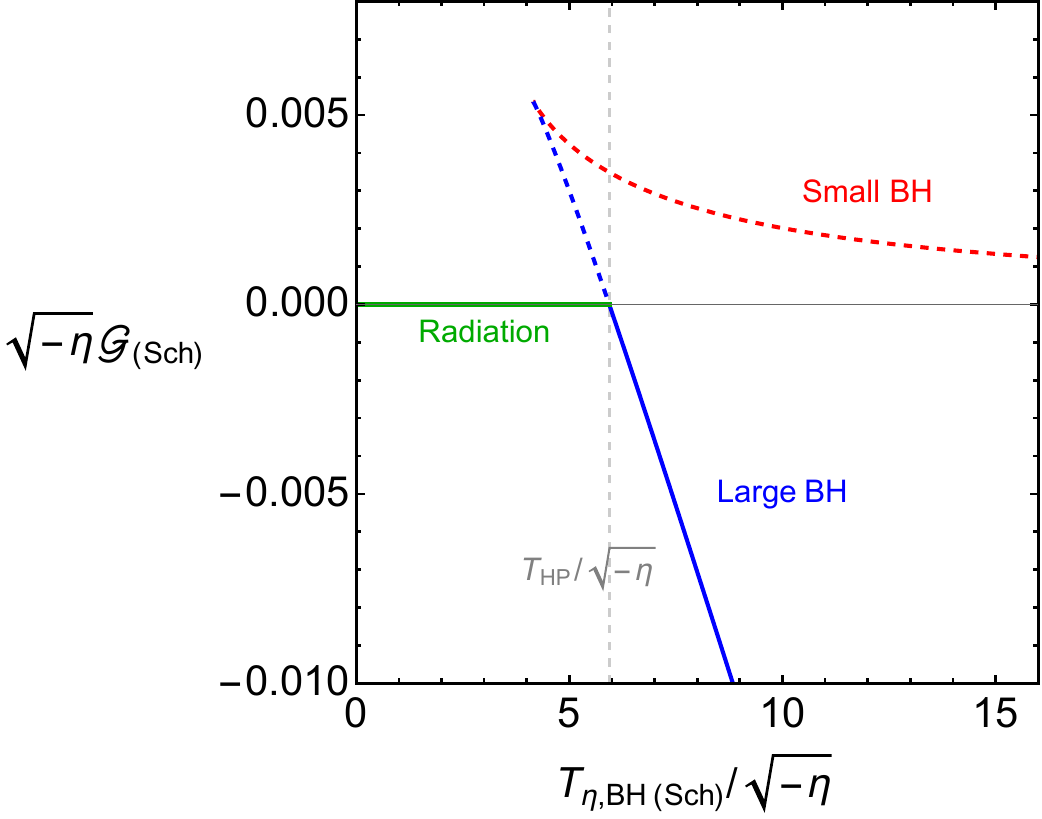}
\caption{The left panel shows the profiles of the heat capacity and Gibbs free energy for negative $\eta$. 
The right panel shows the Gibbs free energy versus the temperature of the black hole with negative $\eta$.}
\label{G}
\end{figure}

Apart from the local stability, the global stability can be determined by analyzing the free energy. 
The phrase ``global'' refers to comparing the black hole with other possible states. 
A black hole is preferred to exist when its free energy must be less than the free energy of a state without a black hole, i.e., radiation or hot gas.
A black hole with free energy greater than radiation's free energy cannot form. Suppose that the radiation has zero free energy, the global stability condition for the black hole is thus the phase with negative free energy. With $\eta$ kept constant, the Gibbs free energy is defined by
\begin{eqnarray}
\displaystyle
\mathcal{G}_\text{(Sch)} 
&=& M-T_{\eta, H(\text{Sch})} S_{\eta, BH(\text{Sch})}, 
\nonumber\\
&=& \frac{-2 - 11\eta A_{h} + 18 \eta^{2} A_{h}^{2} +2(1+3\eta A_{h})\exp\left[\displaystyle\frac{11\eta A_{h}}{4(1+3\eta A_{h})}\right]}{4 \eta \sqrt{\pi A_{h}} (1 + 36\eta A_{h})}.
\label{Gibbs rb}
\end{eqnarray}
The behavior of the free energy is illustrated on the left panel of Fig.~\ref{G}.
Remarkably, the free energy in Eq.~(\ref{Gibbs rb}) can be reduced to one in the GB limit as $\displaystyle\lim_{\eta \rightarrow 0} \mathcal{G}_\text{(Sch)}=\sqrt{A_{h}}/(8\sqrt{\pi})$, which is always positive implying that the Sch black hole is globally unstable based on GB statistical mechanics. For the black hole described by Tsallis statistical mechanics (with negative $\eta$), the free energy can be negative due to the existence of non-extensivity. A lower bound on the area denoted as $A_\mathcal{G}$ of the globally stable black hole is straightforwardly obtained from the vanishing of the numerator of the second line of Eq.~\eqref{Gibbs rb}.
Unfortunately, it is very complicated to analytically solve for $A_\mathcal{G}$ because of dealing with the exponential functions with different bases and exponents.
Instead, we solve numerically and then obtain as
\begin{eqnarray}
\displaystyle
A_\mathcal{G}\approx0.695A_\text{max}.
\end{eqnarray}
Comparing to the bound on $A_h$ in Eq.~\eqref{AC eta}, there always exists the phase of both locally and globally stable black holes with the area: $(A_{C_\eta}<)\,A_\mathcal{G}<A_h<A_\text{max}$.
\par
Furthermore, the behavior of the free energy versus the temperature of the black hole can be illustrated in the right panel of Fig.~\ref{G}. 
From this figure, there exists a cusp corresponding to the area $A_C$ at which the temperature is minimized, the heat capacity diverges and the Gibbs free energy is maximized.
The red and blue dashed lines represent the locally unstable and locally stable but globally unstable phases of the black hole, respectively.  
The radiation or non-black hole and stable black hole phases are represented as the green line of zero free energy and the solid blue line, respectively. 
In addition, there exists a temperature, called the Hawking-Page temperature, denoted by $T_{HP}$, which can be obtained from the condition $\mathcal{G}_\text{(Sch)} = M-T_{HP} S_{\eta, BH\text{(Sch)}} = 0$ as
\begin{equation}
\displaystyle
T_{HP}\approx5.935\sqrt{-\eta}.
\end{equation}
At this temperature, the first-order phase transition from the thermal radiation to the stable large-sized black hole emerges.
It is well-known as the Hawking-Page phase transition \cite{Hawking:1982dh}.
Note that the aforementioned phase transitions occur due to the existence of the non-extensivity.
\par
\begin{figure}[ht]\centering
\includegraphics[width=7.1cm]{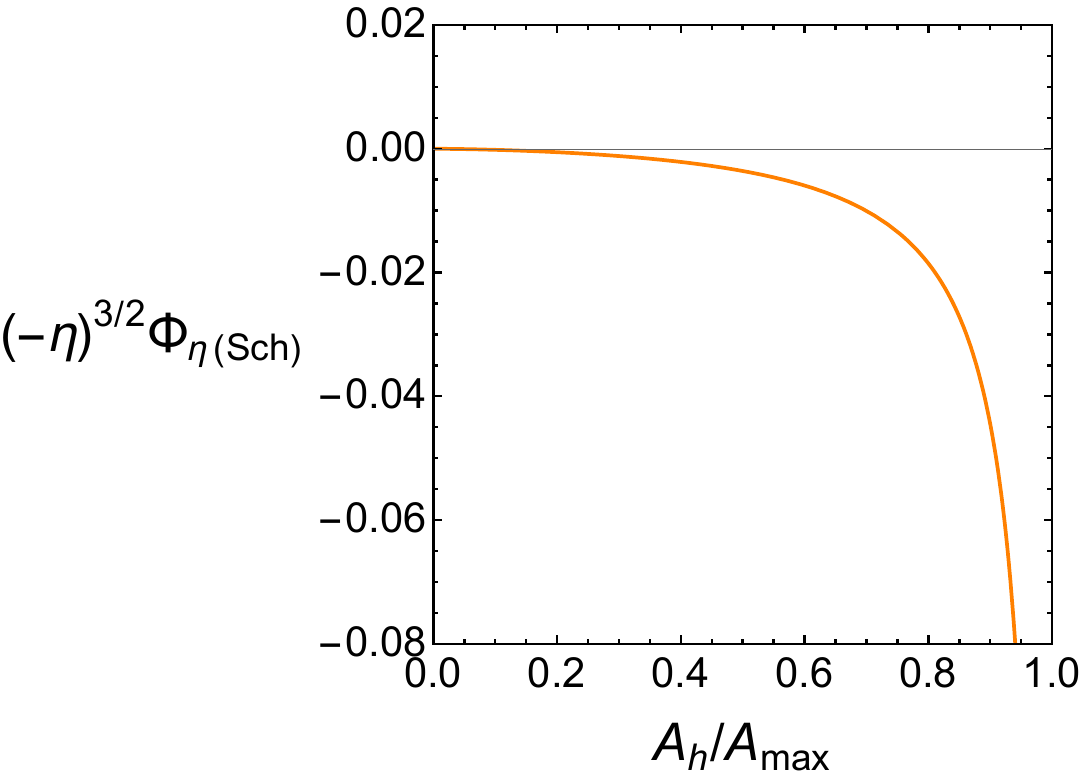}
\qquad
\includegraphics[width=6cm]{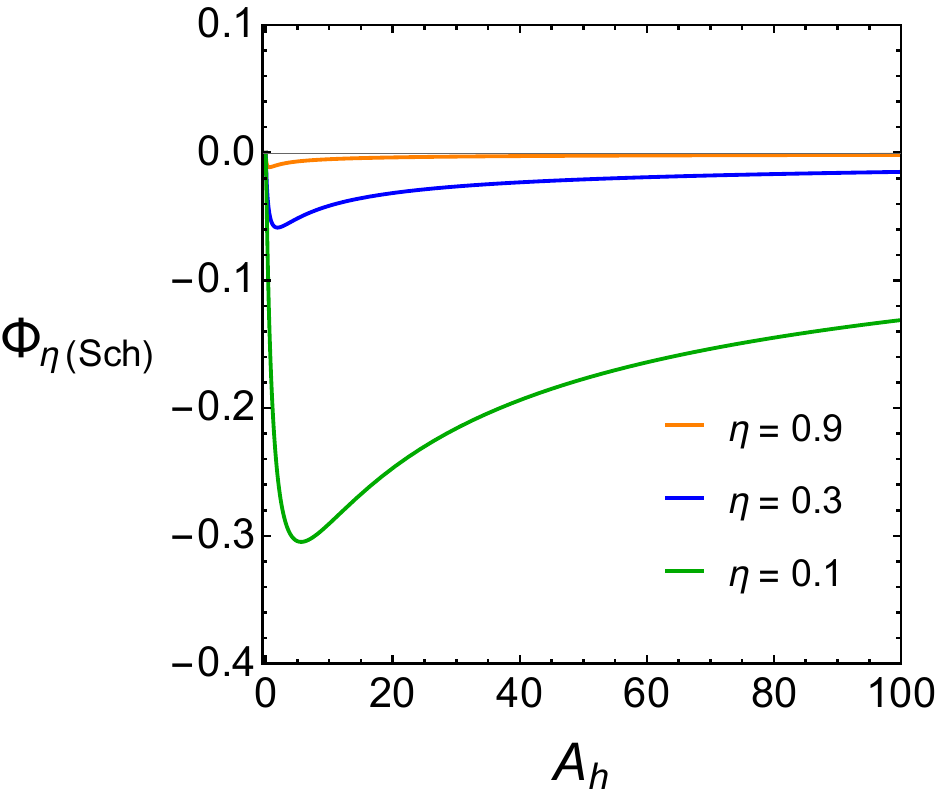}
\caption{The profiles of $\Phi_{\eta\text{(Sch)}}$ with negative $\eta$ (left) and positive $\eta$ (right).}
\label{Phi plot}
\end{figure}

Another possible process, in which the heat transfer exists, is the process of fixing $\Phi_{\eta\text{(Sch)}}$.
The conjugate variable $\Phi_{\eta\text{(Sch)}}$ can be computed using the definition in the second relation of Eq.~\eqref{Sch temp 1} or the Smarr formula~\eqref{Sch modified Smarr 2} as
\begin{eqnarray}
\displaystyle
\Phi_{\eta\text{(Sch)}} 
= \frac{4+23\eta A_{h}-4(1+3\eta A_{h})\exp\left[\displaystyle\frac{11\eta A_{h}}{4(1 + 3\eta A_{h})}\right]}{8 \eta^{2} \sqrt{\pi A_{h}} (1 + 36 \eta A_{h})}.\label{Phi}
\end{eqnarray}
The behavior of $\Phi_{\eta\text{(Sch)}}$ is illustrated in Fig.~\ref{Phi plot}. It is noticed that $\Phi_{\eta\text{(Sch)}}$ always has negative valued for all possible values of $\eta$ and $A_h$. 
\par
According to the fact that  $\Phi_{\eta\text{(Sch)}}$ depends on both $\eta$ and $A_h$, in principle, the behavior of a quantity under the process of fixing $\Phi_{\eta\text{(Sch)}}$ should be expressed in the terms of $\Phi_{\eta\text{(Sch)}}$ and $\eta$ (or $A_h$) obeying the constraint~\eqref{Phi}.
Practically, it is very complicated to solve the above constraint analytically, so a numerical technique may be required to achieve our purpose.
For any change under this process, the derivative $dA_h/d\eta$ must be fixed as
\begin{eqnarray}
\displaystyle
\frac{dA_h}{d\eta}
=2A_h\frac{(3 \eta  A_h+1) (\eta  A_h b_1+8)-e^{\frac{11 \eta  A_h}{4(1+3 \eta  A_h)}} \Big[\eta  A_h \big(72 \eta  A_hb_2+457\big)+8\Big]}{2 \eta  e^{\frac{11 \eta  A_h}{4(1+3 \eta  A_h)}} \Big[\eta  A_h \big(18 \eta  A_h b_2+205\big)+2\Big]-\eta  (3 \eta  A_h+1) \big(\eta  A_hb_3+4\big)}.
\end{eqnarray}
where $b_1=1656 \eta  A_h+455, 
b_2=36 \eta  A_h+25$ and 
$b_3=828 \eta  A_h+409$.
In this part of this work, we will use these manners to deal with computing the relevant thermodynamic quantities.
\par
\begin{figure}[ht]\centering
\includegraphics[width=5.5cm]{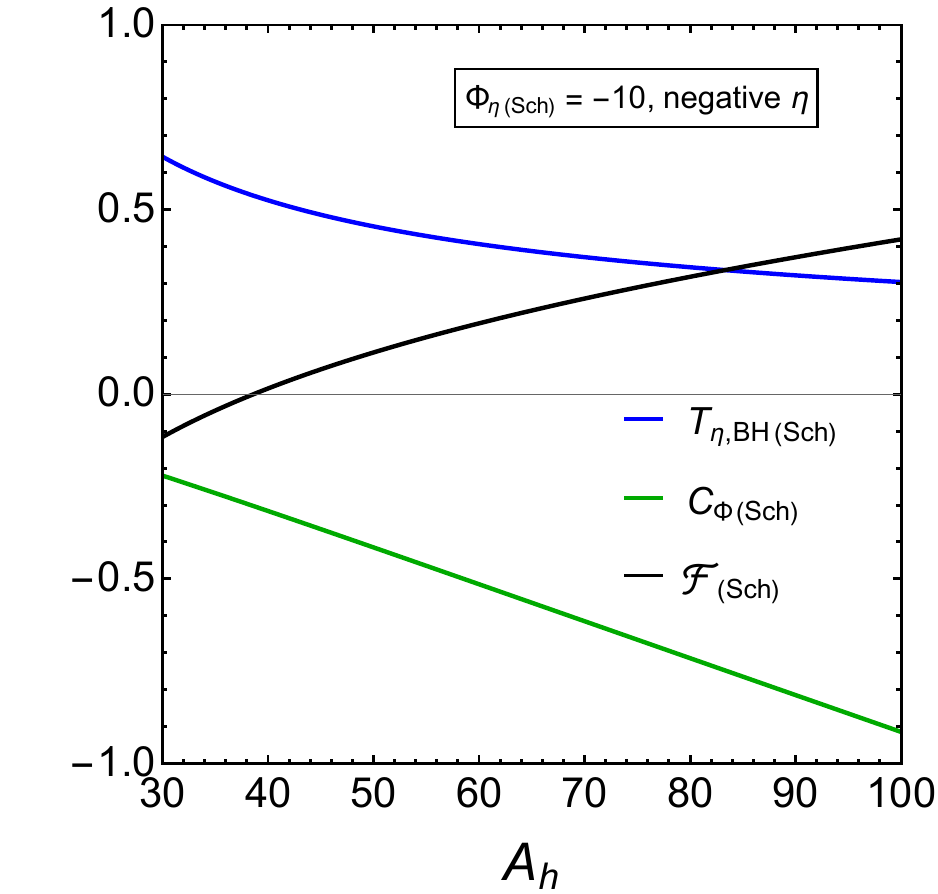}
\qquad
\includegraphics[width=5.5cm]{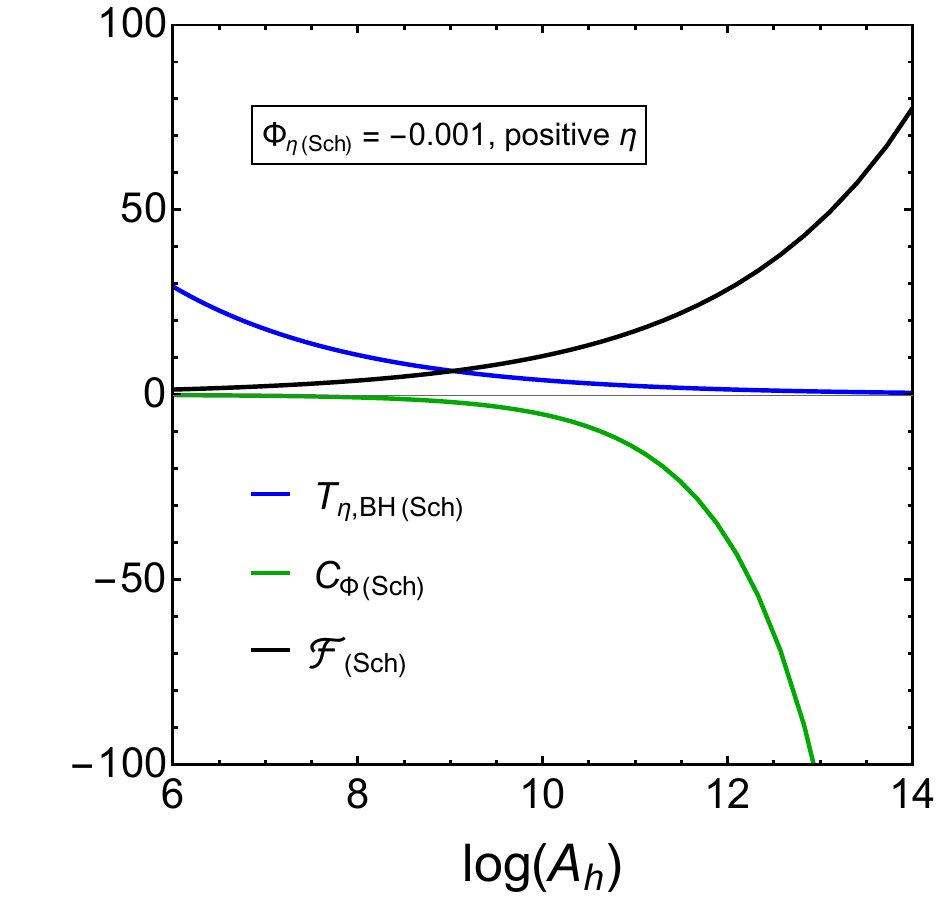}\\\vspace{0.5cm}
\includegraphics[width=7cm]{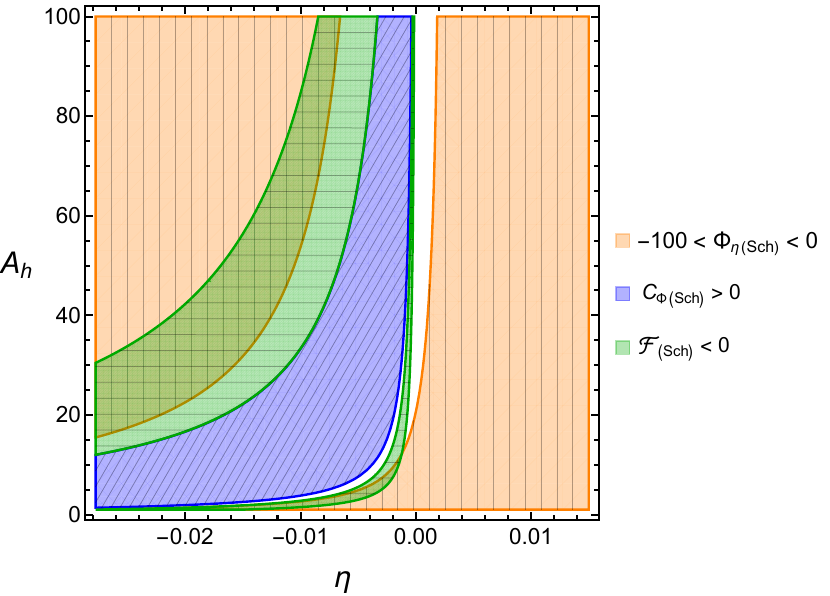}
\caption{The top panels show the profiles of the black hole thermodynamic quantities under $\Phi_{\eta\text{(Sch)}}$-fixed process with negative $\eta$ (left) and positive $\eta$ (right).
The bottom panel shows the regions of existence of $\Phi_{\eta\text{(Sch)}}$, locally stable and globally stable black hole phases.}
\label{pos eta}
\end{figure}

To investigate the thermodynamic stability, let us introduce a new thermodynamic potential $M-\eta\,\Phi_{\eta\text{(Sch)}}$ in which
its change corresponds to
\begin{eqnarray}
\displaystyle
d(M-\eta\,\Phi_{\eta\text{(Sch)}})=T_{\eta,BH\text{(Sch)}}dS_{\eta,BH\text{(Sch)}}-\eta d\Phi_{\eta\text{(Sch)}}.
\end{eqnarray}
Therefore, the heat capacity under the process of fixing $\Phi_{\eta\text{(Sch)}})$ can be computed as follows
\begin{eqnarray}
\displaystyle
C_{\Phi\text{(Sch)}}
&=&\left(\frac{d(M-\eta\,\Phi_{\eta\text{(Sch)}})}{dT_{\eta,BH\text{(Sch)}}}\right)_{\Phi_{\eta\text{(Sch)}}},
\nonumber\\
\displaystyle
&=&-\frac{(1+3\eta A_\eta)\,e^{-\frac{11\eta A_h}{4(1+3\eta A_h)}}}{\eta}\left[\frac{c_1+8\,c_2\,e^{\frac{11\eta A_h}{2(1+3\eta A_h)}}-8\,c_3\,e^{\frac{11\eta A_h}{4(1+3\eta A_h)}}}{c_4-16\,c_5\,e^{\frac{11\eta A_h}{4(1+3\eta A_h)}}}\right],
\end{eqnarray}
where
\begin{eqnarray}
\displaystyle
c_1&=&119232 \eta ^4 A_h^4+55116 \eta ^3 A_h^3+14205 \eta ^2 A_h^2+1744 \eta  A_h+16,\nonumber\\
c_2&=&648 \eta ^3 A_h^3+450 \eta ^2 A_h^2+205 \eta  A_h+2,\nonumber\\
c_3&=&7776 \eta ^4 A_h^4+5508 \eta ^3 A_h^3+2205 \eta ^2 A_h^2+423 \eta  A_h+4,\nonumber\\
c_4&=&59616 \eta ^4 A_h^4+56124 \eta ^3 A_h^3+26181 \eta ^2 A_h^2+3464 \eta  A_h+32,\nonumber\\
c_5&=&1944 \eta ^4 A_h^4+1998 \eta ^3 A_h^3+1065 \eta ^2 A_h^2+211 \eta  A_h+2.\nonumber\label{C Phi}
\end{eqnarray}
The corresponding Helmholtz free energy is given by
\begin{eqnarray}
\displaystyle
\mathcal{F}_\text{(Sch)}
&=&M-\eta\Phi_{\eta\text{(Sch)}}-T_{\eta,BH\text{(Sch)}}S_{\eta,BH\text{(Sch)}},\nonumber\\
\displaystyle
&=&-\frac{8+9\eta A_{h}(5-4\eta A_h)-8(1+3\eta A_{h})\exp\left[\displaystyle\frac{11\eta A_{h}}{4(1 + 3\eta A_{h})}\right]}{8 \eta \sqrt{\pi A_{h}} (1 + 36 \eta A_{h})}.\label{F}
\end{eqnarray}
The behaviors of the thermodynamic quantities are illustrated in the top panels of Fig.~\ref{pos eta}. 
As mentioned, we employ the numerical technique to find the pairs of $(\eta, A_h)$ with fixing $\Phi_{\eta\text{(Sch)}}$.
Then, substituting these pairs into Eqs.~\eqref{TA 1}, \eqref{C SdS rb} and \eqref{F}, one can obtain the temperature, heat capacity, and free energy in terms of a suitable variable along the fixed-$\Phi_{\eta\text{(Sch)}}$ process, respectively. According to the examples in the top panels of Fig.~\ref{pos eta}, the black hole cannot be both locally and globally stable in the process of fixing $\Phi_{\eta\text{(Sch)}}$.
These thermodynamic instabilities are confirmed by the plot in the bottom panel of Fig.~\ref{pos eta}. It is obviously seen that there is no overlapped region at which all the valid $\Phi_{\eta\text{(Sch)}}$, locally stable black hole phase ($C_{\Phi\text{(Sch)}}>0$) and globally stable black hole phase ($\mathcal{F}_\text{(Sch)}<0$) exist.
\par
Before ending this section, let us move our attention to $\Phi_{\eta\text{(Sch)}}$. 
At the small $\eta$ regime, the quantity $\Phi_{\eta}$ in Eq.~\eqref{Phi} can be expanded around $\eta=0$ as
\begin{equation}
\displaystyle
\Phi_{\eta\text{(Sch)}} \approx -\frac{121}{8}\pi r_{bh}^{3} + \frac{55297}{24}\pi^2r_{bh}^2\eta + \mathcal{O} (\eta^{2}).
\end{equation}
The leading term of the above expression is proportional to the three-dimensional volume. 
Accordingly, it is worthwhile to introduce, respectively, the thermodynamic pressure and volume as
\begin{eqnarray}
\displaystyle
P_{\eta\text{(Sch)}} = - \frac{363}{32} \eta,\qquad
\displaystyle
V_{\eta\text{(Sch)}} = -\frac{32}{363}\Phi_{\eta\text{(Sch)}}.
\label{P V}
\end{eqnarray}
At the limit $\eta \rightarrow 0$, one obtains $\displaystyle\lim_{\eta \rightarrow 0} V_{\eta\text{(Sch)}} = \frac{4}{3} \pi r_{bh}^{3}$.
Note that $V_{\eta\text{(Sch)}}$ can behave properly as a positive-valued volume since $\Phi_{\eta\text{(Sch)}}$ is always negative.
Regarding the non-extensive parameter as a state variable, the thermodynamic $P_{\eta\text{(Sch)}}$ behaves as a proper pressure for negative $\eta$ and as a tension for positive $\eta$.
\par
Remarkably, the thermodynamic pressure $P_{\eta\text{(Sch)}}$ is different from the pressure of gas found in Eq.~(\ref{PTsallis}).
It is because the pressure $P_{\eta\text{(Sch)}} = 0$, while $P_{q, loc} \neq 0$, at limit $\eta \rightarrow 0$. Furthermore, according to $P_{\eta\text{(Sch)}}$ and $V_{\eta\text{(Sch)}}$ expressed in Eq.~\eqref{P V}, the Smarr formula and the first law of thermodynamics of the Sch black hole can be rewritten, respectively, as follows
\begin{eqnarray}
\displaystyle
M = 2T_{\eta, H\text{(Sch)}} S_{\eta, BH\text{(Sch)}} - 2P_{\eta\text{(Sch)}} V_{\eta\text{(Sch)}},\\
dM = T_{\eta, H\text{(Sch)}} dS_{\eta, BH\text{(Sch)}} + V_{\eta\text{(Sch)}} dP_{\eta\text{(Sch)}}.
\end{eqnarray}
The above first law is reminiscent of the change of the chemical enthalpy.
Hence, one can interpret the black hole's mass as the enthalpy, and $M-P_{\eta\text{(Sch)}} V_{\eta\text{(Sch)}}=M-\eta \Phi_{\eta\text{(Sch)}}$ as the internal energy.
In addition, $\eta$ and $\Phi_{\eta\text{(Sch)}}$ can also be thought of as the chemical potential and number of particles, respectively.
No matter how $\eta$ and its conjugate are interpreted, the stability analysis done previously will still be reliable.

\section{Conclusion and Discussion}\label{conclusion}
\par
In the first part of this work, we proposed the black hole entropy based on $q$-statistical mechanics. 
Based on traditional GB statistical mechanics, the Bekenstein-Hawking entropy can be derived by treating the system near a black hole as the shell of the indistinguishable classical gas with the maximum number of gas particles, i.e., $N=A_{h}/l_{P}^{2}$. The significant conditions are imposed as follows: 
i) the gas system is in thermal equilibrium with the black hole, i.e. the gas temperature is equal to the Hawking one, 
ii) it approaches very closely the black hole's horizon with a distance of one-tenth of the Planck length. For Tsallis statistical mechanics, it can be thought of as the one-parameter extension of GB statistical mechanics which is characterized by the non-extensive entropy. With Tsallis statistical mechanics, the gas $q$-entropy was derived as found in Eq. (\ref{q-entropy 1}). Moreover, we proposed that the internal energy is written as $U_{q, loc} = 2^{1/3} \Gamma_{q}(3)^{-1/3} U_{loc}$ in order to obtain the unknown function $\mathcal{J}(q)$, from which the term $\Gamma_{q}(3)$ in the gas $q$-entropy can be eliminated. According to the $\Gamma_{q}(3)$-independent entropy of gas, it is possible to formulate the black hole entropy as a homogeneous function. By following a similar manner in deriving the black hole entropy as done for the GB case, the resulting black hole $q$-entropy has been shown in Eq.~\eqref{BH q-entropy 1}. Note that in our derivation, we also employ the formal logarithm map for the sake of factorization of the gas distribution function. With Taylor’s series around the GB limit ($q\to1$), the leading order term of the $q$-entropy is the standard Bekenstein-Hawking one. The non-extensivity of non-extensive statistical mechanics affects the number of microstates of black holes. According to the principle of the maximization of the entropy, the non-extensive parameter $q$ must be positive, and then the non-extensivity can both reduce and increase the number of black hole’s microstates (see Eq.~\eqref{no of micro q}).
\par
Thermodynamic stabilities of the Sch black hole have been investigated in the second part of this work. The thermodynamic phase-space associated with $q$-statistical mechanics can be constructed by adopting Euler’s theorem for the $q$-entropy which is the homogeneous function of degree 1. As a result, the Smarr formula and the first law of thermodynamics were derived in Eqs.~\eqref{Sch modified Smarr 2} and \eqref{Sch Smarr 1 law}, respectively. It is emphasized that the non-extensive parameter $\eta=(1-q)$ automatically plays a role as the thermodynamic variable. To realize the thermodynamic stability, we firstly focused on the scenario that the thermal system associated with Sch black hole undergoes the process of fixing $\eta$. We found that, unlike the consideration based on the GB entropy, the non-extensivity can stabilize the black hole. In particular, the system can be both locally and globally stable when the non-extensive parameter lies in the range $-1/36 < \eta < 0$. Furthermore, we considered another possible process that $\Phi_{\eta\text{(Sch)}}$ is fixed. It is found that for such a process, the black hole cannot be both locally and globally stable confirmed by the plot in the bottom panel of Fig.~\ref{pos eta}.
\par
According to Eq.~\eqref{no of micro q}, the black hole with negative $\eta$ has a number of microstates less than the system described by GB statistical mechanics. In this case, we can imagine that the Planck area units are partially overlapped and then there are over-counting states. As a result, the actual number of microstates is less than the one counted from GB statistical mechanics. Furthermore, in the viewpoint of the total energy, the non-extensive system has non-additive total energy, for example, for $N = 2$, $E = E_{1} + E_{2} - (1-q) \beta_{q} E_{1} E_{2}$. The non-additive term, i.e., $-(1-q) \beta_{q} E_{1} E_{2}$, can be interpreted as the classical correlation between particles due to long-range interaction. This correlation might correspond to the overlapping region of the Planck area units as we have argued previously. The aforementioned argument may provide insight into the thermodynamic nature of black holes via statistical mechanics.

One of the key results of the present work is the outcome of black hole entropy, $q$-entropy, found in Eq.~\eqref{Sbh 1}. 
Up to our knowledge, it is not expressed in the well-known form as found in the literature. 
In fact, this entropy is quite unique due to its complicated form as the exponential function of the horizon area.  
In order to compare our result to the other, it is worthwhile to use an approximation where $\eta \rightarrow 0$ and then analyze the resulting correction by comparing it to one of the quantum corrections.
One of the usual quantum corrections of the area law entropy is the logarithmic correction given by $\alpha \ln (A/4)$ \cite{Banerjee:2008ry,Banerjee:2008cf} where $\alpha$ is a dimensionless constant which is usually $\alpha = -1/2$ or $\alpha = -3/2$ \cite{Kaul:2000kf,Domagala:2004jt}. 
One can see that the quantum correction is always negative which is consistent with our result where the correction can be written as $\frac{\eta}{2} (11 A/4)^2$. 
As a result, it may be inferred that quantum correction influences the number of microstates in the same way as non-extensivity does. 
This may shed light on the connection between the quantum nature of the black hole and the non-extensivity behavior of the black hole.

\par
In the manner presented in this work, the black hole entropy is derived independently of the type of black hole solution and different types of gas near the black hole. Therefore, it is worthwhile to apply this manner to investigate the thermodynamic implications based on non-extensive statistical mechanics for other black holes. This may pave the way to exploring the effect of non-extensivity on black hole thermodynamics. Moreover, the thermodynamic stability analysis in our consideration corresponds to the system undergoing the processes of fixing $\eta$ and its conjugate, $\Phi_{\eta}$. 

It is important to note that our investigation is mainly involved in the development of black hole thermodynamics which seems to be interested only in theoretical aspects. 
In order to capture such properties of the black hole, it is natural to ask whether it is possible to find some observational signatures to identify the black hole's thermodynamic properties. 
In fact, it is possible to characterize the behavior of the quasinormal modes (QNMs) by analyzing the signature in the gravitational waves emitted from a black hole during the ringdown stage \cite{He:2010zb,Liu:2014gvf,Chabab:2016cem,Zou:2017juz}. 
Moreover, some intrinsic properties of the black hole can be identified by using null geodesics of test particles moving near the black hole \cite{Wei:2017mwc,Zhang:2019tzi,Zhang:2019glo,Belhaj:2020nqy,Du:2022quq}. Unfortunately, such direct signatures of black hole thermodynamics are still difficult to detect from recent observations. 
However, recent investigation into the geodesic instability of the test particles, specifically the Lyapunov exponents of both massless and massive particles, serves as an order parameter to investigate the thermodynamic phase transition of the black hole \cite{Guo:2022kio,Yang:2023hci,Lyu:2023sih,Kumara:2024obd,Hale:2024lzh,Promsiri:2024hrl}. 
This might be a possible way to connect the observational signature to the properties of black hole thermodynamics. 
The further investigation of the thermodynamic phase transition of the black hole with Tsallis entropy considered in this work in order to link to the Lyapunov exponents may shed light on the interplay between the non-extensive nature of the black hole and the observational signatures. 
This is an interesting issue to investigate and we will leave this investigation for further work.

\appendix

\section{The first law of thermodynamics and consistent thermodynamic quantities}\label{q-quantity}
\par
The first law of thermodynamics relating the thermodynamic quantities, e.g., the internal energy, can be constructed in the following manner. Let us begin with calculating the total differential of $U_{q}$:
\begin{equation}
\displaystyle
dU_{q} = q\int d \mathcal{V} f^{q-1} E \delta f + \int d \mathcal{V} f^{q} \delta E.
\label{dU_{q} 1}
\end{equation}
Moreover, the total differential of $S_{q}$ can be written as
\begin{equation}
\displaystyle
d S_{q} = \frac{k_{B}}{1-q} \int d \mathcal{V} (qf^{q-1} - 1) \delta f = \frac{k_{B} q}{1-q} \int d \mathcal{V} f^{q-1} \delta f.
\label{dSq}
\end{equation}
The leftover term on the right-hand side in Eq. (\ref{dSq}) is obtained by using the variation of the normalization condition: $\delta \int d \mathcal{V} f = 0 = \int d \mathcal{V} \delta f$. From Eq. (\ref{q-distribution 1}), it can be rearranged as
\begin{equation}
\displaystyle
E = \frac{1}{(1-q) \beta_{q}} (1 - f^{1-q} Z^{1-q}_{q}). \label{Eq}
\end{equation}
According to Eq. (\ref{Eq}), the first term on the right-hand side of Eq. (\ref{dU_{q} 1}) can be expressed as
\begin{eqnarray}
\displaystyle
q \int d \mathcal{V} f^{q-1} E \delta f &=& \frac{q}{(1-q) \beta_{q}} \int d \mathcal{V} f^{q-1} (1 - f^{1-q} Z^{1-q}_{q}) \delta f\nonumber\\ &=& \frac{q}{(1-q) \beta_{q}} \int d \mathcal{V} f^{q-1} \delta f.
\label{Eq2}
\end{eqnarray}
Substituting Eq. (\ref{dSq}) in Eq. (\ref{Eq2}), it reads
\begin{equation}
\displaystyle
q \int d \mathcal{V} f^{q-1} E \delta f = T_{q} dS_{q}.
\label{heat term}
\end{equation}
Obviously, the first term on the right-hand side of Eq. (\ref{dU_{q} 1}) is a heat-term, i.e., $\delta Q = T_{q} dS_{q}$. Immediately, the second (leftover) term can be interpreted as a work term:
\begin{equation}
\displaystyle
\int d \mathcal{V} f^{q} \delta E = \delta W_{q}. \label{work term}
\end{equation}
Substituting Eqs. (\ref{heat term}) and (\ref{work term}) in Eq. (\ref{dU_{q} 1}), the first law of thermodynamics can be expressed as \cite{Tsallis:1998ws}
\begin{equation}
\displaystyle
dU_{q} = T_{q} dS_{q} + \delta W_{q}, \label{q-1st law}
\end{equation}
Specially, for the hydrostatic system described by $(P_{q}, V, T_{q})$, the first law of thermodynamics can be expressed as
\begin{equation}
\displaystyle
dU_{q} = T_{q} dS_{q} - P_{q} dV,
\end{equation}
with setting the work term as $\delta W_{q} = -P_{q} dV$ where $P_{q}$ is the mean $q$-pressure. Furthermore, the consistent thermodynamic quantities are defined by
\begin{equation}
\displaystyle
\begin{array}{lcr}
T_{q} = \displaystyle \left(\frac{\partial U_{q}}{\partial S_{q}}\right)_{V}, \qquad
P_{q} = \displaystyle -\left(\frac{\partial U_{q}}{\partial V}\right)_{S_{q}}.
\end{array}
\end{equation}
\par
Instead of expressing the thermodynamic quantities in terms of the distribution function, e.g., $U_{q} = \int d\mathcal{V} f^{q} E$, those can be written in terms of the partition function as follows. Let us start with considering the internal energy:
\begin{equation}
\displaystyle
U_{q} = \int d \mathcal{V} f^{q} E.
\label{mean energy 1}
\end{equation}
Using Eq. (\ref{q-partition}), the derivative of $Z_{q}$ with respect to $\beta_{q}$ can be obtained as
\begin{equation}
\displaystyle
\frac{\partial Z_{q}}{\partial \beta_{q}} = -\int d \mathcal{V} E (1 - (1-q) \beta_{q} E)^{\frac{q}{1-q}}.
\label{q-dive partition 1}
\end{equation}
Substituting Eq. (\ref{q-dive partition 1}) in Eq. (\ref{mean energy 1}), the internal energy reads
\begin{equation}
\displaystyle
U_{q} = -\frac{1}{Z_{q}^{q}} \frac{\partial Z_{q}}{\partial \beta_{q}} = -\frac{\partial}{\partial \beta_{q}} \ln_{q} Z_{q}.
\label{mean energy 3}
\end{equation}
Note that one obtains $\displaystyle{\lim_{q \rightarrow 1} U_{q} = -\partial \ln_{1} Z/\partial \beta}$ as expected.
\par
Furthermore, the mean $q$-pressure is defined by
\begin{equation}
\displaystyle
P_{q} = \int d \mathcal{V} f^{q} P^{'}_{q}.
\label{mean pressure 1}
\end{equation}
By following the same manner done for the internal energy, we have
\begin{equation}
\displaystyle
\int d \mathcal{V} f^{q} P^{'}_{q} = \frac{1}{\beta_{q}} \frac{\partial}{\partial V} \ln_{q} Z_{q}.
\label{q-dive partition 2}
\end{equation}
Note that we have employed the thermodynamic relation $P^{'}_{q} = -\partial E/\partial V$ in Eq. (\ref{q-dive partition 2}). Substituting Eq. (\ref{q-dive partition 2}) in Eq. (\ref{mean pressure 1}), one hence obtains
\begin{equation}
\displaystyle
P_{q} = \frac{1}{\beta_{q}} \frac{\partial}{\partial V} \ln_{q} Z_{q}.
\label{mean pressure 3}
\end{equation}
Again, we can check that the mean $q$-pressure can be reduced to one in the GB case at limit $q \rightarrow 1$.
\par
For the $q$-entropy related to those quantities expressed in Eqs. (\ref{mean energy 3}) and (\ref{mean pressure 3}) via the first law of thermodynamics, it can be constructed as follows. Let us consider the partition function depending on $\beta_{q}$ and $V$, i.e., $\ln_{q} Z_{q} = \ln_{q} Z_{q} (\beta_{q}, V)$, and hence the total differential of $\ln_{q} Z_{q}$ can be written as follows
\begin{eqnarray}
\displaystyle
d \ln_{q} Z_{q} &=& \left(\frac{\partial}{\partial \beta_{q}} \ln_{q} Z_{q}\right)_{V} d\beta_{q} + \left(\frac{\partial}{\partial V} \ln_{q} Z_{q}\right)_{\beta_{q}} dV, \nonumber\\
\displaystyle
&=& d\left(\beta_{q} \frac{\partial}{\partial \beta_{q}} \ln_{q} Z_{q}\right)_{V} - \beta_{q} d \left(\frac{\partial}{\partial \beta_{q}} \ln_{q} Z_{q}\right)_{V} + \left(\frac{\partial}{\partial V} \ln_{q} Z_{q}\right)_{\beta_{q}} dV, \nonumber\\
\displaystyle
d \left[\left(1 - \beta_{q} \frac{\partial}{\partial \beta_{q}}\right) \ln_{q} Z_{q}\right] &=& \beta_{q} d \left(-\frac{\partial}{\partial \beta_{q}} \ln_{q} Z_{q}\right) + \left(\frac{\partial}{\partial V} \ln_{q} Z_{q}\right)_{\beta_{q}} dV.
\end{eqnarray}
The total differential of the internal energy is obtained as 
\begin{eqnarray}
\displaystyle
dU_{q} = T_{q} d \left[k_{B} \left(1 - \beta_{q} \frac{\partial}{\partial \beta_{q}}\right) \ln_{q} Z_{q}\right] - P_{q} dV.
\label{q-entropy Z 1}
\end{eqnarray}
From Eq. (\ref{q-entropy Z 1}), it is obviously seen that the $q$-entropy can be properly defined by
\begin{equation}
\displaystyle
S_{q} = k_{B} \left(1 - \beta_{q} \frac{\partial}{\partial \beta_{q}}\right) \ln_{q} Z_{q}.
\label{q-entropy Z 2}
\end{equation}
Note that, the $q$-entropy can reduce to one in GB statistics:
\begin{equation}
\displaystyle
\lim_{q \rightarrow 1} S_{q} = k_{B} \left(1 - \beta \frac{\partial}{\partial \beta}\right) \ln Z.
\end{equation}
\par
Besides the internal energy, the other thermodynamic potentials, e.g., Gibbs free energy, can be constructed by using the Legendre transform. As seen from Eq. (\ref{q-1st law}), the quantities $S_{q}$ and $T_{q}$ are the Legendre pair of one another. Accordingly, the Helmholtz free energy can be defined by using the Legendre transform of the internal energy with the Legendre pair $S_{q}$ and $T_{q}$ as \cite{Tsallis:1998ws}
\begin{equation}
\displaystyle
\mathcal{F}_{q} = U_{q} - T_{q} S_{q}. \label{q-free energy 1}
\end{equation}
Remarkably, the thermodynamic Legendre structure remains valid for all values
of $q$. By employing the same manner as done for the those thermodynamic quantities, the Helmholtz free energy can be written in terms of the partition function as
\begin{equation}
\displaystyle
\mathcal{F}_{q}
= - k_{B} T _{q} \ln_{q} Z_{q}. \label{q-free energy 2}
\end{equation}
The first law of thermodynamics with the Helmholtz free energy can be expressed as follows
\begin{eqnarray}
\displaystyle
d\mathcal{F}_{q} &=& dU_{q} - T_{q} dS_{q} - S_{q} dT_{q}
=-S_{q} dT_{q} - P_{q} dV.
\end{eqnarray}
In addition, the corresponding thermodynamic quantities can be defined by
\begin{equation}
\displaystyle
\begin{array}{lcr}
S_{q} = \displaystyle -\left(\frac{\partial \mathcal{F}_{q}}{\partial T_{q}}\right)_{V}, \qquad P_{q} = \displaystyle - \left(\frac{\partial \mathcal{F}_{q}}{\partial V}\right)_{T_{q}}.
\end{array}
\end{equation}
\par
Apart from the Legendre pair $S_{q}$ and $T_{q}$, the leftover pair are $V$ and $P_{q}$. The Gibbs free energy can be constructed by adopting the Legendre transform of $U_{q}$ together with the Legendre pair $S_{q}$ and $T_{q}$, and $V$ and $P_{q}$ as
\begin{equation}
\displaystyle
G_{q} = U_{q} - T_{q} S_{q} + P_{q} V = H_{q} - T_{q} S_{q},
\end{equation}
where $H_{q}$ is a chemical enthalpy defined by
\begin{equation}
\displaystyle
H_{q} = U_{q} + P_{q} V.
\end{equation}
In addition, the first law of thermodynamics with the Gibbs free energy and the chemical enthalpy can be, respectively, written as follows
\begin{equation}
\displaystyle
dG_{q} = -S_{q} dT_{q} + V dP_{q},
\end{equation}
and
\begin{equation}
\displaystyle
dH_{q} = T_{q} dS_{q} + V dP_{q}.
\end{equation}
The thermodynamic quantities corresponding to the Gibbs free energy and the chemical enthalpy are, respectively, defined by
\begin{equation}
\displaystyle
\begin{array}{lcr}
S_{q} = -\left(\displaystyle \frac{\partial G_{q}}{\partial T_{q}}\right)_{P_{q}}, \qquad V = \left(\displaystyle \frac{\partial G_{q}}{\partial P_{q}}\right)_{T_{q}},
\end{array}
\end{equation}
and
\begin{equation}
\displaystyle
\begin{array}{lcr}
T_{q} = \displaystyle \left(\frac{\partial H_{q}}{\partial S_{q}}\right)_{P_{q}}, \qquad V = \displaystyle  \left(\frac{\partial H_{q}}{\partial P_{q}}\right)_{S_{q}}.
\end{array}
\end{equation}

\section*{Acknowledgement}
\par
This research has received funding support from the NSRF via the Program Management Unit for Human Resources \& Institutional Development,
Research and Innovation [grant number B39G670016], and is also supported by the Second Century Fund (C2F), Chulalongkorn University.
It is with great appreciation that the authors thank the referee for his valuable comments.


\nocite{*}
\bibliography{ref}

\begin{thebibliography}{115}%
\makeatletter
\providecommand \@ifxundefined [1]{%
 \@ifx{#1\undefined}
}%
\providecommand \@ifnum [1]{%
 \ifnum #1\expandafter \@firstoftwo
 \else \expandafter \@secondoftwo
 \fi
}%
\providecommand \@ifx [1]{%
 \ifx #1\expandafter \@firstoftwo
 \else \expandafter \@secondoftwo
 \fi
}%
\providecommand \natexlab [1]{#1}%
\providecommand \enquote  [1]{``#1''}%
\providecommand \bibnamefont  [1]{#1}%
\providecommand \bibfnamefont [1]{#1}%
\providecommand \citenamefont [1]{#1}%
\providecommand \href@noop [0]{\@secondoftwo}%
\providecommand \href [0]{\begingroup \@sanitize@url \@href}%
\providecommand \@href[1]{\@@startlink{#1}\@@href}%
\providecommand \@@href[1]{\endgroup#1\@@endlink}%
\providecommand \@sanitize@url [0]{\catcode `\\12\catcode `\$12\catcode
  `\&12\catcode `\#12\catcode `\^12\catcode `\_12\catcode `\%12\relax}%
\providecommand \@@startlink[1]{}%
\providecommand \@@endlink[0]{}%
\providecommand \url  [0]{\begingroup\@sanitize@url \@url }%
\providecommand \@url [1]{\endgroup\@href {#1}{\urlprefix }}%
\providecommand \urlprefix  [0]{URL }%
\providecommand \Eprint [0]{\href }%
\providecommand \doibase [0]{https://doi.org/}%
\providecommand \selectlanguage [0]{\@gobble}%
\providecommand \bibinfo  [0]{\@secondoftwo}%
\providecommand \bibfield  [0]{\@secondoftwo}%
\providecommand \translation [1]{[#1]}%
\providecommand \BibitemOpen [0]{}%
\providecommand \bibitemStop [0]{}%
\providecommand \bibitemNoStop [0]{.\EOS\space}%
\providecommand \EOS [0]{\spacefactor3000\relax}%
\providecommand \BibitemShut  [1]{\csname bibitem#1\endcsname}%
\let\auto@bib@innerbib\@empty
\bibitem [{\citenamefont {Bardeen}\ \emph {et~al.}(1973)\citenamefont
  {Bardeen}, \citenamefont {Carter},\ and\ \citenamefont
  {Hawking}}]{Bardeen:1973gs}%
  \BibitemOpen
  \bibfield  {author} {\bibinfo {author} {\bibfnamefont {J.~M.}\ \bibnamefont
  {Bardeen}}, \bibinfo {author} {\bibfnamefont {B.}~\bibnamefont {Carter}},\
  and\ \bibinfo {author} {\bibfnamefont {S.~W.}\ \bibnamefont {Hawking}},\
  }\bibfield  {title} {\bibinfo {title} {{The Four laws of black hole
  mechanics}},\ }\href {https://doi.org/10.1007/BF01645742} {\bibfield
  {journal} {\bibinfo  {journal} {Commun. Math. Phys.}\ }\textbf {\bibinfo
  {volume} {31}},\ \bibinfo {pages} {161} (\bibinfo {year} {1973})}\BibitemShut
  {NoStop}%
\bibitem [{\citenamefont {Bekenstein}(1973)}]{Bekenstein:1973ur}%
  \BibitemOpen
  \bibfield  {author} {\bibinfo {author} {\bibfnamefont {J.~D.}\ \bibnamefont
  {Bekenstein}},\ }\bibfield  {title} {\bibinfo {title} {{Black holes and
  entropy}},\ }\href {https://doi.org/10.1103/PhysRevD.7.2333} {\bibfield
  {journal} {\bibinfo  {journal} {Phys. Rev. D}\ }\textbf {\bibinfo {volume}
  {7}},\ \bibinfo {pages} {2333} (\bibinfo {year} {1973})}\BibitemShut
  {NoStop}%
\bibitem [{\citenamefont {Bekenstein}(1974)}]{Bekenstein:1974ax}%
  \BibitemOpen
  \bibfield  {author} {\bibinfo {author} {\bibfnamefont {J.~D.}\ \bibnamefont
  {Bekenstein}},\ }\bibfield  {title} {\bibinfo {title} {{Generalized second
  law of thermodynamics in black hole physics}},\ }\href
  {https://doi.org/10.1103/PhysRevD.9.3292} {\bibfield  {journal} {\bibinfo
  {journal} {Phys. Rev. D}\ }\textbf {\bibinfo {volume} {9}},\ \bibinfo {pages}
  {3292} (\bibinfo {year} {1974})}\BibitemShut {NoStop}%
\bibitem [{\citenamefont {Hawking}(1975)}]{Hawking:1975vcx}%
  \BibitemOpen
  \bibfield  {author} {\bibinfo {author} {\bibfnamefont {S.~W.}\ \bibnamefont
  {Hawking}},\ }\bibfield  {title} {\bibinfo {title} {{Particle Creation by
  Black Holes}},\ }\href {https://doi.org/10.1007/BF02345020} {\bibfield
  {journal} {\bibinfo  {journal} {Commun. Math. Phys.}\ }\textbf {\bibinfo
  {volume} {43}},\ \bibinfo {pages} {199} (\bibinfo {year} {1975})},\ \bibinfo
  {note} {[Erratum: Commun.Math.Phys. 46, 206 (1976)]}\BibitemShut {NoStop}%
\bibitem [{\citenamefont {Wallace}(2010)}]{Wallace:2009dc}%
  \BibitemOpen
  \bibfield  {author} {\bibinfo {author} {\bibfnamefont {D.}~\bibnamefont
  {Wallace}},\ }\bibfield  {title} {\bibinfo {title} {{Gravity, Entropy, and
  Cosmology: In Search of Clarity}},\ }\href
  {https://doi.org/10.1093/bjps/axp048} {\bibfield  {journal} {\bibinfo
  {journal} {Brit. J. Phil. Sci.}\ }\textbf {\bibinfo {volume} {61}},\ \bibinfo
  {pages} {513} (\bibinfo {year} {2010})},\ \Eprint
  {https://arxiv.org/abs/0907.0659} {arXiv:0907.0659 [cond-mat.stat-mech]}
  \BibitemShut {NoStop}%
\bibitem [{\citenamefont {Padmanabhan}(2011)}]{Padmanabhan:2010xe}%
  \BibitemOpen
  \bibfield  {author} {\bibinfo {author} {\bibfnamefont {T.}~\bibnamefont
  {Padmanabhan}},\ }\bibfield  {title} {\bibinfo {title} {{Lessons from
  Classical Gravity about the Quantum Structure of Spacetime}},\ }\href
  {https://doi.org/10.1088/1742-6596/306/1/012001} {\bibfield  {journal}
  {\bibinfo  {journal} {J. Phys. Conf. Ser.}\ }\textbf {\bibinfo {volume}
  {306}},\ \bibinfo {pages} {012001} (\bibinfo {year} {2011})},\ \Eprint
  {https://arxiv.org/abs/1012.4476} {arXiv:1012.4476 [gr-qc]} \BibitemShut
  {NoStop}%
\bibitem [{\citenamefont {Gregoris}\ and\ \citenamefont
  {Ong}(2022)}]{Gregoris:2021fiw}%
  \BibitemOpen
  \bibfield  {author} {\bibinfo {author} {\bibfnamefont {D.}~\bibnamefont
  {Gregoris}}\ and\ \bibinfo {author} {\bibfnamefont {Y.~C.}\ \bibnamefont
  {Ong}},\ }\bibfield  {title} {\bibinfo {title} {{Understanding gravitational
  entropy of black holes: A new proposal via curvature invariants}},\ }\href
  {https://doi.org/10.1103/PhysRevD.105.104017} {\bibfield  {journal} {\bibinfo
   {journal} {Phys. Rev. D}\ }\textbf {\bibinfo {volume} {105}},\ \bibinfo
  {pages} {104017} (\bibinfo {year} {2022})},\ \Eprint
  {https://arxiv.org/abs/2109.11968} {arXiv:2109.11968 [gr-qc]} \BibitemShut
  {NoStop}%
\bibitem [{\citenamefont {Strominger}\ and\ \citenamefont
  {Vafa}(1996)}]{Strominger:1996sh}%
  \BibitemOpen
  \bibfield  {author} {\bibinfo {author} {\bibfnamefont {A.}~\bibnamefont
  {Strominger}}\ and\ \bibinfo {author} {\bibfnamefont {C.}~\bibnamefont
  {Vafa}},\ }\bibfield  {title} {\bibinfo {title} {{Microscopic origin of the
  Bekenstein-Hawking entropy}},\ }\href
  {https://doi.org/10.1016/0370-2693(96)00345-0} {\bibfield  {journal}
  {\bibinfo  {journal} {Phys. Lett. B}\ }\textbf {\bibinfo {volume} {379}},\
  \bibinfo {pages} {99} (\bibinfo {year} {1996})},\ \Eprint
  {https://arxiv.org/abs/hep-th/9601029} {arXiv:hep-th/9601029} \BibitemShut
  {NoStop}%
\bibitem [{\citenamefont {Kolekar}\ and\ \citenamefont
  {Padmanabhan}(2011)}]{Kolekar:2010py}%
  \BibitemOpen
  \bibfield  {author} {\bibinfo {author} {\bibfnamefont {S.}~\bibnamefont
  {Kolekar}}\ and\ \bibinfo {author} {\bibfnamefont {T.}~\bibnamefont
  {Padmanabhan}},\ }\bibfield  {title} {\bibinfo {title} {{Ideal Gas in a
  strong Gravitational field: Area dependence of Entropy}},\ }\href
  {https://doi.org/10.1103/PhysRevD.83.064034} {\bibfield  {journal} {\bibinfo
  {journal} {Phys. Rev. D}\ }\textbf {\bibinfo {volume} {83}},\ \bibinfo
  {pages} {064034} (\bibinfo {year} {2011})},\ \Eprint
  {https://arxiv.org/abs/1012.5421} {arXiv:1012.5421 [gr-qc]} \BibitemShut
  {NoStop}%
\bibitem [{\citenamefont {Mirza}\ \emph {et~al.}(2012)\citenamefont {Mirza},
  \citenamefont {Mohammadzadeh},\ and\ \citenamefont {Raissi}}]{Mirza:2011dd}%
  \BibitemOpen
  \bibfield  {author} {\bibinfo {author} {\bibfnamefont {B.}~\bibnamefont
  {Mirza}}, \bibinfo {author} {\bibfnamefont {H.}~\bibnamefont
  {Mohammadzadeh}},\ and\ \bibinfo {author} {\bibfnamefont {Z.}~\bibnamefont
  {Raissi}},\ }\bibfield  {title} {\bibinfo {title} {{Condensation of an ideal
  gas with intermediate statistics on the horizon}},\ }\href
  {https://doi.org/10.1140/epjc/s10052-012-2152-5} {\bibfield  {journal}
  {\bibinfo  {journal} {Eur. Phys. J. C}\ }\textbf {\bibinfo {volume} {72}},\
  \bibinfo {pages} {2152} (\bibinfo {year} {2012})},\ \Eprint
  {https://arxiv.org/abs/1108.6149} {arXiv:1108.6149 [gr-qc]} \BibitemShut
  {NoStop}%
\bibitem [{\citenamefont {Bhattacharya}\ \emph {et~al.}(2017)\citenamefont
  {Bhattacharya}, \citenamefont {Chakraborty},\ and\ \citenamefont
  {Padmanabhan}}]{Bhattacharya:2017bpl}%
  \BibitemOpen
  \bibfield  {author} {\bibinfo {author} {\bibfnamefont {S.}~\bibnamefont
  {Bhattacharya}}, \bibinfo {author} {\bibfnamefont {S.}~\bibnamefont
  {Chakraborty}},\ and\ \bibinfo {author} {\bibfnamefont {T.}~\bibnamefont
  {Padmanabhan}},\ }\bibfield  {title} {\bibinfo {title} {{Entropy of a box of
  gas in an external gravitational field $-$ revisited}},\ }\href
  {https://doi.org/10.1103/PhysRevD.96.084030} {\bibfield  {journal} {\bibinfo
  {journal} {Phys. Rev. D}\ }\textbf {\bibinfo {volume} {96}},\ \bibinfo
  {pages} {084030} (\bibinfo {year} {2017})},\ \Eprint
  {https://arxiv.org/abs/1702.08723} {arXiv:1702.08723 [gr-qc]} \BibitemShut
  {NoStop}%
\bibitem [{\citenamefont {Li}\ \emph {et~al.}(2021)\citenamefont {Li},
  \citenamefont {Wu}, \citenamefont {Xu},\ and\ \citenamefont
  {Yang}}]{Li:2021zuw}%
  \BibitemOpen
  \bibfield  {author} {\bibinfo {author} {\bibfnamefont {D.}~\bibnamefont
  {Li}}, \bibinfo {author} {\bibfnamefont {B.}~\bibnamefont {Wu}}, \bibinfo
  {author} {\bibfnamefont {Z.-M.}\ \bibnamefont {Xu}},\ and\ \bibinfo {author}
  {\bibfnamefont {W.-L.}\ \bibnamefont {Yang}},\ }\bibfield  {title} {\bibinfo
  {title} {{A shell of bosons in spherically symmetric spacetimes}},\ }\href
  {https://doi.org/10.1016/j.physletb.2021.136588} {\bibfield  {journal}
  {\bibinfo  {journal} {Phys. Lett. B}\ }\textbf {\bibinfo {volume} {820}},\
  \bibinfo {pages} {136588} (\bibinfo {year} {2021})},\ \Eprint
  {https://arxiv.org/abs/2106.08653} {arXiv:2106.08653 [gr-qc]} \BibitemShut
  {NoStop}%
\bibitem [{\citenamefont {Sourtzinou}\ and\ \citenamefont
  {Anastopoulos}(2023)}]{Sourtzinou:2022xhd}%
  \BibitemOpen
  \bibfield  {author} {\bibinfo {author} {\bibfnamefont {E.}~\bibnamefont
  {Sourtzinou}}\ and\ \bibinfo {author} {\bibfnamefont {C.}~\bibnamefont
  {Anastopoulos}},\ }\bibfield  {title} {\bibinfo {title} {{Quantum statistical
  mechanics near a black hole horizon}},\ }\href
  {https://doi.org/10.1103/PhysRevD.107.085006} {\bibfield  {journal} {\bibinfo
   {journal} {Phys. Rev. D}\ }\textbf {\bibinfo {volume} {107}},\ \bibinfo
  {pages} {085006} (\bibinfo {year} {2023})},\ \Eprint
  {https://arxiv.org/abs/2203.14319} {arXiv:2203.14319 [gr-qc]} \BibitemShut
  {NoStop}%
\bibitem [{\citenamefont {Pretorius}\ \emph {et~al.}(1998)\citenamefont
  {Pretorius}, \citenamefont {Vollick},\ and\ \citenamefont
  {Israel}}]{Pretorius:1997wr}%
  \BibitemOpen
  \bibfield  {author} {\bibinfo {author} {\bibfnamefont {F.}~\bibnamefont
  {Pretorius}}, \bibinfo {author} {\bibfnamefont {D.}~\bibnamefont {Vollick}},\
  and\ \bibinfo {author} {\bibfnamefont {W.}~\bibnamefont {Israel}},\
  }\bibfield  {title} {\bibinfo {title} {{An Operational approach to black hole
  entropy}},\ }\href {https://doi.org/10.1103/PhysRevD.57.6311} {\bibfield
  {journal} {\bibinfo  {journal} {Phys. Rev. D}\ }\textbf {\bibinfo {volume}
  {57}},\ \bibinfo {pages} {6311} (\bibinfo {year} {1998})},\ \Eprint
  {https://arxiv.org/abs/gr-qc/9712085} {arXiv:gr-qc/9712085} \BibitemShut
  {NoStop}%
\bibitem [{\citenamefont {Oppenheim}(2002)}]{Oppenheim:2001az}%
  \BibitemOpen
  \bibfield  {author} {\bibinfo {author} {\bibfnamefont {J.}~\bibnamefont
  {Oppenheim}},\ }\bibfield  {title} {\bibinfo {title} {{Area scaling entropies
  for gravitating systems}},\ }\href
  {https://doi.org/10.1103/PhysRevD.65.024020} {\bibfield  {journal} {\bibinfo
  {journal} {Phys. Rev. D}\ }\textbf {\bibinfo {volume} {65}},\ \bibinfo
  {pages} {024020} (\bibinfo {year} {2002})},\ \Eprint
  {https://arxiv.org/abs/gr-qc/0105101} {arXiv:gr-qc/0105101} \BibitemShut
  {NoStop}%
\bibitem [{\citenamefont {Tsallis}(1988)}]{Tsallis:1987eu}%
  \BibitemOpen
  \bibfield  {author} {\bibinfo {author} {\bibfnamefont {C.}~\bibnamefont
  {Tsallis}},\ }\bibfield  {title} {\bibinfo {title} {{Possible Generalization
  of Boltzmann-Gibbs Statistics}},\ }\href {https://doi.org/10.1007/BF01016429}
  {\bibfield  {journal} {\bibinfo  {journal} {J. Statist. Phys.}\ }\textbf
  {\bibinfo {volume} {52}},\ \bibinfo {pages} {479} (\bibinfo {year}
  {1988})}\BibitemShut {NoStop}%
\bibitem [{\citenamefont {Louis-Martinez}(2011)}]{Louis-Martinez:2010jvm}%
  \BibitemOpen
  \bibfield  {author} {\bibinfo {author} {\bibfnamefont {D.~J.}\ \bibnamefont
  {Louis-Martinez}},\ }\bibfield  {title} {\bibinfo {title} {{Classical
  relativistic ideal gas in thermodynamic equilibrium in a uniformly
  accelerated reference frame}},\ }\href
  {https://doi.org/10.1088/0264-9381/28/3/035004} {\bibfield  {journal}
  {\bibinfo  {journal} {Class. Quant. Grav.}\ }\textbf {\bibinfo {volume}
  {28}},\ \bibinfo {pages} {035004} (\bibinfo {year} {2011})},\ \Eprint
  {https://arxiv.org/abs/1012.3063} {arXiv:1012.3063 [physics.class-ph]}
  \BibitemShut {NoStop}%
\bibitem [{\citenamefont
  {Jiulin}(2004{\natexlab{a}})}]{jiulin2004nonextensive}%
  \BibitemOpen
  \bibfield  {author} {\bibinfo {author} {\bibfnamefont {D.}~\bibnamefont
  {Jiulin}},\ }\bibfield  {title} {\bibinfo {title} {{The Nonextensive
  parameter and Tsallis distribution for self-gravitating systems}},\ }\href
  {https://doi.org/10.1209/epl/i2004-10145-2} {\bibfield  {journal} {\bibinfo
  {journal} {EPL}\ }\textbf {\bibinfo {volume} {67}},\ \bibinfo {pages} {893}
  (\bibinfo {year} {2004}{\natexlab{a}})},\ \Eprint
  {https://arxiv.org/abs/cond-mat/0409480} {arXiv:cond-mat/0409480}
  \BibitemShut {NoStop}%
\bibitem [{\citenamefont {Silva}\ and\ \citenamefont
  {Lima}(2005)}]{silva2005relativity}%
  \BibitemOpen
  \bibfield  {author} {\bibinfo {author} {\bibfnamefont {R.}~\bibnamefont
  {Silva}}\ and\ \bibinfo {author} {\bibfnamefont {J.~A.~S.}\ \bibnamefont
  {Lima}},\ }\bibfield  {title} {\bibinfo {title} {{Relativity, nonextensivity,
  and extended power law distributions}},\ }\href
  {https://doi.org/10.1103/PhysRevE.72.057101} {\bibfield  {journal} {\bibinfo
  {journal} {Phys. Rev. E}\ }\textbf {\bibinfo {volume} {72}},\ \bibinfo
  {pages} {057101} (\bibinfo {year} {2005})},\ \Eprint
  {https://arxiv.org/abs/cond-mat/0510201} {arXiv:cond-mat/0510201}
  \BibitemShut {NoStop}%
\bibitem [{\citenamefont {Jiulin}(2007)}]{jiulin2007nonextensivity}%
  \BibitemOpen
  \bibfield  {author} {\bibinfo {author} {\bibfnamefont {D.}~\bibnamefont
  {Jiulin}},\ }\bibfield  {title} {\bibinfo {title} {{Nonextensive power law
  distributions and the $q$-kinetic theory for the systems with
  self-gravitating long-range interactions}},\ }\href
  {https://doi.org/10.1007/s10509-007-9611-8} {\bibfield  {journal} {\bibinfo
  {journal} {Astrophysics and Space Science}\ }\textbf {\bibinfo {volume}
  {312}},\ \bibinfo {pages} {47} (\bibinfo {year} {2007})},\ \Eprint
  {https://arxiv.org/abs/cond-mat/0603803} {arXiv:cond-mat/0603803}
  \BibitemShut {NoStop}%
\bibitem [{\citenamefont {Zhipeng}\ \emph {et~al.}(2011)\citenamefont
  {Zhipeng}, \citenamefont {Jiulin},\ and\ \citenamefont
  {Lina}}]{liu2011nonextensivity}%
  \BibitemOpen
  \bibfield  {author} {\bibinfo {author} {\bibfnamefont {L.}~\bibnamefont
  {Zhipeng}}, \bibinfo {author} {\bibfnamefont {D.}~\bibnamefont {Jiulin}},\
  and\ \bibinfo {author} {\bibfnamefont {G.}~\bibnamefont {Lina}},\ }\bibfield
  {title} {\bibinfo {title} {{Nonextensivity and $q$-distribution of a
  relativistic gas under an external electromagnetic field}},\ }\href
  {https://doi.org/10.1007/s11434-011-4750-2} {\bibfield  {journal} {\bibinfo
  {journal} {Chin. Sci. Bull.}\ }\textbf {\bibinfo {volume} {56}},\ \bibinfo
  {pages} {3689} (\bibinfo {year} {2011})},\ \Eprint
  {https://arxiv.org/abs/0802.2492} {arXiv:0802.2492 [cond-mat.stat-mech]}
  \BibitemShut {NoStop}%
\bibitem [{\citenamefont {Zheng}\ \emph {et~al.}(2017)\citenamefont {Zheng},
  \citenamefont {Du},\ and\ \citenamefont {Liang}}]{zheng2017limit}%
  \BibitemOpen
  \bibfield  {author} {\bibinfo {author} {\bibfnamefont {Y.}~\bibnamefont
  {Zheng}}, \bibinfo {author} {\bibfnamefont {J.}~\bibnamefont {Du}},\ and\
  \bibinfo {author} {\bibfnamefont {F.}~\bibnamefont {Liang}},\ }\bibfield
  {title} {\bibinfo {title} {{The limit behavior of the evolution of the
  Tsallis entropy in self-gravitating systems}},\ }\href
  {https://doi.org/10.1209/0295-5075/118/50007} {\bibfield  {journal} {\bibinfo
   {journal} {EPL}\ }\textbf {\bibinfo {volume} {118}},\ \bibinfo {pages}
  {50007} (\bibinfo {year} {2017})},\ \Eprint
  {https://arxiv.org/abs/1710.03567} {arXiv:1710.03567 [cond-mat.stat-mech]}
  \BibitemShut {NoStop}%
\bibitem [{\citenamefont {Sheykhi}(2018)}]{Sheykhi:2018dpn}%
  \BibitemOpen
  \bibfield  {author} {\bibinfo {author} {\bibfnamefont {A.}~\bibnamefont
  {Sheykhi}},\ }\bibfield  {title} {\bibinfo {title} {{Modified Friedmann
  Equations from Tsallis Entropy}},\ }\href
  {https://doi.org/10.1016/j.physletb.2018.08.036} {\bibfield  {journal}
  {\bibinfo  {journal} {Phys. Lett. B}\ }\textbf {\bibinfo {volume} {785}},\
  \bibinfo {pages} {118} (\bibinfo {year} {2018})},\ \Eprint
  {https://arxiv.org/abs/1806.03996} {arXiv:1806.03996 [gr-qc]} \BibitemShut
  {NoStop}%
\bibitem [{\citenamefont {Saridakis}\ \emph {et~al.}(2018)\citenamefont
  {Saridakis}, \citenamefont {Bamba}, \citenamefont {Myrzakulov},\ and\
  \citenamefont {Anagnostopoulos}}]{Saridakis:2018unr}%
  \BibitemOpen
  \bibfield  {author} {\bibinfo {author} {\bibfnamefont {E.~N.}\ \bibnamefont
  {Saridakis}}, \bibinfo {author} {\bibfnamefont {K.}~\bibnamefont {Bamba}},
  \bibinfo {author} {\bibfnamefont {R.}~\bibnamefont {Myrzakulov}},\ and\
  \bibinfo {author} {\bibfnamefont {F.~K.}\ \bibnamefont {Anagnostopoulos}},\
  }\bibfield  {title} {\bibinfo {title} {{Holographic dark energy through
  Tsallis entropy}},\ }\href {https://doi.org/10.1088/1475-7516/2018/12/012}
  {\bibfield  {journal} {\bibinfo  {journal} {JCAP}\ }\textbf {\bibinfo
  {volume} {2018}}\bibfield  {number} {\bibinfo  {number} { (012)}},\ }\Eprint
  {https://arxiv.org/abs/1806.01301} {arXiv:1806.01301 [gr-qc]} \BibitemShut
  {NoStop}%
\bibitem [{\citenamefont {Lymperis}\ and\ \citenamefont
  {Saridakis}(2018)}]{Lymperis:2018iuz}%
  \BibitemOpen
  \bibfield  {author} {\bibinfo {author} {\bibfnamefont {A.}~\bibnamefont
  {Lymperis}}\ and\ \bibinfo {author} {\bibfnamefont {E.~N.}\ \bibnamefont
  {Saridakis}},\ }\bibfield  {title} {\bibinfo {title} {{Modified cosmology
  through nonextensive horizon thermodynamics}},\ }\href
  {https://doi.org/10.1140/epjc/s10052-018-6480-y} {\bibfield  {journal}
  {\bibinfo  {journal} {Eur. Phys. J. C}\ }\textbf {\bibinfo {volume} {78}},\
  \bibinfo {pages} {993} (\bibinfo {year} {2018})},\ \Eprint
  {https://arxiv.org/abs/1806.04614} {arXiv:1806.04614 [gr-qc]} \BibitemShut
  {NoStop}%
\bibitem [{\citenamefont {Jizba}\ and\ \citenamefont
  {Lambiase}(2023)}]{Jizba:2023fkp}%
  \BibitemOpen
  \bibfield  {author} {\bibinfo {author} {\bibfnamefont {P.}~\bibnamefont
  {Jizba}}\ and\ \bibinfo {author} {\bibfnamefont {G.}~\bibnamefont
  {Lambiase}},\ }\bibfield  {title} {\bibinfo {title} {{Constraints on Tsallis
  Cosmology from Big Bang Nucleosynthesis and the Relic Abundance of Cold Dark
  Matter Particles}},\ }\href {https://doi.org/10.3390/e25111495} {\bibfield
  {journal} {\bibinfo  {journal} {Entropy}\ }\textbf {\bibinfo {volume} {25}},\
  \bibinfo {pages} {1495} (\bibinfo {year} {2023})},\ \Eprint
  {https://arxiv.org/abs/2310.19045} {arXiv:2310.19045 [gr-qc]} \BibitemShut
  {NoStop}%
\bibitem [{\citenamefont {Sheykhi}(2020)}]{Sheykhi_2020}%
  \BibitemOpen
  \bibfield  {author} {\bibinfo {author} {\bibfnamefont {A.}~\bibnamefont
  {Sheykhi}},\ }\bibfield  {title} {\bibinfo {title} {New explanation for
  accelerated expansion and flat galactic rotation curves},\ }\bibfield
  {journal} {\bibinfo  {journal} {The European Physical Journal C}\ }\textbf
  {\bibinfo {volume} {80}},\ \href
  {https://doi.org/10.1140/epjc/s10052-019-7599-1}
  {10.1140/epjc/s10052-019-7599-1} (\bibinfo {year} {2020})\BibitemShut
  {NoStop}%
\bibitem [{\citenamefont {Dehpour}(2024)}]{Dehpour:2023dfo}%
  \BibitemOpen
  \bibfield  {author} {\bibinfo {author} {\bibfnamefont {M.}~\bibnamefont
  {Dehpour}},\ }\bibfield  {title} {\bibinfo {title} {{Thermal leptogenesis in
  nonextensive cosmology}},\ }\href
  {https://doi.org/10.1140/epjc/s10052-024-12697-7} {\bibfield  {journal}
  {\bibinfo  {journal} {Eur. Phys. J. C}\ }\textbf {\bibinfo {volume} {84}},\
  \bibinfo {pages} {340} (\bibinfo {year} {2024})},\ \Eprint
  {https://arxiv.org/abs/2401.00229} {arXiv:2401.00229 [hep-ph]} \BibitemShut
  {NoStop}%
\bibitem [{\citenamefont {Bir\'o}\ and\ \citenamefont
  {Czinner}(2013)}]{Biro:2013cra}%
  \BibitemOpen
  \bibfield  {author} {\bibinfo {author} {\bibfnamefont {T.~S.}\ \bibnamefont
  {Bir\'o}}\ and\ \bibinfo {author} {\bibfnamefont {V.~G.}\ \bibnamefont
  {Czinner}},\ }\bibfield  {title} {\bibinfo {title} {{A $q$-parameter bound
  for particle spectra based on black hole thermodynamics with R\'enyi
  entropy}},\ }\href {https://doi.org/10.1016/j.physletb.2013.09.032}
  {\bibfield  {journal} {\bibinfo  {journal} {Phys. Lett. B}\ }\textbf
  {\bibinfo {volume} {726}},\ \bibinfo {pages} {861} (\bibinfo {year}
  {2013})},\ \Eprint {https://arxiv.org/abs/1309.4261} {arXiv:1309.4261
  [gr-qc]} \BibitemShut {NoStop}%
\bibitem [{\citenamefont {Czinner}\ and\ \citenamefont
  {Iguchi}(2016)}]{Czinner:2015eyk}%
  \BibitemOpen
  \bibfield  {author} {\bibinfo {author} {\bibfnamefont {V.~G.}\ \bibnamefont
  {Czinner}}\ and\ \bibinfo {author} {\bibfnamefont {H.}~\bibnamefont
  {Iguchi}},\ }\bibfield  {title} {\bibinfo {title} {{R\'enyi Entropy and the
  Thermodynamic Stability of Black Holes}},\ }\href
  {https://doi.org/10.1016/j.physletb.2015.11.061} {\bibfield  {journal}
  {\bibinfo  {journal} {Phys. Lett. B}\ }\textbf {\bibinfo {volume} {752}},\
  \bibinfo {pages} {306} (\bibinfo {year} {2016})},\ \Eprint
  {https://arxiv.org/abs/1511.06963} {arXiv:1511.06963 [gr-qc]} \BibitemShut
  {NoStop}%
\bibitem [{\citenamefont {Czinner}\ and\ \citenamefont
  {Iguchi}(2017)}]{Czinner:2017tjq}%
  \BibitemOpen
  \bibfield  {author} {\bibinfo {author} {\bibfnamefont {V.~G.}\ \bibnamefont
  {Czinner}}\ and\ \bibinfo {author} {\bibfnamefont {H.}~\bibnamefont
  {Iguchi}},\ }\bibfield  {title} {\bibinfo {title} {{Thermodynamics, stability
  and Hawking\textendash{}Page transition of Kerr black holes from R\'enyi
  statistics}},\ }\href {https://doi.org/10.1140/epjc/s10052-017-5453-x}
  {\bibfield  {journal} {\bibinfo  {journal} {Eur. Phys. J. C}\ }\textbf
  {\bibinfo {volume} {77}},\ \bibinfo {pages} {892} (\bibinfo {year} {2017})},\
  \Eprint {https://arxiv.org/abs/1702.05341} {arXiv:1702.05341 [gr-qc]}
  \BibitemShut {NoStop}%
\bibitem [{\citenamefont {Tannukij}\ \emph {et~al.}(2020)\citenamefont
  {Tannukij}, \citenamefont {Wongjun}, \citenamefont {Hirunsirisawat},
  \citenamefont {Deesuwan},\ and\ \citenamefont {Promsiri}}]{Tannukij:2020njz}%
  \BibitemOpen
  \bibfield  {author} {\bibinfo {author} {\bibfnamefont {L.}~\bibnamefont
  {Tannukij}}, \bibinfo {author} {\bibfnamefont {P.}~\bibnamefont {Wongjun}},
  \bibinfo {author} {\bibfnamefont {E.}~\bibnamefont {Hirunsirisawat}},
  \bibinfo {author} {\bibfnamefont {T.}~\bibnamefont {Deesuwan}},\ and\
  \bibinfo {author} {\bibfnamefont {C.}~\bibnamefont {Promsiri}},\ }\bibfield
  {title} {\bibinfo {title} {{Thermodynamics and phase transition of
  spherically symmetric black hole in de Sitter space from R\'enyi
  statistics}},\ }\href {https://doi.org/10.1140/epjp/s13360-020-00517-2}
  {\bibfield  {journal} {\bibinfo  {journal} {Eur. Phys. J. Plus}\ }\textbf
  {\bibinfo {volume} {135}},\ \bibinfo {pages} {500} (\bibinfo {year}
  {2020})},\ \Eprint {https://arxiv.org/abs/2002.00377} {arXiv:2002.00377
  [gr-qc]} \BibitemShut {NoStop}%
\bibitem [{\citenamefont {Promsiri}\ \emph {et~al.}(2020)\citenamefont
  {Promsiri}, \citenamefont {Hirunsirisawat},\ and\ \citenamefont
  {Liewrian}}]{Promsiri:2020jga}%
  \BibitemOpen
  \bibfield  {author} {\bibinfo {author} {\bibfnamefont {C.}~\bibnamefont
  {Promsiri}}, \bibinfo {author} {\bibfnamefont {E.}~\bibnamefont
  {Hirunsirisawat}},\ and\ \bibinfo {author} {\bibfnamefont {W.}~\bibnamefont
  {Liewrian}},\ }\bibfield  {title} {\bibinfo {title} {{Thermodynamics and Van
  der Waals phase transition of charged black holes in flat spacetime via
  R\'enyi statistics}},\ }\href {https://doi.org/10.1103/PhysRevD.102.064014}
  {\bibfield  {journal} {\bibinfo  {journal} {Phys. Rev. D}\ }\textbf {\bibinfo
  {volume} {102}},\ \bibinfo {pages} {064014} (\bibinfo {year} {2020})},\
  \Eprint {https://arxiv.org/abs/2003.12986} {arXiv:2003.12986 [hep-th]}
  \BibitemShut {NoStop}%
\bibitem [{\citenamefont {Promsiri}\ \emph {et~al.}(2021)\citenamefont
  {Promsiri}, \citenamefont {Hirunsirisawat},\ and\ \citenamefont
  {Liewrian}}]{Promsiri:2021hhv}%
  \BibitemOpen
  \bibfield  {author} {\bibinfo {author} {\bibfnamefont {C.}~\bibnamefont
  {Promsiri}}, \bibinfo {author} {\bibfnamefont {E.}~\bibnamefont
  {Hirunsirisawat}},\ and\ \bibinfo {author} {\bibfnamefont {W.}~\bibnamefont
  {Liewrian}},\ }\bibfield  {title} {\bibinfo {title} {{Solid-liquid phase
  transition and heat engine in an asymptotically flat Schwarzschild black hole
  via the R\'enyi extended phase space approach}},\ }\href
  {https://doi.org/10.1103/PhysRevD.104.064004} {\bibfield  {journal} {\bibinfo
   {journal} {Phys. Rev. D}\ }\textbf {\bibinfo {volume} {104}},\ \bibinfo
  {pages} {064004} (\bibinfo {year} {2021})},\ \Eprint
  {https://arxiv.org/abs/2106.02406} {arXiv:2106.02406 [hep-th]} \BibitemShut
  {NoStop}%
\bibitem [{\citenamefont {Nakarachinda}\ \emph {et~al.}(2021)\citenamefont
  {Nakarachinda}, \citenamefont {Hirunsirisawat}, \citenamefont {Tannukij},\
  and\ \citenamefont {Wongjun}}]{Nakarachinda:2021jxd}%
  \BibitemOpen
  \bibfield  {author} {\bibinfo {author} {\bibfnamefont {R.}~\bibnamefont
  {Nakarachinda}}, \bibinfo {author} {\bibfnamefont {E.}~\bibnamefont
  {Hirunsirisawat}}, \bibinfo {author} {\bibfnamefont {L.}~\bibnamefont
  {Tannukij}},\ and\ \bibinfo {author} {\bibfnamefont {P.}~\bibnamefont
  {Wongjun}},\ }\bibfield  {title} {\bibinfo {title} {{Effective
  thermodynamical system of Schwarzschild\textendash{}de Sitter black holes
  from R\'enyi statistics}},\ }\href
  {https://doi.org/10.1103/PhysRevD.104.064003} {\bibfield  {journal} {\bibinfo
   {journal} {Phys. Rev. D}\ }\textbf {\bibinfo {volume} {104}},\ \bibinfo
  {pages} {064003} (\bibinfo {year} {2021})},\ \Eprint
  {https://arxiv.org/abs/2106.02838} {arXiv:2106.02838 [gr-qc]} \BibitemShut
  {NoStop}%
\bibitem [{\citenamefont {Hirunsirisawat}\ \emph {et~al.}(2022)\citenamefont
  {Hirunsirisawat}, \citenamefont {Nakarachinda},\ and\ \citenamefont
  {Promsiri}}]{Hirunsirisawat:2022fsb}%
  \BibitemOpen
  \bibfield  {author} {\bibinfo {author} {\bibfnamefont {E.}~\bibnamefont
  {Hirunsirisawat}}, \bibinfo {author} {\bibfnamefont {R.}~\bibnamefont
  {Nakarachinda}},\ and\ \bibinfo {author} {\bibfnamefont {C.}~\bibnamefont
  {Promsiri}},\ }\bibfield  {title} {\bibinfo {title} {{Emergent phase,
  thermodynamic geometry, and criticality of charged black holes from R\'enyi
  statistics}},\ }\href {https://doi.org/10.1103/PhysRevD.105.124049}
  {\bibfield  {journal} {\bibinfo  {journal} {Phys. Rev. D}\ }\textbf {\bibinfo
  {volume} {105}},\ \bibinfo {pages} {124049} (\bibinfo {year} {2022})},\
  \Eprint {https://arxiv.org/abs/2204.13023} {arXiv:2204.13023 [hep-th]}
  \BibitemShut {NoStop}%
\bibitem [{\citenamefont {Chunaksorn}\ \emph {et~al.}(2022)\citenamefont
  {Chunaksorn}, \citenamefont {Hirunsirisawat}, \citenamefont {Nakarachinda},
  \citenamefont {Tannukij},\ and\ \citenamefont
  {Wongjun}}]{Chunaksorn:2022whl}%
  \BibitemOpen
  \bibfield  {author} {\bibinfo {author} {\bibfnamefont {P.}~\bibnamefont
  {Chunaksorn}}, \bibinfo {author} {\bibfnamefont {E.}~\bibnamefont
  {Hirunsirisawat}}, \bibinfo {author} {\bibfnamefont {R.}~\bibnamefont
  {Nakarachinda}}, \bibinfo {author} {\bibfnamefont {L.}~\bibnamefont
  {Tannukij}},\ and\ \bibinfo {author} {\bibfnamefont {P.}~\bibnamefont
  {Wongjun}},\ }\bibfield  {title} {\bibinfo {title} {{Thermodynamics of
  asymptotically de Sitter black hole in dRGT massive gravity from R\'enyi
  entropy}},\ }\href {https://doi.org/10.1140/epjc/s10052-022-11110-5}
  {\bibfield  {journal} {\bibinfo  {journal} {Eur. Phys. J. C}\ }\textbf
  {\bibinfo {volume} {82}},\ \bibinfo {pages} {1174} (\bibinfo {year}
  {2022})},\ \Eprint {https://arxiv.org/abs/2208.14770} {arXiv:2208.14770
  [gr-qc]} \BibitemShut {NoStop}%
\bibitem [{\citenamefont {Alonso-Serrano}\ \emph {et~al.}(2021)\citenamefont
  {Alonso-Serrano}, \citenamefont {D\c{a}browski},\ and\ \citenamefont
  {Gohar}}]{Alonso-Serrano:2020hpb}%
  \BibitemOpen
  \bibfield  {author} {\bibinfo {author} {\bibfnamefont {A.}~\bibnamefont
  {Alonso-Serrano}}, \bibinfo {author} {\bibfnamefont {M.~P.}\ \bibnamefont
  {D\c{a}browski}},\ and\ \bibinfo {author} {\bibfnamefont {H.}~\bibnamefont
  {Gohar}},\ }\bibfield  {title} {\bibinfo {title} {{Nonextensive Black Hole
  Entropy and Quantum Gravity Effects at the Last Stages of Evaporation}},\
  }\href {https://doi.org/10.1103/PhysRevD.103.026021} {\bibfield  {journal}
  {\bibinfo  {journal} {Phys. Rev. D}\ }\textbf {\bibinfo {volume} {103}},\
  \bibinfo {pages} {026021} (\bibinfo {year} {2021})},\ \Eprint
  {https://arxiv.org/abs/2009.02129} {arXiv:2009.02129 [gr-qc]} \BibitemShut
  {NoStop}%
\bibitem [{\citenamefont {\c{C}imidiker}\ \emph {et~al.}(2023)\citenamefont
  {\c{C}imidiker}, \citenamefont {D\c{a}browski},\ and\ \citenamefont
  {Gohar}}]{Cimidiker:2023kle}%
  \BibitemOpen
  \bibfield  {author} {\bibinfo {author} {\bibfnamefont {I.}~\bibnamefont
  {\c{C}imidiker}}, \bibinfo {author} {\bibfnamefont {M.~P.}\ \bibnamefont
  {D\c{a}browski}},\ and\ \bibinfo {author} {\bibfnamefont {H.}~\bibnamefont
  {Gohar}},\ }\bibfield  {title} {\bibinfo {title} {{Generalized uncertainty
  principle impact on nonextensive black hole thermodynamics}},\ }\href
  {https://doi.org/10.1088/1361-6382/acdb40} {\bibfield  {journal} {\bibinfo
  {journal} {Class. Quant. Grav.}\ }\textbf {\bibinfo {volume} {40}},\ \bibinfo
  {pages} {145001} (\bibinfo {year} {2023})},\ \Eprint
  {https://arxiv.org/abs/2301.00609} {arXiv:2301.00609 [gr-qc]} \BibitemShut
  {NoStop}%
\bibitem [{\citenamefont {Nojiri}\ \emph {et~al.}(2021)\citenamefont {Nojiri},
  \citenamefont {Odintsov},\ and\ \citenamefont {Faraoni}}]{Nojiri:2021czz}%
  \BibitemOpen
  \bibfield  {author} {\bibinfo {author} {\bibfnamefont {S.}~\bibnamefont
  {Nojiri}}, \bibinfo {author} {\bibfnamefont {S.~D.}\ \bibnamefont
  {Odintsov}},\ and\ \bibinfo {author} {\bibfnamefont {V.}~\bibnamefont
  {Faraoni}},\ }\bibfield  {title} {\bibinfo {title} {{Area-law versus R\'enyi
  and Tsallis black hole entropies}},\ }\href
  {https://doi.org/10.1103/PhysRevD.104.084030} {\bibfield  {journal} {\bibinfo
   {journal} {Phys. Rev. D}\ }\textbf {\bibinfo {volume} {104}},\ \bibinfo
  {pages} {084030} (\bibinfo {year} {2021})},\ \Eprint
  {https://arxiv.org/abs/2109.05315} {arXiv:2109.05315 [gr-qc]} \BibitemShut
  {NoStop}%
\bibitem [{\citenamefont {\c{C}imdiker}\ \emph {et~al.}(2023)\citenamefont
  {\c{C}imdiker}, \citenamefont {D\c{a}browski},\ and\ \citenamefont
  {Gohar}}]{_imdiker_2023}%
  \BibitemOpen
  \bibfield  {author} {\bibinfo {author} {\bibfnamefont {I.}~\bibnamefont
  {\c{C}imdiker}}, \bibinfo {author} {\bibfnamefont {M.~P.}\ \bibnamefont
  {D\c{a}browski}},\ and\ \bibinfo {author} {\bibfnamefont {H.}~\bibnamefont
  {Gohar}},\ }\bibfield  {title} {\bibinfo {title} {{Equilibrium temperature
  for black holes with nonextensive entropy}},\ }\bibfield  {journal} {\bibinfo
   {journal} {The European Physical Journal C}\ }\textbf {\bibinfo {volume}
  {83}},\ \href {https://doi.org/10.1140/epjc/s10052-023-11317-0}
  {10.1140/epjc/s10052-023-11317-0} (\bibinfo {year} {2023})\BibitemShut
  {NoStop}%
\bibitem [{\citenamefont {Nakarachinda}\ \emph {et~al.}(2025)\citenamefont
  {Nakarachinda}, \citenamefont {Promsiri}, \citenamefont {Tannukij},\ and\
  \citenamefont {Wongjun}}]{Nakarachinda:2022gsb}%
  \BibitemOpen
  \bibfield  {author} {\bibinfo {author} {\bibfnamefont {R.}~\bibnamefont
  {Nakarachinda}}, \bibinfo {author} {\bibfnamefont {C.}~\bibnamefont
  {Promsiri}}, \bibinfo {author} {\bibfnamefont {L.}~\bibnamefont {Tannukij}},\
  and\ \bibinfo {author} {\bibfnamefont {P.}~\bibnamefont {Wongjun}},\
  }\bibfield  {title} {\bibinfo {title} {{Thermodynamics of black holes with
  R\'enyi entropy from classical gravity}},\ }\href
  {https://doi.org/10.1016/j.nuclphysb.2025.116796} {\bibfield  {journal}
  {\bibinfo  {journal} {Nucl. Phys. B}\ }\textbf {\bibinfo {volume} {1011}},\
  \bibinfo {pages} {116796} (\bibinfo {year} {2025})},\ \Eprint
  {https://arxiv.org/abs/2211.05989} {arXiv:2211.05989 [gr-qc]} \BibitemShut
  {NoStop}%
\bibitem [{\citenamefont {Nojiri}\ \emph
  {et~al.}(2022{\natexlab{a}})\citenamefont {Nojiri}, \citenamefont
  {Odintsov},\ and\ \citenamefont {Faraoni}}]{Nojiri:2022aof}%
  \BibitemOpen
  \bibfield  {author} {\bibinfo {author} {\bibfnamefont {S.}~\bibnamefont
  {Nojiri}}, \bibinfo {author} {\bibfnamefont {S.~D.}\ \bibnamefont
  {Odintsov}},\ and\ \bibinfo {author} {\bibfnamefont {V.}~\bibnamefont
  {Faraoni}},\ }\bibfield  {title} {\bibinfo {title} {{From nonextensive
  statistics and black hole entropy to the holographic dark universe}},\ }\href
  {https://doi.org/10.1103/PhysRevD.105.044042} {\bibfield  {journal} {\bibinfo
   {journal} {Phys. Rev. D}\ }\textbf {\bibinfo {volume} {105}},\ \bibinfo
  {pages} {044042} (\bibinfo {year} {2022}{\natexlab{a}})},\ \Eprint
  {https://arxiv.org/abs/2201.02424} {arXiv:2201.02424 [gr-qc]} \BibitemShut
  {NoStop}%
\bibitem [{\citenamefont {Nojiri}\ \emph
  {et~al.}(2022{\natexlab{b}})\citenamefont {Nojiri}, \citenamefont
  {Odintsov},\ and\ \citenamefont {Paul}}]{Nojiri:2022dkr}%
  \BibitemOpen
  \bibfield  {author} {\bibinfo {author} {\bibfnamefont {S.}~\bibnamefont
  {Nojiri}}, \bibinfo {author} {\bibfnamefont {S.~D.}\ \bibnamefont
  {Odintsov}},\ and\ \bibinfo {author} {\bibfnamefont {T.}~\bibnamefont
  {Paul}},\ }\bibfield  {title} {\bibinfo {title} {{Early and late universe
  holographic cosmology from a new generalized entropy}},\ }\href
  {https://doi.org/10.1016/j.physletb.2022.137189} {\bibfield  {journal}
  {\bibinfo  {journal} {Phys. Lett. B}\ }\textbf {\bibinfo {volume} {831}},\
  \bibinfo {pages} {137189} (\bibinfo {year} {2022}{\natexlab{b}})},\ \Eprint
  {https://arxiv.org/abs/2205.08876} {arXiv:2205.08876 [gr-qc]} \BibitemShut
  {NoStop}%
\bibitem [{\citenamefont {Nojiri}\ \emph
  {et~al.}(2022{\natexlab{c}})\citenamefont {Nojiri}, \citenamefont
  {Odintsov},\ and\ \citenamefont {Faraoni}}]{Nojiri:2022sfd}%
  \BibitemOpen
  \bibfield  {author} {\bibinfo {author} {\bibfnamefont {S.}~\bibnamefont
  {Nojiri}}, \bibinfo {author} {\bibfnamefont {S.~D.}\ \bibnamefont
  {Odintsov}},\ and\ \bibinfo {author} {\bibfnamefont {V.}~\bibnamefont
  {Faraoni}},\ }\bibfield  {title} {\bibinfo {title} {{Alternative entropies
  and consistent black hole thermodynamics}},\ }\href
  {https://doi.org/10.1142/S0219887822502103} {\bibfield  {journal} {\bibinfo
  {journal} {Int. J. Geom. Meth. Mod. Phys.}\ }\textbf {\bibinfo {volume}
  {19}},\ \bibinfo {pages} {2250210} (\bibinfo {year} {2022}{\natexlab{c}})},\
  \Eprint {https://arxiv.org/abs/2207.07905} {arXiv:2207.07905 [gr-qc]}
  \BibitemShut {NoStop}%
\bibitem [{\citenamefont {Nojiri}\ \emph
  {et~al.}(2022{\natexlab{d}})\citenamefont {Nojiri}, \citenamefont
  {Odintsov},\ and\ \citenamefont {Faraoni}}]{Nojiri:2022ljp}%
  \BibitemOpen
  \bibfield  {author} {\bibinfo {author} {\bibfnamefont {S.}~\bibnamefont
  {Nojiri}}, \bibinfo {author} {\bibfnamefont {S.~D.}\ \bibnamefont
  {Odintsov}},\ and\ \bibinfo {author} {\bibfnamefont {V.}~\bibnamefont
  {Faraoni}},\ }\bibfield  {title} {\bibinfo {title} {{New Entropies, Black
  Holes, and Holographic Dark Energy}},\ }\href
  {https://doi.org/10.1007/s10511-023-09759-1} {\bibfield  {journal} {\bibinfo
  {journal} {Astrophysics}\ }\textbf {\bibinfo {volume} {65}},\ \bibinfo
  {pages} {534} (\bibinfo {year} {2022}{\natexlab{d}})},\ \Eprint
  {https://arxiv.org/abs/2208.10235} {arXiv:2208.10235 [gr-qc]} \BibitemShut
  {NoStop}%
\bibitem [{\citenamefont {Odintsov}\ and\ \citenamefont
  {Paul}(2023)}]{Odintsov:2022qnn}%
  \BibitemOpen
  \bibfield  {author} {\bibinfo {author} {\bibfnamefont {S.~D.}\ \bibnamefont
  {Odintsov}}\ and\ \bibinfo {author} {\bibfnamefont {T.}~\bibnamefont
  {Paul}},\ }\bibfield  {title} {\bibinfo {title} {{A non-singular generalized
  entropy and its implications on bounce cosmology}},\ }\href
  {https://doi.org/10.1016/j.dark.2022.101159} {\bibfield  {journal} {\bibinfo
  {journal} {Phys. Dark Univ.}\ }\textbf {\bibinfo {volume} {39}},\ \bibinfo
  {pages} {101159} (\bibinfo {year} {2023})},\ \Eprint
  {https://arxiv.org/abs/2212.05531} {arXiv:2212.05531 [gr-qc]} \BibitemShut
  {NoStop}%
\bibitem [{\citenamefont {D'Agostino}\ and\ \citenamefont
  {Luciano}(2024)}]{DAgostino:2024sgm}%
  \BibitemOpen
  \bibfield  {author} {\bibinfo {author} {\bibfnamefont {R.}~\bibnamefont
  {D'Agostino}}\ and\ \bibinfo {author} {\bibfnamefont {G.~G.}\ \bibnamefont
  {Luciano}},\ }\bibfield  {title} {\bibinfo {title} {{Lagrangian formulation
  of the Tsallis entropy}},\ }\href
  {https://doi.org/10.1016/j.physletb.2024.138987} {\bibfield  {journal}
  {\bibinfo  {journal} {Phys. Lett. B}\ }\textbf {\bibinfo {volume} {857}},\
  \bibinfo {pages} {138987} (\bibinfo {year} {2024})},\ \Eprint
  {https://arxiv.org/abs/2408.13638} {arXiv:2408.13638 [gr-qc]} \BibitemShut
  {NoStop}%
\bibitem [{\citenamefont
  {Volovik}(2025)}]{volovik2024tsalliscirtoentropyblackhole}%
  \BibitemOpen
  \bibfield  {author} {\bibinfo {author} {\bibfnamefont {G.~E.}\ \bibnamefont
  {Volovik}},\ }\bibfield  {title} {\bibinfo {title} {{Tsallis-Cirto entropy of
  black hole and black hole atom}},\ }\href
  {https://doi.org/10.1134/S0021364024603464} {\bibfield  {journal} {\bibinfo
  {journal} {JETP Letters}\ }\textbf {\bibinfo {volume} {121}},\ \bibinfo
  {pages} {243} (\bibinfo {year} {2025})},\ \Eprint
  {https://arxiv.org/abs/2409.15362} {arXiv:2409.15362 [physics.gen-ph]}
  \BibitemShut {NoStop}%
\bibitem [{\citenamefont {Manoharan}\ and\ \citenamefont
  {Shaji}(2025)}]{manoharan2025reconcilingfractionalentropyblack}%
  \BibitemOpen
  \bibfield  {author} {\bibinfo {author} {\bibfnamefont {M.~T.}\ \bibnamefont
  {Manoharan}}\ and\ \bibinfo {author} {\bibfnamefont {N.}~\bibnamefont
  {Shaji}},\ }\bibfield  {title} {\bibinfo {title} {{Reconciling fractional
  entropy and black hole entropy compositions}},\ }\href
  {https://doi.org/10.1140/epjc/s10052-025-14107-y} {\bibfield  {journal}
  {\bibinfo  {journal} {Eur. Phys. J. C}\ }\textbf {\bibinfo {volume} {85}},\
  \bibinfo {pages} {373} (\bibinfo {year} {2025})},\ \Eprint
  {https://arxiv.org/abs/2501.18151} {arXiv:2501.18151 [gr-qc]} \BibitemShut
  {NoStop}%
\bibitem [{\citenamefont {Lu}\ \emph {et~al.}(2025)\citenamefont {Lu},
  \citenamefont {Di~Gennaro},\ and\ \citenamefont {Ong}}]{Lu:2024ppa}%
  \BibitemOpen
  \bibfield  {author} {\bibinfo {author} {\bibfnamefont {H.}~\bibnamefont
  {Lu}}, \bibinfo {author} {\bibfnamefont {S.}~\bibnamefont {Di~Gennaro}},\
  and\ \bibinfo {author} {\bibfnamefont {Y.~C.}\ \bibnamefont {Ong}},\
  }\bibfield  {title} {\bibinfo {title} {{Generalized entropy implies
  varying-G: Horizon area dependent field equations and black hole-cosmology
  coupling}},\ }\href {https://doi.org/10.1016/j.aop.2024.169914} {\bibfield
  {journal} {\bibinfo  {journal} {Annals Phys.}\ }\textbf {\bibinfo {volume}
  {474}},\ \bibinfo {pages} {169914} (\bibinfo {year} {2025})},\ \Eprint
  {https://arxiv.org/abs/2407.00484} {arXiv:2407.00484 [gr-qc]} \BibitemShut
  {NoStop}%
\bibitem [{\citenamefont {R{\'e}nyi}(1959)}]{Rnyi1959OnTD}%
  \BibitemOpen
  \bibfield  {author} {\bibinfo {author} {\bibfnamefont {A.}~\bibnamefont
  {R{\'e}nyi}},\ }\bibfield  {title} {\bibinfo {title} {On the dimension and
  entropy of probability distributions},\ }\href@noop {} {\bibfield  {journal}
  {\bibinfo  {journal} {Acta Mathematica Academiae Scientiarum Hungarica}\
  }\textbf {\bibinfo {volume} {10}},\ \bibinfo {pages} {193} (\bibinfo {year}
  {1959})}\BibitemShut {NoStop}%
\bibitem [{\citenamefont {Barrow}(2020)}]{Barrow:2020tzx}%
  \BibitemOpen
  \bibfield  {author} {\bibinfo {author} {\bibfnamefont {J.~D.}\ \bibnamefont
  {Barrow}},\ }\bibfield  {title} {\bibinfo {title} {{The Area of a Rough Black
  Hole}},\ }\href {https://doi.org/10.1016/j.physletb.2020.135643} {\bibfield
  {journal} {\bibinfo  {journal} {Phys. Lett. B}\ }\textbf {\bibinfo {volume}
  {808}},\ \bibinfo {pages} {135643} (\bibinfo {year} {2020})},\ \Eprint
  {https://arxiv.org/abs/2004.09444} {arXiv:2004.09444 [gr-qc]} \BibitemShut
  {NoStop}%
\bibitem [{\citenamefont {Sharma}\ and\ \citenamefont
  {Mittal}(1975)}]{sharma1975new}%
  \BibitemOpen
  \bibfield  {author} {\bibinfo {author} {\bibfnamefont {B.~D.}\ \bibnamefont
  {Sharma}}\ and\ \bibinfo {author} {\bibfnamefont {D.~P.}\ \bibnamefont
  {Mittal}},\ }\bibfield  {title} {\bibinfo {title} {New non-additive measures
  of entropy for discrete probability distributions},\ }\href@noop {}
  {\bibfield  {journal} {\bibinfo  {journal} {J. Math. Sci}\ }\textbf {\bibinfo
  {volume} {10}},\ \bibinfo {pages} {28} (\bibinfo {year} {1975})}\BibitemShut
  {NoStop}%
\bibitem [{\citenamefont {Bir\'o}\ and\ \citenamefont
  {V\'an}(2011)}]{Biro:2011ncf}%
  \BibitemOpen
  \bibfield  {author} {\bibinfo {author} {\bibfnamefont {T.~S.}\ \bibnamefont
  {Bir\'o}}\ and\ \bibinfo {author} {\bibfnamefont {P.}~\bibnamefont {V\'an}},\
  }\bibfield  {title} {\bibinfo {title} {{Zeroth law compatibility of
  nonadditive thermodynamics}},\ }\href
  {https://doi.org/10.1103/PhysRevE.83.061147} {\bibfield  {journal} {\bibinfo
  {journal} {Phys. Rev. E}\ }\textbf {\bibinfo {volume} {83}},\ \bibinfo
  {pages} {061147} (\bibinfo {year} {2011})}\BibitemShut {NoStop}%
\bibitem [{\citenamefont {Maleki}\ \emph {et~al.}(2024)\citenamefont {Maleki},
  \citenamefont {Ebadi},\ and\ \citenamefont
  {Mohammadzadeh}}]{doi:10.1142/S0217751X24500192}%
  \BibitemOpen
  \bibfield  {author} {\bibinfo {author} {\bibfnamefont {M.}~\bibnamefont
  {Maleki}}, \bibinfo {author} {\bibfnamefont {Z.}~\bibnamefont {Ebadi}},\ and\
  \bibinfo {author} {\bibfnamefont {H.}~\bibnamefont {Mohammadzadeh}},\
  }\bibfield  {title} {\bibinfo {title} {{Nonextensive statistics and the
  entropy on the horizon}},\ }\href {https://doi.org/10.1142/S0217751X24500192}
  {\bibfield  {journal} {\bibinfo  {journal} {Int. J. Mod. Phys. A}\ }\textbf
  {\bibinfo {volume} {39}},\ \bibinfo {pages} {2450019} (\bibinfo {year}
  {2024})}\BibitemShut {NoStop}%
\bibitem [{\citenamefont {Padmanabhan}(1989)}]{Padmanabhan:1989qn}%
  \BibitemOpen
  \bibfield  {author} {\bibinfo {author} {\bibfnamefont {T.}~\bibnamefont
  {Padmanabhan}},\ }\bibfield  {title} {\bibinfo {title} {{Phase volume
  occupied by a test particle around an incipient black hole}},\ }\href
  {https://doi.org/10.1016/0375-9601(89)90562-8} {\bibfield  {journal}
  {\bibinfo  {journal} {Phys. Lett. A}\ }\textbf {\bibinfo {volume} {136}},\
  \bibinfo {pages} {203} (\bibinfo {year} {1989})}\BibitemShut {NoStop}%
\bibitem [{\citenamefont {Poisson}(2009)}]{Poisson:2009pwt}%
  \BibitemOpen
  \bibfield  {author} {\bibinfo {author} {\bibfnamefont {E.}~\bibnamefont
  {Poisson}},\ }\href {https://doi.org/10.1017/CBO9780511606601} {\emph
  {\bibinfo {title} {{A Relativist's Toolkit: The Mathematics of Black-Hole
  Mechanics}}}}\ (\bibinfo  {publisher} {Cambridge University Press},\ \bibinfo
  {year} {2009})\BibitemShut {NoStop}%
\bibitem [{\citenamefont {Tolman}\ and\ \citenamefont
  {Ehrenfest}(1930)}]{PhysRev.36.1791}%
  \BibitemOpen
  \bibfield  {author} {\bibinfo {author} {\bibfnamefont {R.~C.}\ \bibnamefont
  {Tolman}}\ and\ \bibinfo {author} {\bibfnamefont {P.}~\bibnamefont
  {Ehrenfest}},\ }\bibfield  {title} {\bibinfo {title} {Temperature equilibrium
  in a static gravitational field},\ }\href
  {https://doi.org/10.1103/PhysRev.36.1791} {\bibfield  {journal} {\bibinfo
  {journal} {Phys. Rev.}\ }\textbf {\bibinfo {volume} {36}},\ \bibinfo {pages}
  {1791} (\bibinfo {year} {1930})}\BibitemShut {NoStop}%
\bibitem [{\citenamefont {Pathria}\ and\ \citenamefont
  {Beale}(2021)}]{pathria2021}%
  \BibitemOpen
  \bibfield  {author} {\bibinfo {author} {\bibfnamefont {R.}~\bibnamefont
  {Pathria}}\ and\ \bibinfo {author} {\bibfnamefont {P.}~\bibnamefont
  {Beale}},\ }\href {https://books.google.co.th/books?id=esr2DwAAQBAJ} {\emph
  {\bibinfo {title} {Statistical Mechanics}}}\ (\bibinfo  {publisher} {Elsevier
  Science},\ \bibinfo {year} {2021})\BibitemShut {NoStop}%
\bibitem [{\citenamefont {Calmet}\ \emph {et~al.}(2004)\citenamefont {Calmet},
  \citenamefont {Graesser},\ and\ \citenamefont {Hsu}}]{Calmet:2004mp}%
  \BibitemOpen
  \bibfield  {author} {\bibinfo {author} {\bibfnamefont {X.}~\bibnamefont
  {Calmet}}, \bibinfo {author} {\bibfnamefont {M.}~\bibnamefont {Graesser}},\
  and\ \bibinfo {author} {\bibfnamefont {S.~D.~H.}\ \bibnamefont {Hsu}},\
  }\bibfield  {title} {\bibinfo {title} {{Minimum length from quantum mechanics
  and general relativity}},\ }\href
  {https://doi.org/10.1103/PhysRevLett.93.211101} {\bibfield  {journal}
  {\bibinfo  {journal} {Phys. Rev. Lett.}\ }\textbf {\bibinfo {volume} {93}},\
  \bibinfo {pages} {211101} (\bibinfo {year} {2004})},\ \Eprint
  {https://arxiv.org/abs/hep-th/0405033} {arXiv:hep-th/0405033} \BibitemShut
  {NoStop}%
\bibitem [{\citenamefont {Kubiznak}\ \emph {et~al.}(2017)\citenamefont
  {Kubiznak}, \citenamefont {Mann},\ and\ \citenamefont
  {Teo}}]{Kubiznak:2016qmn}%
  \BibitemOpen
  \bibfield  {author} {\bibinfo {author} {\bibfnamefont {D.}~\bibnamefont
  {Kubiznak}}, \bibinfo {author} {\bibfnamefont {R.~B.}\ \bibnamefont {Mann}},\
  and\ \bibinfo {author} {\bibfnamefont {M.}~\bibnamefont {Teo}},\ }\bibfield
  {title} {\bibinfo {title} {{Black hole chemistry: thermodynamics with
  Lambda}},\ }\href {https://doi.org/10.1088/1361-6382/aa5c69} {\bibfield
  {journal} {\bibinfo  {journal} {Class. Quant. Grav.}\ }\textbf {\bibinfo
  {volume} {34}},\ \bibinfo {pages} {063001} (\bibinfo {year} {2017})},\
  \Eprint {https://arxiv.org/abs/1608.06147} {arXiv:1608.06147 [hep-th]}
  \BibitemShut {NoStop}%
\bibitem [{\citenamefont {{Tsallis}}\ \emph {et~al.}(2003)\citenamefont
  {{Tsallis}}, \citenamefont {{Baldovin}}, \citenamefont {{Cerbino}},\ and\
  \citenamefont {{Pierobon}}}]{2003cond.mat..9093T}%
  \BibitemOpen
  \bibfield  {author} {\bibinfo {author} {\bibfnamefont {C.}~\bibnamefont
  {{Tsallis}}}, \bibinfo {author} {\bibfnamefont {F.}~\bibnamefont
  {{Baldovin}}}, \bibinfo {author} {\bibfnamefont {R.}~\bibnamefont
  {{Cerbino}}},\ and\ \bibinfo {author} {\bibfnamefont {P.}~\bibnamefont
  {{Pierobon}}},\ }\bibfield  {title} {\bibinfo {title} {{Introduction to
  Nonextensive Statistical Mechanics and Thermodynamics}},\ }\href
  {https://doi.org/10.48550/arXiv.cond-mat/0309093} {\bibfield  {journal}
  {\bibinfo  {journal} {arXiv e-prints}\ ,\ \bibinfo {eid} {cond-mat/0309093}}
  (\bibinfo {year} {2003})},\ \Eprint {https://arxiv.org/abs/cond-mat/0309093}
  {arXiv:cond-mat/0309093 [cond-mat.stat-mech]} \BibitemShut {NoStop}%
\bibitem [{\citenamefont {Gell-Mann}\ and\ \citenamefont
  {Tsallis}(2004)}]{gell2004nonextensive}%
  \BibitemOpen
  \bibfield  {author} {\bibinfo {author} {\bibfnamefont {M.}~\bibnamefont
  {Gell-Mann}}\ and\ \bibinfo {author} {\bibfnamefont {C.}~\bibnamefont
  {Tsallis}},\ }\href@noop {} {\emph {\bibinfo {title} {Nonextensive entropy:
  interdisciplinary applications}}}\ (\bibinfo  {publisher} {Oxford University
  Press},\ \bibinfo {year} {2004})\BibitemShut {NoStop}%
\bibitem [{\citenamefont {Curado}\ and\ \citenamefont
  {Tsallis}(1991)}]{Curado:1991jc}%
  \BibitemOpen
  \bibfield  {author} {\bibinfo {author} {\bibfnamefont {E.~M.~F.}\
  \bibnamefont {Curado}}\ and\ \bibinfo {author} {\bibfnamefont
  {C.}~\bibnamefont {Tsallis}},\ }\bibfield  {title} {\bibinfo {title}
  {{Generalized statistical mechanics: Connection with thermodynamics}},\
  }\href@noop {} {\bibfield  {journal} {\bibinfo  {journal} {J. Phys. A}\
  }\textbf {\bibinfo {volume} {24}},\ \bibinfo {pages} {L69} (\bibinfo {year}
  {1991})},\ \bibinfo {note} {[Erratum: J.Phys.A 25, 1019 (1992)]}\BibitemShut
  {NoStop}%
\bibitem [{\citenamefont {Tsallis}\ \emph {et~al.}(1998)\citenamefont
  {Tsallis}, \citenamefont {Mendes},\ and\ \citenamefont
  {Plastino}}]{Tsallis:1998ws}%
  \BibitemOpen
  \bibfield  {author} {\bibinfo {author} {\bibfnamefont {C.}~\bibnamefont
  {Tsallis}}, \bibinfo {author} {\bibfnamefont {R.~S.}\ \bibnamefont
  {Mendes}},\ and\ \bibinfo {author} {\bibfnamefont {A.~R.}\ \bibnamefont
  {Plastino}},\ }\bibfield  {title} {\bibinfo {title} {{The Role of constraints
  within generalized nonextensive statistics}},\ }\href
  {https://doi.org/10.1016/S0378-4371(98)00437-3} {\bibfield  {journal}
  {\bibinfo  {journal} {Physica A}\ }\textbf {\bibinfo {volume} {261}},\
  \bibinfo {pages} {534} (\bibinfo {year} {1998})}\BibitemShut {NoStop}%
\bibitem [{\citenamefont {Taruya}\ and\ \citenamefont
  {Sakagami}(2002)}]{Taruya:2001mv}%
  \BibitemOpen
  \bibfield  {author} {\bibinfo {author} {\bibfnamefont {A.}~\bibnamefont
  {Taruya}}\ and\ \bibinfo {author} {\bibfnamefont {M.-a.}\ \bibnamefont
  {Sakagami}},\ }\bibfield  {title} {\bibinfo {title} {{Gravothermal
  catastrophe and generalized entropy of selfgravitating systems}},\ }\href
  {https://doi.org/10.1016/S0378-4371(01)00622-7} {\bibfield  {journal}
  {\bibinfo  {journal} {Physica A}\ }\textbf {\bibinfo {volume} {307}},\
  \bibinfo {pages} {185} (\bibinfo {year} {2002})},\ \Eprint
  {https://arxiv.org/abs/cond-mat/0107494} {arXiv:cond-mat/0107494}
  \BibitemShut {NoStop}%
\bibitem [{\citenamefont {Taruya}\ and\ \citenamefont
  {Sakagami}(2003{\natexlab{a}})}]{Taruya:2002bh}%
  \BibitemOpen
  \bibfield  {author} {\bibinfo {author} {\bibfnamefont {A.}~\bibnamefont
  {Taruya}}\ and\ \bibinfo {author} {\bibfnamefont {M.}~\bibnamefont
  {Sakagami}},\ }\bibfield  {title} {\bibinfo {title} {{Gravothermal
  catastrophe and Tsallis' generalized entropy of selfgravitating systems. 2.
  Thermodynamic properties of stellar polytrope}},\ }\href
  {https://doi.org/10.1016/S0378-4371(02)01342-0} {\bibfield  {journal}
  {\bibinfo  {journal} {Physica A}\ }\textbf {\bibinfo {volume} {318}},\
  \bibinfo {pages} {387} (\bibinfo {year} {2003}{\natexlab{a}})},\ \Eprint
  {https://arxiv.org/abs/cond-mat/0204315} {arXiv:cond-mat/0204315}
  \BibitemShut {NoStop}%
\bibitem [{\citenamefont {Taruya}\ and\ \citenamefont
  {Sakagami}(2003{\natexlab{b}})}]{Taruya:2002jz}%
  \BibitemOpen
  \bibfield  {author} {\bibinfo {author} {\bibfnamefont {A.}~\bibnamefont
  {Taruya}}\ and\ \bibinfo {author} {\bibfnamefont {M.-a.}\ \bibnamefont
  {Sakagami}},\ }\bibfield  {title} {\bibinfo {title} {{Gravothermal
  catastrophe and Tsallis' generalized entropy of selfgravitating systems. 3.
  Equilibrium structure using normalized q values}},\ }\href
  {https://doi.org/10.1016/S0378-4371(03)00039-6} {\bibfield  {journal}
  {\bibinfo  {journal} {Physica A}\ }\textbf {\bibinfo {volume} {322}},\
  \bibinfo {pages} {285} (\bibinfo {year} {2003}{\natexlab{b}})},\ \Eprint
  {https://arxiv.org/abs/cond-mat/0211305} {arXiv:cond-mat/0211305}
  \BibitemShut {NoStop}%
\bibitem [{\citenamefont {Sakagami}\ and\ \citenamefont
  {Taruya}(2004)}]{Sakagami:2003qs}%
  \BibitemOpen
  \bibfield  {author} {\bibinfo {author} {\bibfnamefont {M.-a.}\ \bibnamefont
  {Sakagami}}\ and\ \bibinfo {author} {\bibfnamefont {A.}~\bibnamefont
  {Taruya}},\ }\bibfield  {title} {\bibinfo {title} {{Selfgravitating stellar
  systems and nonextensive thermostatistics}},\ }\href
  {https://doi.org/10.1007/s00161-003-0168-7} {\bibfield  {journal} {\bibinfo
  {journal} {Contin. Mech. Thermodyn.}\ }\textbf {\bibinfo {volume} {16}},\
  \bibinfo {pages} {279} (\bibinfo {year} {2004})},\ \Eprint
  {https://arxiv.org/abs/cond-mat/0310082} {arXiv:cond-mat/0310082}
  \BibitemShut {NoStop}%
\bibitem [{\citenamefont {Zavala}\ \emph {et~al.}(2006)\citenamefont {Zavala},
  \citenamefont {Nunez}, \citenamefont {Sussman}, \citenamefont
  {Cabral-Rosetti},\ and\ \citenamefont {Matos}}]{Zavala:2006de}%
  \BibitemOpen
  \bibfield  {author} {\bibinfo {author} {\bibfnamefont {J.}~\bibnamefont
  {Zavala}}, \bibinfo {author} {\bibfnamefont {D.}~\bibnamefont {Nunez}},
  \bibinfo {author} {\bibfnamefont {R.~A.}\ \bibnamefont {Sussman}}, \bibinfo
  {author} {\bibfnamefont {L.~G.}\ \bibnamefont {Cabral-Rosetti}},\ and\
  \bibinfo {author} {\bibfnamefont {T.}~\bibnamefont {Matos}},\ }\bibfield
  {title} {\bibinfo {title} {{Stellar polytropes and NFW halo model: Stellar
  polytropes and Navarro-Frenk-White halo models: comparison with
  observations}},\ }\href {https://doi.org/10.1088/1475-7516/2006/06/008}
  {\bibfield  {journal} {\bibinfo  {journal} {JCAP}\ }\textbf {\bibinfo
  {volume} {2006}}\bibfield  {number} {\bibinfo  {number} { (008)}},\ }\Eprint
  {https://arxiv.org/abs/astro-ph/0605665} {arXiv:astro-ph/0605665}
  \BibitemShut {NoStop}%
\bibitem [{\citenamefont {Coraddu}\ \emph {et~al.}(1999)\citenamefont
  {Coraddu}, \citenamefont {Kaniadakis}, \citenamefont {Lavagno}, \citenamefont
  {Lissia}, \citenamefont {Mezzorani},\ and\ \citenamefont
  {Quarati}}]{Coraddu:1998yb}%
  \BibitemOpen
  \bibfield  {author} {\bibinfo {author} {\bibfnamefont {M.}~\bibnamefont
  {Coraddu}}, \bibinfo {author} {\bibfnamefont {G.}~\bibnamefont {Kaniadakis}},
  \bibinfo {author} {\bibfnamefont {A.}~\bibnamefont {Lavagno}}, \bibinfo
  {author} {\bibfnamefont {M.}~\bibnamefont {Lissia}}, \bibinfo {author}
  {\bibfnamefont {G.}~\bibnamefont {Mezzorani}},\ and\ \bibinfo {author}
  {\bibfnamefont {P.}~\bibnamefont {Quarati}},\ }\bibfield  {title} {\bibinfo
  {title} {{Thermal distributions in stellar plasmas, nuclear reactions and
  solar neutrinos}},\ }\href {https://doi.org/10.1590/S0103-97331999000100014}
  {\bibfield  {journal} {\bibinfo  {journal} {Braz. J. Phys.}\ }\textbf
  {\bibinfo {volume} {29}},\ \bibinfo {pages} {153} (\bibinfo {year} {1999})},\
  \Eprint {https://arxiv.org/abs/nucl-th/9811081} {arXiv:nucl-th/9811081}
  \BibitemShut {NoStop}%
\bibitem [{\citenamefont {Chakrabarti}\ \emph {et~al.}(2010)\citenamefont
  {Chakrabarti}, \citenamefont {Chandrashekar},\ and\ \citenamefont
  {Naina~Mohammed}}]{Chakrabarti_2010}%
  \BibitemOpen
  \bibfield  {author} {\bibinfo {author} {\bibfnamefont {R.}~\bibnamefont
  {Chakrabarti}}, \bibinfo {author} {\bibfnamefont {R.}~\bibnamefont
  {Chandrashekar}},\ and\ \bibinfo {author} {\bibfnamefont {S.}~\bibnamefont
  {Naina~Mohammed}},\ }\bibfield  {title} {\bibinfo {title} {Nonextensive
  statistics of the classical relativistic ideal gas},\ }\href
  {https://doi.org/10.1016/j.physa.2009.12.040} {\bibfield  {journal} {\bibinfo
   {journal} {Physica A: Statistical Mechanics and its Applications}\ }\textbf
  {\bibinfo {volume} {389}},\ \bibinfo {pages} {1571^^e2^^80^^931584} (\bibinfo
  {year} {2010})}\BibitemShut {NoStop}%
\bibitem [{\citenamefont {Chandrashekar}\ and\ \citenamefont
  {Naina~Mohammed}(2011)}]{Chandrashekar_2011}%
  \BibitemOpen
  \bibfield  {author} {\bibinfo {author} {\bibfnamefont {R.}~\bibnamefont
  {Chandrashekar}}\ and\ \bibinfo {author} {\bibfnamefont {S.~S.}\ \bibnamefont
  {Naina~Mohammed}},\ }\bibfield  {title} {\bibinfo {title} {A class of
  energy-based ensembles in tsallis statistics},\ }\href
  {https://doi.org/10.1088/1742-5468/2011/05/p05018} {\bibfield  {journal}
  {\bibinfo  {journal} {Journal of Statistical Mechanics: Theory and
  Experiment}\ }\textbf {\bibinfo {volume} {2011}},\ \bibinfo {pages} {P05018}
  (\bibinfo {year} {2011})}\BibitemShut {NoStop}%
\bibitem [{\citenamefont {Tsallis}(2002)}]{Tsallis:2002tp}%
  \BibitemOpen
  \bibfield  {author} {\bibinfo {author} {\bibfnamefont {C.}~\bibnamefont
  {Tsallis}},\ }\bibfield  {title} {\bibinfo {title} {{Entropic nonextensivity:
  A Possible measure of complexity}},\ }\href
  {https://doi.org/10.1016/S0960-0779(01)00019-4} {\bibfield  {journal}
  {\bibinfo  {journal} {Chaos Solitons Fractals}\ }\textbf {\bibinfo {volume}
  {13}},\ \bibinfo {pages} {371} (\bibinfo {year} {2002})},\ \Eprint
  {https://arxiv.org/abs/cond-mat/0010150} {arXiv:cond-mat/0010150}
  \BibitemShut {NoStop}%
\bibitem [{\citenamefont {Wilk}\ and\ \citenamefont
  {Wlodarczyk}(2004)}]{Wilk:2002uf}%
  \BibitemOpen
  \bibfield  {author} {\bibinfo {author} {\bibfnamefont {G.}~\bibnamefont
  {Wilk}}\ and\ \bibinfo {author} {\bibfnamefont {Z.}~\bibnamefont
  {Wlodarczyk}},\ }\bibfield  {title} {\bibinfo {title} {{Nonextensive
  information entropy for stochastic networks}},\ }\href@noop {} {\bibfield
  {journal} {\bibinfo  {journal} {Acta Phys. Polon. B}\ }\textbf {\bibinfo
  {volume} {35}},\ \bibinfo {pages} {871} (\bibinfo {year} {2004})},\ \Eprint
  {https://arxiv.org/abs/cond-mat/0212056} {arXiv:cond-mat/0212056}
  \BibitemShut {NoStop}%
\bibitem [{\citenamefont {Wilk}\ and\ \citenamefont
  {W\l{}odarczyk}(2015)}]{Wilk:2014zka}%
  \BibitemOpen
  \bibfield  {author} {\bibinfo {author} {\bibfnamefont {G.}~\bibnamefont
  {Wilk}}\ and\ \bibinfo {author} {\bibfnamefont {Z.}~\bibnamefont
  {W\l{}odarczyk}},\ }\bibfield  {title} {\bibinfo {title} {{Quasi-power law
  ensembles}},\ }\href {https://doi.org/10.5506/APhysPolB.46.1103} {\bibfield
  {journal} {\bibinfo  {journal} {Acta Phys. Polon. B}\ }\textbf {\bibinfo
  {volume} {46}},\ \bibinfo {pages} {1103} (\bibinfo {year} {2015})},\ \Eprint
  {https://arxiv.org/abs/1501.01936} {arXiv:1501.01936 [cond-mat.stat-mech]}
  \BibitemShut {NoStop}%
\bibitem [{\citenamefont {Jiulin}(2004{\natexlab{b}})}]{Jiulin:2004fi}%
  \BibitemOpen
  \bibfield  {author} {\bibinfo {author} {\bibfnamefont {D.}~\bibnamefont
  {Jiulin}},\ }\bibfield  {title} {\bibinfo {title} {{Nonextensive in
  nonequilibrium plasma systems with Coulombian long-range interactions}},\
  }\href {https://doi.org/10.1016/j.physleta.2004.07.010} {\bibfield  {journal}
  {\bibinfo  {journal} {Phys. Lett. A}\ }\textbf {\bibinfo {volume} {329}},\
  \bibinfo {pages} {262} (\bibinfo {year} {2004}{\natexlab{b}})},\ \Eprint
  {https://arxiv.org/abs/cond-mat/0404602} {arXiv:cond-mat/0404602}
  \BibitemShut {NoStop}%
\bibitem [{\citenamefont {Jiulin}(2004{\natexlab{c}})}]{Jiulin:2004bg}%
  \BibitemOpen
  \bibfield  {author} {\bibinfo {author} {\bibfnamefont {D.}~\bibnamefont
  {Jiulin}},\ }\bibfield  {title} {\bibinfo {title} {{The Nonextensive
  parameter and Tsallis distribution for self-gravitating systems}},\ }\href
  {https://doi.org/10.1209/epl/i2004-10145-2} {\bibfield  {journal} {\bibinfo
  {journal} {EPL}\ }\textbf {\bibinfo {volume} {67}},\ \bibinfo {pages} {893}
  (\bibinfo {year} {2004}{\natexlab{c}})},\ \Eprint
  {https://arxiv.org/abs/cond-mat/0409480} {arXiv:cond-mat/0409480}
  \BibitemShut {NoStop}%
\bibitem [{\citenamefont {Mart\'{i}nez}\ \emph {et~al.}(2000)\citenamefont
  {Mart\'{i}nez}, \citenamefont {Nicol\'{a}s}, \citenamefont {Pennini},\ and\
  \citenamefont {Plastino}}]{Mart_nez_2000}%
  \BibitemOpen
  \bibfield  {author} {\bibinfo {author} {\bibfnamefont {S.}~\bibnamefont
  {Mart\'{i}nez}}, \bibinfo {author} {\bibfnamefont {F.}~\bibnamefont
  {Nicol\'{a}s}}, \bibinfo {author} {\bibfnamefont {F.}~\bibnamefont
  {Pennini}},\ and\ \bibinfo {author} {\bibfnamefont {A.}~\bibnamefont
  {Plastino}},\ }\bibfield  {title} {\bibinfo {title} {Tsallis’ entropy
  maximization procedure revisited},\ }\href
  {https://doi.org/10.1016/s0378-4371(00)00359-9} {\bibfield  {journal}
  {\bibinfo  {journal} {Physica A: Statistical Mechanics and its Applications}\
  }\textbf {\bibinfo {volume} {286}},\ \bibinfo {pages} {489^^e2^^80^^93502}
  (\bibinfo {year} {2000})}\BibitemShut {NoStop}%
\bibitem [{\citenamefont {Lenzi}\ \emph {et~al.}(2000)\citenamefont {Lenzi},
  \citenamefont {Mendes},\ and\ \citenamefont {da~Silva}}]{Lenzi:2000gj}%
  \BibitemOpen
  \bibfield  {author} {\bibinfo {author} {\bibfnamefont {E.~K.}\ \bibnamefont
  {Lenzi}}, \bibinfo {author} {\bibfnamefont {R.~S.}\ \bibnamefont {Mendes}},\
  and\ \bibinfo {author} {\bibfnamefont {L.~R.}\ \bibnamefont {da~Silva}},\
  }\bibfield  {title} {\bibinfo {title} {{Statistical mechanics based on Renyi
  entropy}},\ }\href {https://doi.org/10.1016/S0378-4371(00)00007-8} {\bibfield
   {journal} {\bibinfo  {journal} {Physica A}\ }\textbf {\bibinfo {volume}
  {280}},\ \bibinfo {pages} {337} (\bibinfo {year} {2000})}\BibitemShut
  {NoStop}%
\bibitem [{\citenamefont {Ferri}\ \emph
  {et~al.}(2005{\natexlab{a}})\citenamefont {Ferri}, \citenamefont
  {Mart\'{i}nez},\ and\ \citenamefont {Plastino}}]{Ferri:2005yf}%
  \BibitemOpen
  \bibfield  {author} {\bibinfo {author} {\bibfnamefont {G.~L.}\ \bibnamefont
  {Ferri}}, \bibinfo {author} {\bibfnamefont {S.}~\bibnamefont
  {Mart\'{i}nez}},\ and\ \bibinfo {author} {\bibfnamefont {A.}~\bibnamefont
  {Plastino}},\ }\bibfield  {title} {\bibinfo {title} {{The role of constraints
  in Tsallis' nonextensive treatment revisited}},\ }\href@noop {} {\bibfield
  {journal} {\bibinfo  {journal} {Physica A}\ }\textbf {\bibinfo {volume}
  {347}},\ \bibinfo {pages} {205} (\bibinfo {year}
  {2005}{\natexlab{a}})}\BibitemShut {NoStop}%
\bibitem [{\citenamefont {Abe}\ \emph {et~al.}(2001)\citenamefont {Abe},
  \citenamefont {Mart\'{i}nez}, \citenamefont {Pennini},\ and\ \citenamefont
  {Plastino}}]{Abe_2001}%
  \BibitemOpen
  \bibfield  {author} {\bibinfo {author} {\bibfnamefont {S.}~\bibnamefont
  {Abe}}, \bibinfo {author} {\bibfnamefont {S.}~\bibnamefont {Mart\'{i}nez}},
  \bibinfo {author} {\bibfnamefont {F.}~\bibnamefont {Pennini}},\ and\ \bibinfo
  {author} {\bibfnamefont {A.}~\bibnamefont {Plastino}},\ }\bibfield  {title}
  {\bibinfo {title} {Classical gas in nonextensive optimal lagrange multipliers
  formalism},\ }\href {https://doi.org/10.1016/s0375-9601(00)00780-5}
  {\bibfield  {journal} {\bibinfo  {journal} {Physics Letters A}\ }\textbf
  {\bibinfo {volume} {278}},\ \bibinfo {pages} {249^^e2^^80^^93254} (\bibinfo
  {year} {2001})}\BibitemShut {NoStop}%
\bibitem [{\citenamefont {Ferri}\ \emph
  {et~al.}(2005{\natexlab{b}})\citenamefont {Ferri}, \citenamefont
  {Mart\'{i}nez},\ and\ \citenamefont {Plastino}}]{Ferri_2005}%
  \BibitemOpen
  \bibfield  {author} {\bibinfo {author} {\bibfnamefont {G.~L.}\ \bibnamefont
  {Ferri}}, \bibinfo {author} {\bibfnamefont {S.}~\bibnamefont
  {Mart\'{i}nez}},\ and\ \bibinfo {author} {\bibfnamefont {A.}~\bibnamefont
  {Plastino}},\ }\bibfield  {title} {\bibinfo {title} {Equivalence of the four
  versions of tsallis’s statistics},\ }\href
  {https://doi.org/10.1088/1742-5468/2005/04/p04009} {\bibfield  {journal}
  {\bibinfo  {journal} {Journal of Statistical Mechanics: Theory and
  Experiment}\ }\textbf {\bibinfo {volume} {2005}},\ \bibinfo {pages} {P04009}
  (\bibinfo {year} {2005}{\natexlab{b}})}\BibitemShut {NoStop}%
\bibitem [{\citenamefont {Wang}\ \emph
  {et~al.}(2002{\natexlab{a}})\citenamefont {Wang}, \citenamefont {Pezeril},
  \citenamefont {Nivanen},\ and\ \citenamefont {{Le
  M^^c3^^a9haut^^c3^^a9}}}]{WANG2002131}%
  \BibitemOpen
  \bibfield  {author} {\bibinfo {author} {\bibfnamefont {Q.~A.}\ \bibnamefont
  {Wang}}, \bibinfo {author} {\bibfnamefont {M.}~\bibnamefont {Pezeril}},
  \bibinfo {author} {\bibfnamefont {L.}~\bibnamefont {Nivanen}},\ and\ \bibinfo
  {author} {\bibfnamefont {A.}~\bibnamefont {{Le M^^c3^^a9haut^^c3^^a9}}},\
  }\bibfield  {title} {\bibinfo {title} {Nonextensive distribution and
  factorization of the joint probability},\ }\href
  {https://doi.org/10.1016/S0960-0779(00)00244-7} {\bibfield  {journal}
  {\bibinfo  {journal} {Chaos, Solitons \& Fractals}\ }\textbf {\bibinfo
  {volume} {13}},\ \bibinfo {pages} {131} (\bibinfo {year}
  {2002}{\natexlab{a}})}\BibitemShut {NoStop}%
\bibitem [{\citenamefont {Wang}\ \emph
  {et~al.}(2002{\natexlab{b}})\citenamefont {Wang}, \citenamefont {Nivanen},
  \citenamefont {M$eacute$haut$eacute$},\ and\ \citenamefont
  {Pezeril}}]{Wang_2002}%
  \BibitemOpen
  \bibfield  {author} {\bibinfo {author} {\bibfnamefont {Q.~A.}\ \bibnamefont
  {Wang}}, \bibinfo {author} {\bibfnamefont {L.}~\bibnamefont {Nivanen}},
  \bibinfo {author} {\bibfnamefont {A.~L.}\ \bibnamefont
  {M$eacute$haut$eacute$}},\ and\ \bibinfo {author} {\bibfnamefont
  {M.}~\bibnamefont {Pezeril}},\ }\bibfield  {title} {\bibinfo {title} {On the
  generalized entropy pseudoadditivity for complex systems},\ }\href
  {https://doi.org/10.1088/0305-4470/35/33/304} {\bibfield  {journal} {\bibinfo
   {journal} {Journal of Physics A: Mathematical and General}\ }\textbf
  {\bibinfo {volume} {35}},\ \bibinfo {pages} {7003^^e2^^80^^937007} (\bibinfo
  {year} {2002}{\natexlab{b}})}\BibitemShut {NoStop}%
\bibitem [{\citenamefont {Naik}\ and\ \citenamefont
  {Haubold}(2016)}]{axioms5030024}%
  \BibitemOpen
  \bibfield  {author} {\bibinfo {author} {\bibfnamefont {S.~R.}\ \bibnamefont
  {Naik}}\ and\ \bibinfo {author} {\bibfnamefont {H.~J.}\ \bibnamefont
  {Haubold}},\ }\bibfield  {title} {\bibinfo {title} {{On the $q$-Laplace
  Transform and Related Special Functions}},\ }\bibfield  {journal} {\bibinfo
  {journal} {Axioms}\ }\textbf {\bibinfo {volume} {5}},\ \href
  {https://doi.org/10.3390/axioms5030024} {10.3390/axioms5030024} (\bibinfo
  {year} {2016})\BibitemShut {NoStop}%
\bibitem [{\citenamefont {Lynden-Bell}\ \emph {et~al.}(1968)\citenamefont
  {Lynden-Bell}, \citenamefont {Wood},\ and\ \citenamefont
  {Royal}}]{Lynden-Bell:1968awc}%
  \BibitemOpen
  \bibfield  {author} {\bibinfo {author} {\bibfnamefont {D.}~\bibnamefont
  {Lynden-Bell}}, \bibinfo {author} {\bibfnamefont {R.}~\bibnamefont {Wood}},\
  and\ \bibinfo {author} {\bibfnamefont {A.}~\bibnamefont {Royal}},\ }\bibfield
   {title} {\bibinfo {title} {{The Gravo-Thermal Catastrophe in Isothermal
  Spheres and the Onset of Red-Giant Structure for Stellar Systems}},\ }\href
  {https://doi.org/10.1093/mnras/138.4.495} {\bibfield  {journal} {\bibinfo
  {journal} {Mon. Not. Roy. Astron. Soc.}\ }\textbf {\bibinfo {volume} {138}},\
  \bibinfo {pages} {495} (\bibinfo {year} {1968})}\BibitemShut {NoStop}%
\bibitem [{\citenamefont {Chunaksorn}\ \emph {et~al.}(2024)\citenamefont
  {Chunaksorn}, \citenamefont {Nakarachinda},\ and\ \citenamefont
  {Wongjun}}]{chunaksorn2024qequilibriumgasspacetimemultihorizon}%
  \BibitemOpen
  \bibfield  {author} {\bibinfo {author} {\bibfnamefont {P.}~\bibnamefont
  {Chunaksorn}}, \bibinfo {author} {\bibfnamefont {R.}~\bibnamefont
  {Nakarachinda}},\ and\ \bibinfo {author} {\bibfnamefont {P.}~\bibnamefont
  {Wongjun}},\ }\href {https://arxiv.org/abs/2402.13742} {\bibinfo {title}
  {$q$-equilibrium of gas in spacetime of multi-horizon black holes}} (\bibinfo
  {year} {2024}),\ \Eprint {https://arxiv.org/abs/2402.13742} {arXiv:2402.13742
  [gr-qc]} \BibitemShut {NoStop}%
\bibitem [{\citenamefont {Tsallis}\ and\ \citenamefont
  {Cirto}(2013)}]{Tsallis_2013}%
  \BibitemOpen
  \bibfield  {author} {\bibinfo {author} {\bibfnamefont {C.}~\bibnamefont
  {Tsallis}}\ and\ \bibinfo {author} {\bibfnamefont {L.~J.~L.}\ \bibnamefont
  {Cirto}},\ }\bibfield  {title} {\bibinfo {title} {Black hole thermodynamical
  entropy},\ }\bibfield  {journal} {\bibinfo  {journal} {The European Physical
  Journal C}\ }\textbf {\bibinfo {volume} {73}},\ \href
  {https://doi.org/10.1140/epjc/s10052-013-2487-6}
  {10.1140/epjc/s10052-013-2487-6} (\bibinfo {year} {2013})\BibitemShut
  {NoStop}%
\bibitem [{\citenamefont {Tsallis}(2019)}]{article}%
  \BibitemOpen
  \bibfield  {author} {\bibinfo {author} {\bibfnamefont {C.}~\bibnamefont
  {Tsallis}},\ }\bibfield  {title} {\bibinfo {title} {Black hole entropy: A
  closer look},\ }\href {https://doi.org/10.3390/e22010017} {\bibfield
  {journal} {\bibinfo  {journal} {Entropy}\ }\textbf {\bibinfo {volume} {22}},\
  \bibinfo {pages} {17} (\bibinfo {year} {2019})}\BibitemShut {NoStop}%
\bibitem [{\citenamefont {Pessoa}\ and\ \citenamefont
  {Arderucio~Costa}(2020)}]{Pessoa:2020cih}%
  \BibitemOpen
  \bibfield  {author} {\bibinfo {author} {\bibfnamefont {P.}~\bibnamefont
  {Pessoa}}\ and\ \bibinfo {author} {\bibfnamefont {B.}~\bibnamefont
  {Arderucio~Costa}},\ }\bibfield  {title} {\bibinfo {title} {{Comment on
  \textquotedblleft{}Black Hole Entropy: A Closer Look\textquotedblright{}}},\
  }\href {https://doi.org/10.3390/e22101110} {\bibfield  {journal} {\bibinfo
  {journal} {Entropy}\ }\textbf {\bibinfo {volume} {22}},\ \bibinfo {pages}
  {1110} (\bibinfo {year} {2020})},\ \Eprint {https://arxiv.org/abs/2008.13744}
  {arXiv:2008.13744 [gr-qc]} \BibitemShut {NoStop}%
\bibitem [{\citenamefont {Tsallis}(2021)}]{Tsallis:2021mvq}%
  \BibitemOpen
  \bibfield  {author} {\bibinfo {author} {\bibfnamefont {C.}~\bibnamefont
  {Tsallis}},\ }\bibfield  {title} {\bibinfo {title} {{Reply to Pessoa, P.;
  Arderucio Costa, B. Comment on \textquotedblleft{}Tsallis, C. Black Hole
  Entropy: A Closer Look. Entropy 2020, 22, 17\textquotedblright{}}},\ }\href
  {https://doi.org/10.3390/e23050630} {\bibfield  {journal} {\bibinfo
  {journal} {Entropy}\ }\textbf {\bibinfo {volume} {23}},\ \bibinfo {pages}
  {630} (\bibinfo {year} {2021})}\BibitemShut {NoStop}%
\bibitem [{\citenamefont {Pessoa}\ \emph {et~al.}(2022)\citenamefont {Pessoa},
  \citenamefont {Arderucio~Costa},\ and\ \citenamefont
  {Press\'e}}]{Pessoa:2022fwo}%
  \BibitemOpen
  \bibfield  {author} {\bibinfo {author} {\bibfnamefont {P.}~\bibnamefont
  {Pessoa}}, \bibinfo {author} {\bibfnamefont {B.}~\bibnamefont
  {Arderucio~Costa}},\ and\ \bibinfo {author} {\bibfnamefont {S.}~\bibnamefont
  {Press\'e}},\ }\bibfield  {title} {\bibinfo {title} {{Revisiting Claims in
  ``Black Hole Entropy: A Closer Look''}},\ }\href@noop {} {\  (\bibinfo {year}
  {2022})},\ \Eprint {https://arxiv.org/abs/2210.00324} {arXiv:2210.00324
  [gr-qc]} \BibitemShut {NoStop}%
\bibitem [{\citenamefont {Smarr}(1973)}]{PhysRevLett.30.71}%
  \BibitemOpen
  \bibfield  {author} {\bibinfo {author} {\bibfnamefont {L.}~\bibnamefont
  {Smarr}},\ }\bibfield  {title} {\bibinfo {title} {Mass formula for kerr black
  holes},\ }\href {https://doi.org/10.1103/PhysRevLett.30.71} {\bibfield
  {journal} {\bibinfo  {journal} {Phys. Rev. Lett.}\ }\textbf {\bibinfo
  {volume} {30}},\ \bibinfo {pages} {71} (\bibinfo {year} {1973})}\BibitemShut
  {NoStop}%
\bibitem [{\citenamefont {Hawking}\ and\ \citenamefont
  {Page}(1983)}]{Hawking:1982dh}%
  \BibitemOpen
  \bibfield  {author} {\bibinfo {author} {\bibfnamefont {S.~W.}\ \bibnamefont
  {Hawking}}\ and\ \bibinfo {author} {\bibfnamefont {D.~N.}\ \bibnamefont
  {Page}},\ }\bibfield  {title} {\bibinfo {title} {{Thermodynamics of Black
  Holes in anti-De Sitter Space}},\ }\href {https://doi.org/10.1007/BF01208266}
  {\bibfield  {journal} {\bibinfo  {journal} {Commun. Math. Phys.}\ }\textbf
  {\bibinfo {volume} {87}},\ \bibinfo {pages} {577} (\bibinfo {year}
  {1983})}\BibitemShut {NoStop}%
\bibitem [{\citenamefont {Banerjee}\ and\ \citenamefont
  {Majhi}(2008{\natexlab{a}})}]{Banerjee:2008ry}%
  \BibitemOpen
  \bibfield  {author} {\bibinfo {author} {\bibfnamefont {R.}~\bibnamefont
  {Banerjee}}\ and\ \bibinfo {author} {\bibfnamefont {B.~R.}\ \bibnamefont
  {Majhi}},\ }\bibfield  {title} {\bibinfo {title} {{Quantum Tunneling and Back
  Reaction}},\ }\href {https://doi.org/10.1016/j.physletb.2008.02.044}
  {\bibfield  {journal} {\bibinfo  {journal} {Phys. Lett. B}\ }\textbf
  {\bibinfo {volume} {662}},\ \bibinfo {pages} {62} (\bibinfo {year}
  {2008}{\natexlab{a}})},\ \Eprint {https://arxiv.org/abs/0801.0200}
  {arXiv:0801.0200 [hep-th]} \BibitemShut {NoStop}%
\bibitem [{\citenamefont {Banerjee}\ and\ \citenamefont
  {Majhi}(2008{\natexlab{b}})}]{Banerjee:2008cf}%
  \BibitemOpen
  \bibfield  {author} {\bibinfo {author} {\bibfnamefont {R.}~\bibnamefont
  {Banerjee}}\ and\ \bibinfo {author} {\bibfnamefont {B.~R.}\ \bibnamefont
  {Majhi}},\ }\bibfield  {title} {\bibinfo {title} {{Quantum Tunneling Beyond
  Semiclassical Approximation}},\ }\href
  {https://doi.org/10.1088/1126-6708/2008/06/095} {\bibfield  {journal}
  {\bibinfo  {journal} {JHEP}\ }\textbf {\bibinfo {volume} {06}},\ \bibinfo
  {pages} {095}},\ \Eprint {https://arxiv.org/abs/0805.2220} {arXiv:0805.2220
  [hep-th]} \BibitemShut {NoStop}%
\bibitem [{\citenamefont {Kaul}\ and\ \citenamefont
  {Majumdar}(2000)}]{Kaul:2000kf}%
  \BibitemOpen
  \bibfield  {author} {\bibinfo {author} {\bibfnamefont {R.~K.}\ \bibnamefont
  {Kaul}}\ and\ \bibinfo {author} {\bibfnamefont {P.}~\bibnamefont
  {Majumdar}},\ }\bibfield  {title} {\bibinfo {title} {{Logarithmic correction
  to the Bekenstein-Hawking entropy}},\ }\href
  {https://doi.org/10.1103/PhysRevLett.84.5255} {\bibfield  {journal} {\bibinfo
   {journal} {Phys. Rev. Lett.}\ }\textbf {\bibinfo {volume} {84}},\ \bibinfo
  {pages} {5255} (\bibinfo {year} {2000})},\ \Eprint
  {https://arxiv.org/abs/gr-qc/0002040} {arXiv:gr-qc/0002040} \BibitemShut
  {NoStop}%
\bibitem [{\citenamefont {Domagala}\ and\ \citenamefont
  {Lewandowski}(2004)}]{Domagala:2004jt}%
  \BibitemOpen
  \bibfield  {author} {\bibinfo {author} {\bibfnamefont {M.}~\bibnamefont
  {Domagala}}\ and\ \bibinfo {author} {\bibfnamefont {J.}~\bibnamefont
  {Lewandowski}},\ }\bibfield  {title} {\bibinfo {title} {{Black hole entropy
  from quantum geometry}},\ }\href
  {https://doi.org/10.1088/0264-9381/21/22/014} {\bibfield  {journal} {\bibinfo
   {journal} {Class. Quant. Grav.}\ }\textbf {\bibinfo {volume} {21}},\
  \bibinfo {pages} {5233} (\bibinfo {year} {2004})},\ \Eprint
  {https://arxiv.org/abs/gr-qc/0407051} {arXiv:gr-qc/0407051} \BibitemShut
  {NoStop}%
\bibitem [{\citenamefont {He}\ \emph {et~al.}(2010)\citenamefont {He},
  \citenamefont {Wang}, \citenamefont {Cai},\ and\ \citenamefont
  {Lin}}]{He:2010zb}%
  \BibitemOpen
  \bibfield  {author} {\bibinfo {author} {\bibfnamefont {X.}~\bibnamefont
  {He}}, \bibinfo {author} {\bibfnamefont {B.}~\bibnamefont {Wang}}, \bibinfo
  {author} {\bibfnamefont {R.-G.}\ \bibnamefont {Cai}},\ and\ \bibinfo {author}
  {\bibfnamefont {C.-Y.}\ \bibnamefont {Lin}},\ }\bibfield  {title} {\bibinfo
  {title} {{Signature of the black hole phase transition in quasinormal
  modes}},\ }\href {https://doi.org/10.1016/j.physletb.2010.04.006} {\bibfield
  {journal} {\bibinfo  {journal} {Phys. Lett. B}\ }\textbf {\bibinfo {volume}
  {688}},\ \bibinfo {pages} {230} (\bibinfo {year} {2010})},\ \Eprint
  {https://arxiv.org/abs/1002.2679} {arXiv:1002.2679 [hep-th]} \BibitemShut
  {NoStop}%
\bibitem [{\citenamefont {Liu}\ \emph {et~al.}(2014)\citenamefont {Liu},
  \citenamefont {Zou},\ and\ \citenamefont {Wang}}]{Liu:2014gvf}%
  \BibitemOpen
  \bibfield  {author} {\bibinfo {author} {\bibfnamefont {Y.}~\bibnamefont
  {Liu}}, \bibinfo {author} {\bibfnamefont {D.-C.}\ \bibnamefont {Zou}},\ and\
  \bibinfo {author} {\bibfnamefont {B.}~\bibnamefont {Wang}},\ }\bibfield
  {title} {\bibinfo {title} {{Signature of the Van der Waals like small-large
  charged AdS black hole phase transition in quasinormal modes}},\ }\href
  {https://doi.org/10.1007/JHEP09(2014)179} {\bibfield  {journal} {\bibinfo
  {journal} {JHEP}\ }\textbf {\bibinfo {volume} {09}},\ \bibinfo {pages}
  {179}},\ \Eprint {https://arxiv.org/abs/1405.2644} {arXiv:1405.2644 [hep-th]}
  \BibitemShut {NoStop}%
\bibitem [{\citenamefont {Chabab}\ \emph {et~al.}(2016)\citenamefont {Chabab},
  \citenamefont {El~Moumni}, \citenamefont {Iraoui},\ and\ \citenamefont
  {Masmar}}]{Chabab:2016cem}%
  \BibitemOpen
  \bibfield  {author} {\bibinfo {author} {\bibfnamefont {M.}~\bibnamefont
  {Chabab}}, \bibinfo {author} {\bibfnamefont {H.}~\bibnamefont {El~Moumni}},
  \bibinfo {author} {\bibfnamefont {S.}~\bibnamefont {Iraoui}},\ and\ \bibinfo
  {author} {\bibfnamefont {K.}~\bibnamefont {Masmar}},\ }\bibfield  {title}
  {\bibinfo {title} {{Behavior of quasinormal modes and high dimension
  RN\textendash{}AdS black hole phase transition}},\ }\href
  {https://doi.org/10.1140/epjc/s10052-016-4518-6} {\bibfield  {journal}
  {\bibinfo  {journal} {Eur. Phys. J. C}\ }\textbf {\bibinfo {volume} {76}},\
  \bibinfo {pages} {676} (\bibinfo {year} {2016})},\ \Eprint
  {https://arxiv.org/abs/1606.08524} {arXiv:1606.08524 [hep-th]} \BibitemShut
  {NoStop}%
\bibitem [{\citenamefont {Zou}\ \emph {et~al.}(2017)\citenamefont {Zou},
  \citenamefont {Liu},\ and\ \citenamefont {Yue}}]{Zou:2017juz}%
  \BibitemOpen
  \bibfield  {author} {\bibinfo {author} {\bibfnamefont {D.-C.}\ \bibnamefont
  {Zou}}, \bibinfo {author} {\bibfnamefont {Y.}~\bibnamefont {Liu}},\ and\
  \bibinfo {author} {\bibfnamefont {R.-H.}\ \bibnamefont {Yue}},\ }\bibfield
  {title} {\bibinfo {title} {{Behavior of quasinormal modes and Van der
  Waals-like phase transition of charged AdS black holes in massive gravity}},\
  }\href {https://doi.org/10.1140/epjc/s10052-017-4937-z} {\bibfield  {journal}
  {\bibinfo  {journal} {Eur. Phys. J. C}\ }\textbf {\bibinfo {volume} {77}},\
  \bibinfo {pages} {365} (\bibinfo {year} {2017})},\ \Eprint
  {https://arxiv.org/abs/1702.08118} {arXiv:1702.08118 [gr-qc]} \BibitemShut
  {NoStop}%
\bibitem [{\citenamefont {Wei}\ and\ \citenamefont {Liu}(2018)}]{Wei:2017mwc}%
  \BibitemOpen
  \bibfield  {author} {\bibinfo {author} {\bibfnamefont {S.-W.}\ \bibnamefont
  {Wei}}\ and\ \bibinfo {author} {\bibfnamefont {Y.-X.}\ \bibnamefont {Liu}},\
  }\bibfield  {title} {\bibinfo {title} {{Photon orbits and thermodynamic phase
  transition of $d$-dimensional charged AdS black holes}},\ }\href
  {https://doi.org/10.1103/PhysRevD.97.104027} {\bibfield  {journal} {\bibinfo
  {journal} {Phys. Rev. D}\ }\textbf {\bibinfo {volume} {97}},\ \bibinfo
  {pages} {104027} (\bibinfo {year} {2018})},\ \Eprint
  {https://arxiv.org/abs/1711.01522} {arXiv:1711.01522 [gr-qc]} \BibitemShut
  {NoStop}%
\bibitem [{\citenamefont {Zhang}\ \emph {et~al.}(2019)\citenamefont {Zhang},
  \citenamefont {Han}, \citenamefont {Jiang},\ and\ \citenamefont
  {Liu}}]{Zhang:2019tzi}%
  \BibitemOpen
  \bibfield  {author} {\bibinfo {author} {\bibfnamefont {M.}~\bibnamefont
  {Zhang}}, \bibinfo {author} {\bibfnamefont {S.-Z.}\ \bibnamefont {Han}},
  \bibinfo {author} {\bibfnamefont {J.}~\bibnamefont {Jiang}},\ and\ \bibinfo
  {author} {\bibfnamefont {W.-B.}\ \bibnamefont {Liu}},\ }\bibfield  {title}
  {\bibinfo {title} {{Circular orbit of a test particle and phase transition of
  a black hole}},\ }\href {https://doi.org/10.1103/PhysRevD.99.065016}
  {\bibfield  {journal} {\bibinfo  {journal} {Phys. Rev. D}\ }\textbf {\bibinfo
  {volume} {99}},\ \bibinfo {pages} {065016} (\bibinfo {year} {2019})},\
  \Eprint {https://arxiv.org/abs/1903.08293} {arXiv:1903.08293 [hep-th]}
  \BibitemShut {NoStop}%
\bibitem [{\citenamefont {Zhang}\ and\ \citenamefont
  {Guo}(2020)}]{Zhang:2019glo}%
  \BibitemOpen
  \bibfield  {author} {\bibinfo {author} {\bibfnamefont {M.}~\bibnamefont
  {Zhang}}\ and\ \bibinfo {author} {\bibfnamefont {M.}~\bibnamefont {Guo}},\
  }\bibfield  {title} {\bibinfo {title} {{Can shadows reflect phase structures
  of black holes?}},\ }\href {https://doi.org/10.1140/epjc/s10052-020-8389-5}
  {\bibfield  {journal} {\bibinfo  {journal} {Eur. Phys. J. C}\ }\textbf
  {\bibinfo {volume} {80}},\ \bibinfo {pages} {790} (\bibinfo {year} {2020})},\
  \Eprint {https://arxiv.org/abs/1909.07033} {arXiv:1909.07033 [gr-qc]}
  \BibitemShut {NoStop}%
\bibitem [{\citenamefont {Belhaj}\ \emph {et~al.}(2020)\citenamefont {Belhaj},
  \citenamefont {Chakhchi}, \citenamefont {El~Moumni}, \citenamefont
  {Khalloufi},\ and\ \citenamefont {Masmar}}]{Belhaj:2020nqy}%
  \BibitemOpen
  \bibfield  {author} {\bibinfo {author} {\bibfnamefont {A.}~\bibnamefont
  {Belhaj}}, \bibinfo {author} {\bibfnamefont {L.}~\bibnamefont {Chakhchi}},
  \bibinfo {author} {\bibfnamefont {H.}~\bibnamefont {El~Moumni}}, \bibinfo
  {author} {\bibfnamefont {J.}~\bibnamefont {Khalloufi}},\ and\ \bibinfo
  {author} {\bibfnamefont {K.}~\bibnamefont {Masmar}},\ }\bibfield  {title}
  {\bibinfo {title} {{Thermal Image and Phase Transitions of Charged AdS Black
  Holes using Shadow Analysis}},\ }\href
  {https://doi.org/10.1142/S0217751X20501705} {\bibfield  {journal} {\bibinfo
  {journal} {Int. J. Mod. Phys. A}\ }\textbf {\bibinfo {volume} {35}},\
  \bibinfo {pages} {2050170} (\bibinfo {year} {2020})},\ \Eprint
  {https://arxiv.org/abs/2005.05893} {arXiv:2005.05893 [gr-qc]} \BibitemShut
  {NoStop}%
\bibitem [{\citenamefont {Du}\ \emph {et~al.}(2023)\citenamefont {Du},
  \citenamefont {Li}, \citenamefont {Liu},\ and\ \citenamefont
  {Zhang}}]{Du:2022quq}%
  \BibitemOpen
  \bibfield  {author} {\bibinfo {author} {\bibfnamefont {Y.-Z.}\ \bibnamefont
  {Du}}, \bibinfo {author} {\bibfnamefont {H.-F.}\ \bibnamefont {Li}}, \bibinfo
  {author} {\bibfnamefont {F.}~\bibnamefont {Liu}},\ and\ \bibinfo {author}
  {\bibfnamefont {L.-C.}\ \bibnamefont {Zhang}},\ }\bibfield  {title} {\bibinfo
  {title} {{Photon orbits and phase transition for non-linear charged anti-de
  Sitter black holes}},\ }\href {https://doi.org/10.1007/JHEP01(2023)137}
  {\bibfield  {journal} {\bibinfo  {journal} {JHEP}\ }\textbf {\bibinfo
  {volume} {01}},\ \bibinfo {pages} {137}},\ \Eprint
  {https://arxiv.org/abs/2204.01007} {arXiv:2204.01007 [hep-th]} \BibitemShut
  {NoStop}%
\bibitem [{\citenamefont {Guo}\ \emph {et~al.}(2022)\citenamefont {Guo},
  \citenamefont {Lu}, \citenamefont {Mu},\ and\ \citenamefont
  {Wang}}]{Guo:2022kio}%
  \BibitemOpen
  \bibfield  {author} {\bibinfo {author} {\bibfnamefont {X.}~\bibnamefont
  {Guo}}, \bibinfo {author} {\bibfnamefont {Y.}~\bibnamefont {Lu}}, \bibinfo
  {author} {\bibfnamefont {B.}~\bibnamefont {Mu}},\ and\ \bibinfo {author}
  {\bibfnamefont {P.}~\bibnamefont {Wang}},\ }\bibfield  {title} {\bibinfo
  {title} {{Probing phase structure of black holes with Lyapunov exponents}},\
  }\href {https://doi.org/10.1007/JHEP08(2022)153} {\bibfield  {journal}
  {\bibinfo  {journal} {JHEP}\ }\textbf {\bibinfo {volume} {08}},\ \bibinfo
  {pages} {153}},\ \Eprint {https://arxiv.org/abs/2205.02122} {arXiv:2205.02122
  [gr-qc]} \BibitemShut {NoStop}%
\bibitem [{\citenamefont {Yang}\ \emph {et~al.}(2023)\citenamefont {Yang},
  \citenamefont {Tao}, \citenamefont {Mu},\ and\ \citenamefont
  {He}}]{Yang:2023hci}%
  \BibitemOpen
  \bibfield  {author} {\bibinfo {author} {\bibfnamefont {S.}~\bibnamefont
  {Yang}}, \bibinfo {author} {\bibfnamefont {J.}~\bibnamefont {Tao}}, \bibinfo
  {author} {\bibfnamefont {B.}~\bibnamefont {Mu}},\ and\ \bibinfo {author}
  {\bibfnamefont {A.}~\bibnamefont {He}},\ }\bibfield  {title} {\bibinfo
  {title} {{Lyapunov exponents and phase transitions of Born-Infeld AdS black
  holes}},\ }\href {https://doi.org/10.1088/1475-7516/2023/07/045} {\bibfield
  {journal} {\bibinfo  {journal} {JCAP}\ }\textbf {\bibinfo {volume} {07}},\
  \bibinfo {pages} {045}},\ \Eprint {https://arxiv.org/abs/2304.01877}
  {arXiv:2304.01877 [gr-qc]} \BibitemShut {NoStop}%
\bibitem [{\citenamefont {Lyu}\ \emph {et~al.}(2024)\citenamefont {Lyu},
  \citenamefont {Tao},\ and\ \citenamefont {Wang}}]{Lyu:2023sih}%
  \BibitemOpen
  \bibfield  {author} {\bibinfo {author} {\bibfnamefont {X.}~\bibnamefont
  {Lyu}}, \bibinfo {author} {\bibfnamefont {J.}~\bibnamefont {Tao}},\ and\
  \bibinfo {author} {\bibfnamefont {P.}~\bibnamefont {Wang}},\ }\bibfield
  {title} {\bibinfo {title} {{Probing the thermodynamics of charged Gauss
  Bonnet AdS black holes with the Lyapunov exponent}},\ }\href
  {https://doi.org/10.1140/epjc/s10052-024-13354-9} {\bibfield  {journal}
  {\bibinfo  {journal} {Eur. Phys. J. C}\ }\textbf {\bibinfo {volume} {84}},\
  \bibinfo {pages} {974} (\bibinfo {year} {2024})},\ \Eprint
  {https://arxiv.org/abs/2312.11912} {arXiv:2312.11912 [gr-qc]} \BibitemShut
  {NoStop}%
\bibitem [{\citenamefont {Kumara}\ \emph {et~al.}(2024)\citenamefont {Kumara},
  \citenamefont {Punacha},\ and\ \citenamefont {Ali}}]{Kumara:2024obd}%
  \BibitemOpen
  \bibfield  {author} {\bibinfo {author} {\bibfnamefont {A.~N.}\ \bibnamefont
  {Kumara}}, \bibinfo {author} {\bibfnamefont {S.}~\bibnamefont {Punacha}},\
  and\ \bibinfo {author} {\bibfnamefont {M.~S.}\ \bibnamefont {Ali}},\
  }\bibfield  {title} {\bibinfo {title} {{Lyapunov exponents and phase
  structure of Lifshitz and hyperscaling violating black holes}},\ }\href
  {https://doi.org/10.1088/1475-7516/2024/07/061} {\bibfield  {journal}
  {\bibinfo  {journal} {JCAP}\ }\textbf {\bibinfo {volume} {07}},\ \bibinfo
  {pages} {061}},\ \Eprint {https://arxiv.org/abs/2401.05181} {arXiv:2401.05181
  [gr-qc]} \BibitemShut {NoStop}%
\bibitem [{\citenamefont {Hale}\ \emph {et~al.}(2024)\citenamefont {Hale},
  \citenamefont {Kubiznak},\ and\ \citenamefont
  {Men\v{s}\'\i{}kov\'a}}]{Hale:2024lzh}%
  \BibitemOpen
  \bibfield  {author} {\bibinfo {author} {\bibfnamefont {T.}~\bibnamefont
  {Hale}}, \bibinfo {author} {\bibfnamefont {D.}~\bibnamefont {Kubiznak}},\
  and\ \bibinfo {author} {\bibfnamefont {J.}~\bibnamefont
  {Men\v{s}\'\i{}kov\'a}},\ }\bibfield  {title} {\bibinfo {title} {{Optical
  properties of black holes in regularized Maxwell theory}},\ }\href
  {https://doi.org/10.1103/PhysRevD.109.084061} {\bibfield  {journal} {\bibinfo
   {journal} {Phys. Rev. D}\ }\textbf {\bibinfo {volume} {109}},\ \bibinfo
  {pages} {084061} (\bibinfo {year} {2024})},\ \Eprint
  {https://arxiv.org/abs/2401.16259} {arXiv:2401.16259 [gr-qc]} \BibitemShut
  {NoStop}%
\bibitem [{\citenamefont {Promsiri}\ \emph {et~al.}(2025)\citenamefont
  {Promsiri}, \citenamefont {Horinouchi},\ and\ \citenamefont
  {Hirunsirisawat}}]{Promsiri:2024hrl}%
  \BibitemOpen
  \bibfield  {author} {\bibinfo {author} {\bibfnamefont {C.}~\bibnamefont
  {Promsiri}}, \bibinfo {author} {\bibfnamefont {W.}~\bibnamefont
  {Horinouchi}},\ and\ \bibinfo {author} {\bibfnamefont {E.}~\bibnamefont
  {Hirunsirisawat}},\ }\bibfield  {title} {\bibinfo {title} {{Observing black
  hole phase transitions in extended phase space and holographic thermodynamics
  approaches from optical features}},\ }\href
  {https://doi.org/10.1140/epjc/s10052-025-14221-x} {\bibfield  {journal}
  {\bibinfo  {journal} {Eur. Phys. J. C}\ }\textbf {\bibinfo {volume} {85}},\
  \bibinfo {pages} {484} (\bibinfo {year} {2025})},\ \Eprint
  {https://arxiv.org/abs/2409.01582} {arXiv:2409.01582 [gr-qc]} \BibitemShut
  {NoStop}%
\end{thebibliography}%

\end{document}